\RequirePackage{fix-cm}
\documentclass[onecolumn,authoryear]{els-mrw}

\usepackage{amssymb,amsmath,graphicx,bm,mathrsfs,aas_macros,soul,physics,comment}
\usepackage{subcaption}
\usepackage{float}
\usepackage{helvet}
\usepackage{xcolor}
\makeatletter
\def\@authfoot{}
\makeatother
\newcommand{\be}{\begin{equation}} 
\newcommand{\ee}{\end{equation}}
\newcommand{\bea}{\begin{eqnarray}}
\newcommand{\eea}{\end{eqnarray}}

\newcommand{\vecp}{{\bm p}}

\newcommand{\eg}{{\it e.g.}}

\usepackage[breaklinks,colorlinks=true]{hyperref}
\hypersetup{
  colorlinks=true,
  linkcolor=black,
  citecolor=blue,
  urlcolor=blue
}

\begin{document}

\chapter{Superfluidity and Vortex Dynamics
 in Neutron Stars}\label{chap1}

\author[1]{Bennett Link}
\address[1]{\orgname{Department of Physics},
\orgdiv{Montana State University},
\orgaddress{Bozeman, MT 59717, USA}}

\author[2,3]{Armen Sedrakian}
\address[2]{\orgname{Institute of Theoretical Physics}, \orgdiv{University of Wroc\l{}aw}, \orgaddress{50-204 Wroc\l{}aw, Poland}}
\address[3]{\orgname{Frankfurt Institute for Advanced Studies}, \orgaddress{60438 Frankfurt am Main, Germany}}


\maketitle


\begin{glossary}[Nomenclature]
  \begin{tabular}{@{}lp{34pc}@{}}
$\rho_s $:&effective one-component neutron-superfluid mass density in Section 3\\
$\rho_{\alpha\beta}$:& multifluid entrainment matrix coefficients in Section 4\\
$\boldsymbol v_s $:& one-component neutron-superfluid velocity\\
$\boldsymbol v_c $:& crust/charged (normal)-component velocity\\
$\boldsymbol v_n,\boldsymbol v_p $:& neutron/proton condensate velocities in the core\\
$\boldsymbol v_v $:& vortex-line velocity\\
$\kappa_{\rm GL}$:& Ginzburg--Landau parameter\\
$\omega_{\rm lag}$:& rotational lag\\
$\bm{\omega}$:& vorticity,   $\omega_j,\quad j = s, T, \dots$
    mode frequency\\
$\kappa $:& neutron circulation quantum\\    
$\Phi_0 $:& flux quantum,  $\Phi_n $: neutron-vortex fractional flux\\
$ H_{cm},H_{c1},H_{c2}$:& thermodynamic, lower, and upper critical
    fields\\
$\boldsymbol H, \boldsymbol B$:& magnetic-field intensity and magnetic induction\\ 
${\boldsymbol \Omega}_c, {\boldsymbol \Omega}_s$:& angular velocities, $I_c$, $I_s$: moments of inertia of the crust/charged component and neutron superfluid, respectively.\\
$\beta$, $\beta'$: &mutual friction coefficients, $\theta$:
 dissipation angle
 \end{tabular}
\end{glossary}

\begin{abstract}[Abstract]
  Neutron stars contain several forms of quantum condensed matter
  whose microscopic properties control macroscopic rotational dynamics
  and magnetic behavior of these fascinating objects. This review
  surveys superfluidity and superconductivity in compact stars, with
  emphasis on phenomena associated with quantized vorticity and
  magnetic-flux structures, and the possible connections to observed
  phenomena. We first summarize the microphysics of nucleonic pairing,
  including spin-singlet $^1S_0$ neutron pairing in the inner crust,
  proton superconductivity in the outer core, and spin-triplet
  $^3P_2$--$^3F_2$ neutron pairing at higher densities, together with
  the principal many-body uncertainties affecting the corresponding
  pairing gaps. We then discuss the dynamics of neutron vortices,
  including pinning, vortex creep, and dissipative motion, and the
  role of vortex dynamics in angular-momentum exchange between the
  superfluid and the observable crustal component. We give special
  attention to proton flux tubes in type-II superconducting cores, the
  possible realization of type-I superconductivity, and
  vortex--flux-tube interactions. We also review collective rotational
  phenomena, including Tkachenko oscillations of the vortex lattice
  and free precession, and their possible relation to long-term
  variability in pulsar timing. Finally, we discuss the possible
  deconfinement of hadronic matter into quark matter, the formation of
  color-superconducting phases, and the topological defects associated
  with these phases, together with their possible observational
  consequences. Throughout the review, we identify key open questions
  connecting microscopic pairing, mesoscopic defect dynamics, and
  observable neutron-star phenomena.
\end{abstract}

\begin{BoxTypeA}[chap:superfluid:keypoints]{Key points}
\begin{itemize}
    \item Neutron-star interiors contain quantum condensates: neutron superfluids in the crust and core, proton superconductors in the outer core, and possibly hyperonic or color-superconducting condensates at higher densities.

    \item The density dependence and symmetry of nucleonic pairing determine the relevant superfluid domains: neutrons pair mainly in the ${}^1S_0$ channel at low density and in the anisotropic ${}^3P_2$--${}^3F_2$ channel at higher density, while protons form a ${}^1S_0$ superconductor in the core.

    \item Quantized neutron vortices carry angular momentum in the rotating superfluid. Their pinning, unpinning, creep, and dissipative motion control angular-momentum exchange with the crust and provide the microphysical basis for glitches, post-glitch relaxation, and internal heating.

    \item Proton superconductivity organizes magnetic flux either into quantized flux tubes in type-II regions or into macroscopic normal domains in type-I regions. Entrainment magnetizes neutron vortices and couples vortex dynamics to magnetic-field evolution through vortex--flux-tube interactions.

    \item Interfaces between distinct condensates, magnetar-strength fields, Tkachenko modes, and free precession connect microscopic pairing and vortex physics to observable phenomena such as thermal evolution, magnetic activity, long-term timing variability, and possibly continuous gravitational-wave emission.
\end{itemize}
\end{BoxTypeA}

\section{Introduction}
\label{sec:introduction}

Neutron stars are a striking manifestation of fermionic degeneracy at a macroscopic scale. Observations of neutron stars in all parts of the electromagnetic spectrum, and gravitational waves, bear the imprints of the distinct quantum fluids these objects are predicted to contain. Neutrinos play an important role in neutron-star cooling.

At temperatures well below the relevant Fermi
energies, attractive components of the nuclear interaction induce
nucleonic pairing through the Bardeen-Cooper-Schrieffer (BCS)
mechanism: neutrons form Cooper pairs in which the constituent fermion pairs are correlated on scales much larger
than the interparticle distances. This leads to superfluidity and the creation of 
{\sl vortex lines} that govern dissipative coupling between the superfluid and the normal matter. The same pairing mechanism 
for protons leads to 
superconductivity in the outer core of the star. It is now well
established theoretically, and there is considerable  observational support
for, these condensates altering the thermal, magnetic, transport, and
rotational properties of compact stars. This, in turn, forms the
physical basis for understanding various phenomena, ranging from the
long-term cooling history of neutron stars  to rotational
irregularities in pulsars observed as glitches superimposed on their
secular deceleration.

This article focuses on the aspects of pairing physics that are most
directly connected with neutron-star rotational dynamics and magnetic
properties. We begin with the microphysics of nucleonic pairing in
compact stars, emphasizing the density dependence of the dominant
pairing channels,\footnote{We use the standard spectroscopic notation
$^{2S+1}L_J$, where $S$, $L$, and $J$ denote, respectively, the total
spin, orbital angular momentum, and total angular momentum of the
nucleon pair.  The orbital states $L=0,1,2,3,\ldots$ are denoted by
$S,P,D,F,\ldots$.  Thus, $^1S_0$ denotes a spin-singlet state with
$S=L=J=0$, whereas $^3P_2$--$^3F_2$ denotes the tensor-coupled
spin-triplet channels with $S=1$, $J=2$, and $L=1$ or $3$.} the role of medium-induced corrections, and the
present uncertainties in proton $^1S_0$ and neutron $^3P_2$--$^3F_2$
gaps. We then turn to quantized vorticity. A rotating neutron
superfluid is threaded by vortex lines, topological defects with quantized circulation,
much like terrestrial superfluids such as liquid helium or ultra-cold
atomic vapors. Their interaction with the crustal lattice and magnetic
flux tubes provides the microscopic origin of friction between the
superfluid and the normal component, the phenomenon of vortex creep,
and ultimately the angular-momentum exchange between stellar
components.

The proton superconducting component introduces a second class of
topological defects that form a network of quantized flux tubes. In
type-II regions of the core, magnetic flux is carried by quantized
flux tubes, whose interaction with neutron vortices couples the
magnetic and rotational evolution of the star. We also summarize the
alternative possibility of type-I superconductivity, in which magnetic
flux resides in normal domains embedded in a superconducting
background.  We discuss the issue of interfaces between
low-density $S$-wave and high-density $P$-wave neutron condensates,
which may support a Josephson effect, or interfaces between baryonic and quark
matter, which raises the problem of connecting the topological defects
in both phases.

Finally, we discuss collective and global rotational phenomena:
Tkachenko oscillations of the vortex lattice, free precession in a
multicomponent superfluid star, and their possible relevance for
long-term quasiperiodic variations in pulsar timing. We highlight that
the microscopic pairing, vortex dynamics, and superconducting magnetic
structures cannot be treated as separate topics: together they
determine how neutron stars store, transfer  and dissipate angular
momentum over timescales ranging from glitch rises (seconds) to
long-term variability  (years).

Historically, superfluidity in neutron stars was suggested even before
the discovery of
pulsars~\citep{Bohr1958,Migdal1959,Ginzburg1965}. Many of the key
ideas were developed shortly thereafter, including the role of
quantized vortices and their connection to pulsar
glitches~\citep{Baym1969a,Baym1969b}, microscopic nucleonic
pairing~\citep{Hoffberg1970,Clark1970}, vortex lattice
modes~\citep{Ruderman1970}, and pinning and creep
in analogy with type-II superconductors~\citep{Anderson1975}.
A number of complementary reviews on the subject have appeared over
the past decade or so, and we recommend them to readers seeking a
broader overview. These include reviews of superfluidity and
neutron-star cooling~\citep{Page2013}, superfluidity in the
neutron-star crust~\citep{Pethick2010}, superfluidity and vortex
dynamics~\citep{Haskell2017},  the interplay between laboratory and
stellar condensates~\citep{Graber2017}, and microscopic pairing models~\citep{Sedrakian2019}.

This article is organized as
follows. Section~\ref{sec:pairing_microphysics} reviews the
microphysics of nucleonic and other fermionic pairing in compact
stars, emphasizing the dominant pairing channels and the principal
uncertainties in the corresponding
gaps. Section~\ref{sec:vortex_creep} discusses quantized neutron
vortices, pinning, creep, dissipative motion, and angular-momentum
exchange with the normal component.  Section~\ref{sec:flux_tubes}
examines magnetic structures in superconducting neutron-star cores,
including type-II flux tubes, possible type-I domains,
vortex--flux-tube interactions, magnetar-strength fields, and
interface phenomena. Section~\ref{sec:tkachenko_precession} reviews
collective rotational and oscillatory phenomena, including free
precession, Tkachenko modes, superfluid stellar oscillations, and
magneto-elastic crustal modes. Section~\ref{sec:quark_vorticity}
summarizes quantum vorticity and color-magnetic defects in paired
quark matter. Section~\ref{sec:conclusions} concludes with an outlook
and open problems.

\section{Microphysics of pairing in compact stars}
\label{sec:pairing_microphysics}

The interiors of neutron stars form one of the densest environments in
the current Universe, and consequently a natural laboratory for the
realization of fermionic superfluidity and superconductivity in
strongly interacting matter. The high density renders nucleonic matter
highly degenerate, with the Fermi energy being much larger than the
actual temperature of the stellar environment shortly after its birth
and rapid cooling. In the inner crust, dripped neutrons coexist with
neutron-rich nuclear clusters and an ultrarelativistic,
charge-neutralizing electron gas. Microscopic models often favor
proton numbers near the shell closures $Z\simeq40$ or $50$, although
the cluster composition and mass number vary with density and with the
treatment of shell and pasta effects. In the outer core extending
approximately to twice the nuclear saturation density, neutrons
coexist with a small $\sim 5$--$10\%$ fraction of protons and leptons
(electrons and, at higher densities, also muons) under $\beta$
equilibrium~\citep{ShapiroTeukolsky1983,Glendenning_book,weber_book}. Because
the nucleon--nucleon interaction contains attractive components in
several partial-wave channels, neutrons and protons can form Cooper
pairs near their respective Fermi surfaces, in close analogy with the
BCS mechanism of
superconductivity~\citep{Abrikosov:Fundamentals,Fetter1971}.

In a superconductor the elementary fermionic
excitations of the paired medium are quasiparticles: 
they are coherent superpositions of
particle and hole states rather than individual fermions. 
Nucleon pairing opens an energy gap 
$\Delta$ in their quasiparticle excitation spectrum, separating the paired ground state from the manifold of excited states; breaking a pair, which produces two quasiparticle excitations and dissipates energy, therefore costs at least $2\Delta$, while the same gap enforces coherence of the paired state over a length scale much larger than the interparticle spacing. The magnitude and structure of this gap are determined by the dominant attractive partial-wave component of the nuclear interaction at a given density.

\begin{figure}[tb]
\centering
    \includegraphics[width=0.7\textwidth,angle=0]{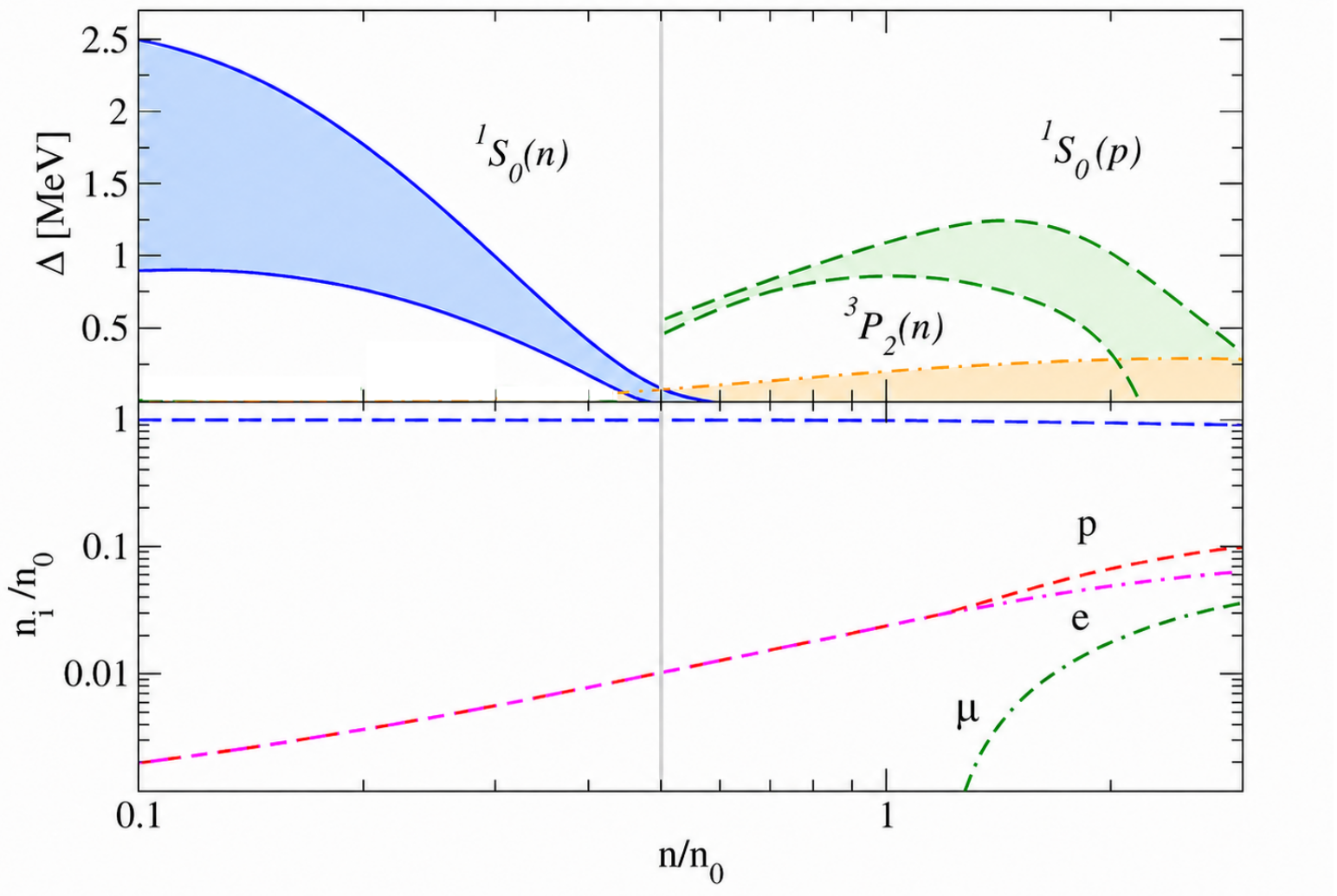} 
  \caption{\textit{Upper panel}---
  Density dependence of the nucleonic pairing gaps;  \textit{Lower panel}--- particle composition in neutron-star matter.  The shaded bands indicate representative ranges of the neutron $^{1}S_{0}$ gap at subnuclear densities, the proton $^{1}S_{0}$ gap in the core, and the neutron $^{3}P_{2}$--$^{3}F_{2}$ gap at higher densities, illustrating the substantial theoretical uncertainty in microscopic calculations of the pairing gaps. The corresponding number densities of neutrons ($n$), protons ($p$), electrons $e$, and muons $\mu$, $n_i/n_0$, are shown on the logarithmic scale, where $n_0$ is the nuclear saturation density and the horizontal axis gives the total baryon density $n/n_0$. The vertical line marks the approximate crust–core transition: neutron $^{1}S_{0}$ pairing extends from the inner crust into the vicinity of this transition, whereas proton $^{1}S_{0}$ and neutron $^{3}P_{2}$--$^{3}F_{2}$ pairing characterize different density regions of the stellar core.  Adapted from~\cite{Sedrakian2019}.}
\label{fig:gaps}
  \end{figure}

At subnuclear and near-nuclear densities\footnote{The nuclear (saturation) density is defined as the characteristic baryon density of the interior of atomic nuclei, $n_0\simeq 0.16~{\rm fm}^{-3}$ (corresponding to a mass density $\rho_0\simeq 2.7\times10^{14}~{\rm g\,cm}^{-3}$), at which the energy
per nucleon of approximately isospin-symmetric nuclear matter is
minimized.}, the dominant attractive interaction between same-isospin nucleons occurs in the spin-singlet $^1S_0$ channel. Opposite-isospin neutron-proton pairing is strongly suppressed by large isospin asymmetry\footnote{Isospin is an approximate internal symmetry of the strong interaction
that treats the proton and neutron as two states of the same nucleon,
distinguished by the third component of isospin, conventionally
assigned as $I_3=+1/2$ for the proton and $I_3=-1/2$ for the neutron.
}. Consequently, neutrons in the
inner crust and low-density outer core are expected to form a $^1S_0$
superfluid, while protons in the outer core form a $^1S_0$
superconductor, being clustered in nuclei in the crust (with the
exception of pasta phases, where they can form a continuum as in the
core). At higher densities — corresponding to the deeper layers of the
core -- the $^1S_0$ neutron interaction becomes repulsive, and neutron
pairing is expected to proceed mainly in the coupled $^3P_2$--$^3F_2$
channel. The latter corresponds to a spin-triplet, anisotropic pairing
state. This feature implies a
matrix-valued order parameter, which renders the quantitative
properties of the $P$-wave superfluid considerably more uncertain than
those of $S$-wave pairing; see Fig.~\ref{fig:gaps}.

The characteristic scale of nucleonic pairing gaps is typically of
order $\Delta \sim 0.1$--$1$ MeV, although the precise magnitude and
density dependence are sensitive to many-body effects. The strength of
the pairing is determined by the product of the density of states and
the pairing interaction. The density dependence of the density of
states for Fermi momentum $p_F$ and nucleon effective mass $m^*$ scales as $m^*p_F$, generally increasing with density according to microscopic calculations.  The
gap typically peaks at the point when the weakening of the attractive
interaction with increasing density overwhelms the increase in the
density of states. Since the internal temperatures of mature neutron
stars are generally much smaller than 1 MeV -- in the core and inner
crust by two orders of magnitude -- nucleonic pairing has profound
consequences for the low-energy equilibrium and transport properties
of dense matter. For example, pairing suppresses the specific heat of
the nucleonic component and the dominant neutrino-emission processes
involving nucleonic quasiparticle excitations close to the Fermi
surface. Close to the critical temperature, pairing also opens
additional neutrino channels associated with Cooper-pair breaking and
formation. The quantitative impact of these effects is determined
primarily by the density dependence and angular structure of the
pairing gaps in the case of $P$-wave superfluid~\citep{Page2013}.

The inner core of neutron stars may contain new baryonic degrees of
freedom once the density becomes sufficiently high. In addition to
nucleons, strange baryons, plausibly the $\Lambda$, $\Xi$, and
$\Sigma$ hyperons, as well as excited baryonic states such as the
spin-$3/2$ $\Delta$ resonances may nucleate.  Their appearance is
favored by the growth of the baryon and lepton chemical potentials
with density: at some threshold it becomes energetically advantageous
to replace energetic nucleons and leptons from the tops of their
respective Fermi surfaces by heavier baryons carrying strangeness of
suitable charge. The detailed onset densities remain model dependent
because they are sensitive to poorly constrained hyperon-nucleon,
hyperon-hyperon, and $\Delta$-nucleon interactions in dense
matter~\citep{Vidana2018,Tolos2020,Sedrakian2023}.

Among hyperons, the $\Lambda$ -- being electrically neutral,
relatively light, and possessing an attractive potential in symmetric
nuclear matter (and by model-dependent extension, also in neutron-rich
matter) -- is expected to appear first. This is indeed the case in
many hyperonic equations of state. Once a sizable $\Lambda$ population
nucleates, $\Lambda$ hyperons may form Cooper pairs and develop their
own superfluid condensate, coexisting with the neutron superfluid. The
plausibility of $\Lambda\Lambda$ pairing is supported by the existence
of double-$\Lambda$ hypernuclei and by the empirically attractive
low-energy $\Lambda\Lambda$ interaction, although the strength of this
attraction in dense matter is still uncertain and expected to be
small; consequently the gaps are expected to be comparable to or
smaller than nucleonic gaps. Cross-species pairing, such as
neutron--$\Lambda$ pairing, is suppressed both by mismatched Fermi
momenta and by the substantial (on the scale of the pairing gap)
difference in single-particle spectra (predominantly the mass and
chemical potential differences) of species. Naturally, hyperonic
pairing can qualitatively influence neutron-star cooling and transport
in a manner analogous to nucleonic superfluidity: it suppresses
neutrino-emitting reactions involving paired hyperons, modifies
specific heat and transport coefficients, and opens pair-breaking and
pair-formation neutrino channels near the critical temperature.

At still higher densities, hadrons may dissolve into deconfined quark
matter. Because quark matter at low temperature possesses Fermi
surfaces, any attractive interaction in a suitable quark--quark
channel can trigger BCS pairing~\citep{Alford2008}.  In QCD
the attractive color-antitriplet channel leads to color
superconductivity, in which quarks form diquark condensates. The
precise phase realized in neutron-star cores depends on the density,
strange-quark mass, requirements of electric and color neutrality, and
$\beta$ equilibrium. Among the most widely discussed possibilities are
the two-flavor color-superconducting phase (2SC)~\citep{Bailin1984}
and the color--flavor-locked phase (CFL)~\citep{Alford1999}.

In the 2SC phase, up and down quarks of two colors (red and green)
pair in an antisymmetric color and flavor channel, while the third
color (blue) remains unpaired or only weakly paired.  Strange quarks
may be absent over the lower range of chemical potentials relevant to
the 2SC phase or may be present as an unpaired or separately paired
component as their density and Fermi momentum are far too different to
participate in the dominant $ud$ condensate.  The 2SC phase is
therefore a natural candidate for quark matter at intermediate
densities, where deconfinement may have occurred but the strange quark
is not yet sufficiently abundant to join fully symmetric pairing. The
condensate breaks the original color gauge symmetry, with only part of
the quarks acquiring a gap in the spectrum. As a consequence, the
changes in the equation of state, neutrino emissivity, and transport
properties may be less pronounced than in a fully gapped phase, since
some quasiparticles may remain ungapped~\citep{Alford2008}.
Therefore, 2SC matter can retain relatively efficient cooling channels
unless additional pairing gaps appear in the residual modes. In
phenomenological compact-star models, a 2SC shell surrounding a denser
inner core is often considered~\citep{Alford2017}, and recent studies continue to explore
whether such phases are compatible with present mass--radius and
tidal-deformability constraints~\citep{Han2019,Gholami2025,Christian2025}.

In the asymptotically high-density limit, the favored state is expected to be the CFL phase, in which up, down, and strange quarks of all three colors participate in a highly symmetric pairing
pattern. Color and flavor rotations are locked together, and all quark
quasiparticles acquire pairing gaps in the asymptotically free
limit~\citep{Alford1999}. The CFL phase is electrically neutral without requiring a large
electron population, and it breaks baryon number symmetry, implying
that CFL quark matter is also a superfluid~\citep{Iida2002}. Its low-energy sector
contains collective modes associated with the broken symmetries, which
can dominate specific heat, transport, and neutrino emission when
fermionic excitations are strongly suppressed. In neutron stars,
however, the realization of CFL matter is hindered by the finite
strange-quark mass and the mismatch between the chemical potentials of
quarks of various flavors. Even so, CFL matter remains a viable
candidate for the deepest quark-matter cores of sufficiently massive
compact stars, and phase sequences such as hadronic matter
$\rightarrow$ 2SC $\rightarrow$ CFL are frequently discussed in
hybrid-star modeling.

\subsection{BCS formulation of nucleonic pairing}

We start the discussion with nucleonic pairing, which can be
straightforwardly extended to pairing of other types of baryons,
whereas quark pairing requires an extension to the case of
relativistic systems.

The pairing theory can be formulated using various methods, including
canonical (Bogoliubov) transformations, the Green's functions method,
or the original BCS variational Ansatz~\citep{Fetter1971,Dean2003,Sedrakian2019}. All these methods, at the mean-field level, lead to a self-consistent gap equation for the gap function $\Delta(p)$ which, for an isotropic $S$-wave condensate,
satisfies
\be
\label{eq:gap_swave_micro}
\Delta(p)
= - \frac{1}{V} \sum_{\vecp'} v_0(p,p') \frac{\Delta(p')}{2E(p')} \left[1-2f(p')\right],
\ee
where $v_0(p,p')$ is the interaction matrix element in the $^1S_0$
partial wave, $V$ is the normalization volume,
\be
E(p)=\sqrt{\varepsilon(p)^2+\Delta(p)^2}
\ee
is the quasiparticle excitation energy, $\varepsilon(p)$ is the
single-particle energy measured relative to the chemical potential in
the normal (unpaired) state, and
\be
f(p)=\frac{1}{e^{E(p)/T}+1}
\ee
is the Fermi distribution. The gap equation~\eqref{eq:gap_swave_micro}
is a nonlinear integral equation whose solution for a general
interaction potential needs to be carried out numerically, typically
via an iterative procedure.

In weak-coupling treatments, the dominant contribution to the gap
equation arises from momenta close to the Fermi surface, where the
integrand of Eq.~\eqref{eq:gap_swave_micro} is strongly peaked. In
condensed-matter systems and ultra-cold atoms the interaction can
frequently be approximated by a constant, in which case the
zero-temperature solution is analytic
\be \label{eq:Gap_Tzero}
\Delta = 2 \omega_c \exp\!\left(-\frac{1}{\vert v_0(p_F,p_F)\vert \, N(p_F)}\right),
\ee
where the cutoff $\omega_c$ is system dependent; in the neutron-star
context $\omega_c \simeq \epsilon_F$ -- the Fermi energy; $N(p_F)$ is
the density of states at the Fermi surface.  The critical temperature
is related to the zero-temperature gap by the familiar formula
$T_c \simeq 0.57\,\Delta(T=0)$ for isotropic BCS pairing $(k_B=1)$.

When realistic high-precision nucleon--nucleon interactions are used,
mean-field calculations of the $^1S_0$ gap in dilute neutron matter
show substantial consistency among different potentials, provided that
the relevant momentum range remains constrained by elastic scattering
data~\citep{Elgaroy1996,Dean2003,Gandolfi2015}.  Similar agreement is
obtained for proton $^1S_0$ pairing at the level of the bare
interaction. Effective interactions, such as Skyrme and, to a lesser
extent, Gogny forces, have also been widely employed in neutron-star
applications and often reproduce the qualitative form of the pairing
gaps~\citep{Sedrakian2003,Margueron2007,Chamel2010}.

The situation becomes more complex for neutron pairing at supranuclear
densities. In this regime the dominant attractive channel is the
coupled $^3P_2$--$^3F_2$ channel, whose gap equation involves several
angular components and tensor-coupled partial
waves~\citep{Baldo1998a,Zverev2003,SchwenkFriman2004}.  The resulting
order parameter is anisotropic, and several competing magnetic
substates may be realized. Even at the mean-field level, the
calculation is therefore more demanding than for $^1S_0$ pairing. In
addition, the relevant nucleon--nucleon interaction is less well
constrained by scattering data at the corresponding energies: elastic
neutron--neutron phase shifts are not available above laboratory
energies of $\sim 350$ MeV, making the neutron--neutron interaction in
this regime less constrained and the predicted triplet gap highly
model dependent. A further complication at large densities arises from
the presence of three-body forces, whose net contribution is model
dependent and often suppresses the pairing.

\subsection{Medium corrections beyond mean field}

Although the BCS approximation provides the conceptual starting point,
quantitative predictions for pairing gaps in neutron-star matter require
corrections arising from the strongly interacting medium. Two classes of
effects are especially important.

First, correlations in the nucleon self-energy modify the
single-particle spectrum entering the gap equation. The momentum
dependence of the self-energy is conventionally expressed through a nucleon effective mass $m^*$, which differs from the bare mass $m$ and alters the density of states near the
Fermi surface. In neutron matter at low and intermediate densities,
$m^*/m$ is commonly found in the range $0.6$--$0.8$ in the core and
closer to unity in the inner crust, although the precise value depends
on the underlying many-body scheme and adopted
interaction~\citep{Baldo1992,MuetherDickhoff2005,Ding2016a}. Since the
BCS gap depends exponentially on the density of states (as can be seen
from the zero-temperature formula~\eqref{eq:Gap_Tzero})
such changes can have a substantial impact on the magnitude of the
gap. The energy dependence of the self-energy is encoded in the
quasiparticle residue, or wave-function renormalization factor, whose
effect is channel- and scheme-dependent and results in a moderate to
strong suppression of the gap~\citep{MuetherDickhoff2005,Ding2016a}.

Second, the effective pairing interaction is modified by long-range
density and spin-density fluctuations of the
medium~\citep{ClarkKallman1976,Wambach1993}. These modifications,
termed induced interactions, arise from the polarization of the
surrounding nuclear medium. Here, polarization means the response of the many-body system to the presence/motion of a nucleon: the
surrounding Fermi sea is locally disturbed, producing fluctuations in
the particle density and spin density. A second nucleon then interacts
not only directly with the first one, but also with the disturbance
that it induces in the medium. Microscopically, this medium response is
described by particle--hole excitations and is represented
diagrammatically by particle--hole polarization diagrams. Depending on
the channel, these fluctuations can either enhance or screen the
underlying interaction. In dilute neutron matter, the net polarization
effect reduces the attraction in the $^1S_0$ pairing channel and
thereby suppresses the mean-field pairing gap significantly.
A broad body of work based on
Fermi-liquid methods and microscopic many-body calculations indicates
that the bare BCS neutron gap, typically of order several MeV at its
peak, is reduced by factors of up to roughly five once the effective
mass and medium polarization effects are
included~\citep{Wambach1993,Schwenk2003,Urban2020}.

The inclusion of such medium corrections has wide implications for
compact-star phenomenology, extending well beyond a mere refinement of the pairing problem. The gap magnitude and its density dependence feed back directly into observable quantities, creating a useful diagnostic loop: observations can in principle constrain the pairing gaps themselves. A notable example is the putative  cooling of the Cassiopeia~A supernova remnant. Its rapid temperature decline, directly observed across successive Chandra observations since the neutron star's discovery in 1999, has been interpreted as evidence for the onset of neutron superfluidity in the core. This interpretation can place model-dependent constraints on the triplet gap~\citep{Page2011,Leinson2022}. More generally, the
density region where the $^1S_0$ neutron gap is nonzero determines the extent of the crustal and low-density core superfluid domains, while its maximum value controls the characteristic temperature scale for thermal suppression of baryonic excitations. Similarly, proton pairing in the outer core influences neutrino-emission rates and therefore the thermal evolution of neutron stars.

Several complementary many-body frameworks have been developed beyond
mean-field pairing theory. The correlated-basis-function method
incorporates short-range correlations before pairing is added,
enabling summation of broad classes of cluster diagrams in both
particle-particle and particle-hole channels~\citep{Krotscheck1980,Chen1986,Fabrocini2008}.
Quantum Monte Carlo methods compute the pairing gap from odd-even energy differences in finite periodic systems, yielding important benchmarks for the $^1S_0$ gap in dilute neutron matter~\citep{Gezerlis2008,Gandolfi2022}. The self-consistent Green-function approach goes beyond the quasiparticle approximation by incorporating the full off-shell self-energy, making it
particularly suited to quantifying spectral fragmentation and finite
quasiparticle lifetimes~\citep{Ding2016a,Rios2018}.
Despite their different formulations, CBF, Monte Carlo, and SCGF
calculations broadly agree that the $^1S_0$ neutron gap is sizable but
substantially reduced relative to the bare BCS estimate, an important
result for neutron-star modeling, since the $^1S_0$ condensate is the
best-controlled superfluid component microscopically.

\subsection{Proton superconductivity in the neutron-star core}

Protons form only a small fraction of the baryonic matter in the outer
core, yet their pairing properties are of considerable astrophysical
importance. In conventional neutron-star compositions, protons occupy a
much smaller Fermi sphere than neutrons and pair predominantly in the
$^1S_0$ channel~\citep{ChaoClark1972,Takatsuka1973,Elgaroy1996,Baldo1992}.
The proton pairing gap is influenced not only by proton--proton
interactions but also by the surrounding neutron medium, so the problem
is not equivalent to that of dilute symmetric nuclear matter
\citep{Baldo2007a,Guo2019,Lim2021}.

Several medium effects may be particularly pronounced for proton pairing.
The induced interaction receives substantial contributions from the
polarization of the neutron background, and short-range correlations may
be stronger because the dilute proton component is embedded in a dense
neutron fluid~\citep{Baldo2007a,Guo2019}. Moreover, the proton
single-particle spectral properties can deviate substantially from those
of well-defined quasiparticles~\citep{Baldo2007a}. These features make
the proton gap one of the less settled ingredients in microscopic
compact-star modeling, and its uncertainty propagates directly into
predictions for neutron-star cooling, since proton superconductivity
suppresses modified Urca and related charged-current neutrino processes
over large regions of the core~\citep{Elgaroy1996,Page2013}.

Essentially the same framework applies to $\Lambda\Lambda$ pairing,
but two additional factors of uncertainty arise. First, the
$\Lambda\Lambda$ interaction in free space is only weakly constrained
--- primarily by double-$\Lambda$ hypernuclear data --- and its
behavior in dense matter is even more uncertain~\citep{Balberg1998,WangShen2010,Tolos2020}.
Second, medium-polarization and self-energy corrections analogous to
those discussed above for nucleons are largely unexplored in the
hyperonic sector. Microscopic calculations nonetheless suggest that
$\Lambda$ pairing gaps may be of order a fraction of an MeV, and that
$\Lambda$ superfluidity can influence neutron-star cooling in a manner
qualitatively similar to nucleonic
superfluidity~\citep{Balberg1998,Takatsuka2006,WangShen2010,Raduta2018}.

The situation is even more uncertain for other hyperon species.
$\Xi$--$\Xi$ interactions appear to be attractive, suggesting that
$\Xi$ hyperons could in principle form Cooper pairs, but the interaction
remains poorly determined both experimentally and theoretically~\citep{Raduta2018}.
For $\Sigma$ hyperons, current evidence points to repulsive interactions
in dense matter, making $\Sigma$ pairing unlikely, though this conclusion
is not firmly
established~\citep{Tolos2020,Sedrakian2023}.

\subsection{Triplet neutron pairing at high density}

At densities above the domain of $^1S_0$ neutron superfluidity, pairing
may persist in the coupled $^3P_2$--$^3F_2$ channel
\citep{Hoffberg1970,Tamagaki1970,Takatsuka1972}. As already noted
in Section~\ref{sec:pairing_microphysics}, this state differs
qualitatively from the isotropic $S$-wave condensate: it involves
spin-triplet Cooper pairs with an anisotropic gap that may possess nodes
or deep minima, strongly affecting low-temperature thermodynamic and
transport properties~\citep{Khodel1998,Zverev2003}.
The multi-component nature of the order parameter admits several
candidate phases with comparable free energies, whose selection depends
on temperature, density, and magnetic field
\citep{Khodel2001,Zverev2003,Mizushima2017,Mizushima2021}.

The triplet gap remains one of the central open problems in the
microphysics of neutron-star interiors. The uncertainties --- poor
constraints on the high-momentum nucleon--nucleon interaction, strong
medium suppression relative to bare BCS predictions, and the
near-degeneracy of competing order-parameter phases --- are larger than
for any other pairing channel discussed in this section
\citep{SchwenkFriman2004,Ding2016a,Drischler2017,Krotscheck2024}.
Their resolution will likely require both improved nuclear-force models
and tighter observational constraints from pulsar timing and cooling data.

\subsection{Status and open issues}

The microscopic theory of nucleonic pairing in compact stars has reached
varying levels of maturity in different density regimes. The $^1S_0$
neutron gap in dilute neutron matter is comparatively well understood:
modern many-body methods agree that the gap is substantial but reduced
relative to simple BCS predictions by self-energy and polarization
effects. The proton $^1S_0$ gap is less firmly established because of
the complex influence of the surrounding neutron medium. The
high-density neutron $^3P_2$--$^3F_2$ gap remains the most uncertain
of the standard nucleonic pairing channels, both in magnitude and in
the structure of the favored condensate. In the hyperonic sector,
$\Lambda$ pairing is plausible but poorly constrained, while the status
of $\Xi$ and $\Sigma$ pairing remains essentially open.

Further progress requires improved treatments of induced interactions,
off-shell self-energy effects, and short-range correlations within
consistent many-body frameworks. Non-adiabatic effects, analogous to
those familiar from strong-coupling Eliashberg theory in condensed-matter
systems, may also become relevant for pairing mediated or renormalized
by collective modes of dense nuclear matter. These issues bear directly
on neutron-star phenomenology, since pairing gaps enter the modeling of
cooling, neutrino emission, transport, and the rotational dynamics of
compact stars.

The principal open questions in this area can be summarized as follows:
\begin{itemize}
    \item What is the quantitative impact of medium polarization on the
    proton $^1S_0$ gap, and how sensitively does it depend on the
    proton fraction and the neutron-matter equation of state?

 \item Does the $^3P_2-^3F_2$ neutron condensate survive strong
    medium suppression at the densities relevant to neutron-star cores,
    and if so, which magnetic substate or mixture of substates is
    realized?

  \item How large are non-adiabatic and strong-coupling corrections to
    nucleonic pairing, and can Eliashberg-type treatments of the
    energy dependence of the gap be systematically extended to dense
    nuclear matter?

    \item What are the pairing gaps of $\Lambda$, $\Xi$, and $\Sigma$
    hyperons in dense matter, and what is their net effect on neutron-star
    cooling and rotation?

  \item Can neutron-star observations--in particular cooling curves,
    glitch statistics, post-glitch relaxation, and gravitational-wave
    constraints on the equation of state and stellar moment of
    inertia--place meaningful bounds on pairing gaps across the
    different density regimes?
  \end{itemize}

\section{Superfluid rotation and vortex motion}
\label{sec:vortex_creep}

\begin{figure}[t]
\centering\includegraphics[width=.6\linewidth]{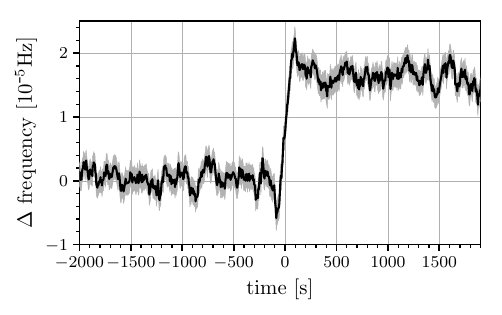}
\caption{The 2016 Vela pulsar glitch at high time resolution. Shown is
  the change in spin frequency relative to the pre-glitch timing model
  versus time. The spin-up was constrained to occur in less than
  $12.6\,\mathrm{s}$. Adapted from \cite{Ashton2019}. }
\label{glitch}
\end{figure}

A long-standing mystery is the cause of spin glitches observed in
neutron stars. These events involve fractional changes in the star's
spin frequency of typically $\Delta\Omega/\Omega\sim 10^{-6}$; see
Fig.~\ref{glitch}. Glitches and post-glitch recovery have been ascribed to variable coupling between the neutron star crust and its interior
superfluid~\citep{Anderson1975,Alpar1984a,Link1996,Haskell2015,Larson2002}.  
We now discuss features of superfluid dynamics and some possible effects on 
the rotational dynamics of neutron stars, beginning with a description of the quantum vortices that permeate a rotating superfluid.

\subsection{Superfluid rotation and angular momentum}
\label{superfluid-rotation}

The remarkable properties of a superfluid arise from its \textit{phase
coherence}. We write the superfluid order parameter as
\be
\Psi(\bm{r})=\sqrt{\rho_s(\bm{r})}\,e^{i\chi(\bm{r})},
\label{sf_wavefunction}
\ee
where $\rho_s$ is the superfluid mass density and $\chi$ is the
superfluid phase. With this normalization, the mass-current density is
\bea
\bm{j}(\bm{r})
&=&\frac{\hbar}{2Mi}\left[
\Psi^*(\bm{r})\nabla\Psi(\bm{r})
-\Psi(\bm{r})\nabla\Psi^*(\bm{r})\right]\nonumber\\
&=&\frac{\rho_s\hbar}{M}\nabla\chi(\bm{r})
=\rho_s\bm{v}_s(\bm{r}),
\eea
which implies
\be
\bm{v}_s(\bm{r})=\frac{\hbar}{M}\nabla\chi(\bm{r}).
\ee
Here $M=2m_n$ is the mass of a neutron Cooper pair and $m_n$ is the
neutron mass. Note that here we use an effective one-component
description of a superfluid with density $\rho_s$; the multifluid generalization is
introduced in  Sec.~\ref{sec:flux_tubes}.

Wherever the phase is single-valued and smooth,
\be
\nabla\times\bm{v}_s=0.
\ee
Consequently, the circulation vanishes for any contractible closed
contour that does not enclose a vortex. In this local sense, the
superfluid flow is \textit{irrotational}.

Above a critical value of the rotation rate of the superfluid's
container, the superfluid must acquire finite angular momentum
\citep{Landau1980}. The superfluid rotates by establishing an array of
quantized vortices, within which superfluidity is destroyed, and about
which the superfluid circulates. The circulation about any one vortex
now takes a fixed value:
\bea\label{eq:circulation}
\oint_C \bm{v}_s\cdot d\bm{l}
=\frac{\hbar}{M} (\Delta\chi)_C
=n\kappa,
\qquad
\kappa\equiv\frac{h}{M}=\frac{\pi\hbar}{m_n},
\qquad n\in\mathbb{Z}.
\eea
Here $(\Delta\chi)_C$ denotes the net change of the condensate phase
upon one complete traversal of contour $C$; single-valuedness of the order parameter requires 
$(\Delta\chi)_C=2\pi n$, where $n$ is the integer winding number.
For a singly-quantized, isolated 
vortex ($n=1$), the superfluid
velocity field is purely azimuthal and falls off as
\bea \label{eq:v_rn}
\bm{v}_s(d)=\frac{\kappa}{2\pi d}\,\hat{\bm{\phi}},
\eea
where $d$ is the distance from the vortex axis and $\hat{\bm{\phi}}$
is oriented in the right-handed sense about the circulation vector
$\bm{\kappa} = \kappa \bm{\hat \kappa} $. The phase winding of such a vortex is $2\pi$, and each Cooper pair (of mass $2m_n$) carries $\hbar$ of angular
momentum about the vortex. Superfluid vortices are stable,
topologically protected configurations. Dissipation associated with
vortex motion arises through interactions with quasiparticle
excitations of the condensate and other unpaired components, such as
charged leptons.

\begin{figure*}[t!]
    \centering
    \begin{subfigure}[t]{0.45\textwidth}
        \centering
        \includegraphics[height=4cm,width=7cm ] {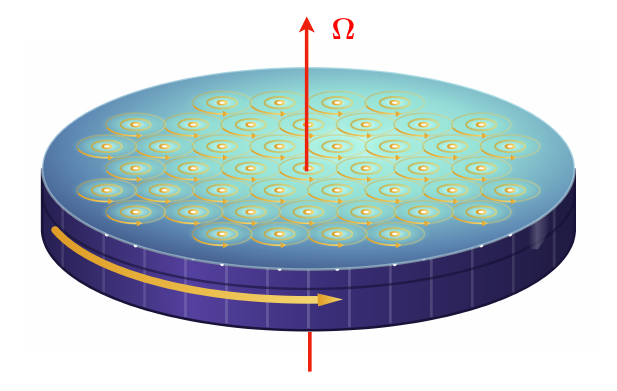}
        \caption{Schematic illustration of a rotating superfluid
          containing a vortex lattice. The quantized vortices form an
          approximately triangular array. Although the flow around
          each individual vortex is locally nonuniform, its
          coarse-grained average approaches rigid-body rotation on
          length scales much larger than the intervortex spacing.}
        \label{bucket}
    \end{subfigure}%
    \hfill
    \begin{subfigure}[t]{0.45\textwidth}
        \centering
        \includegraphics[height=6cm,width=7cm]{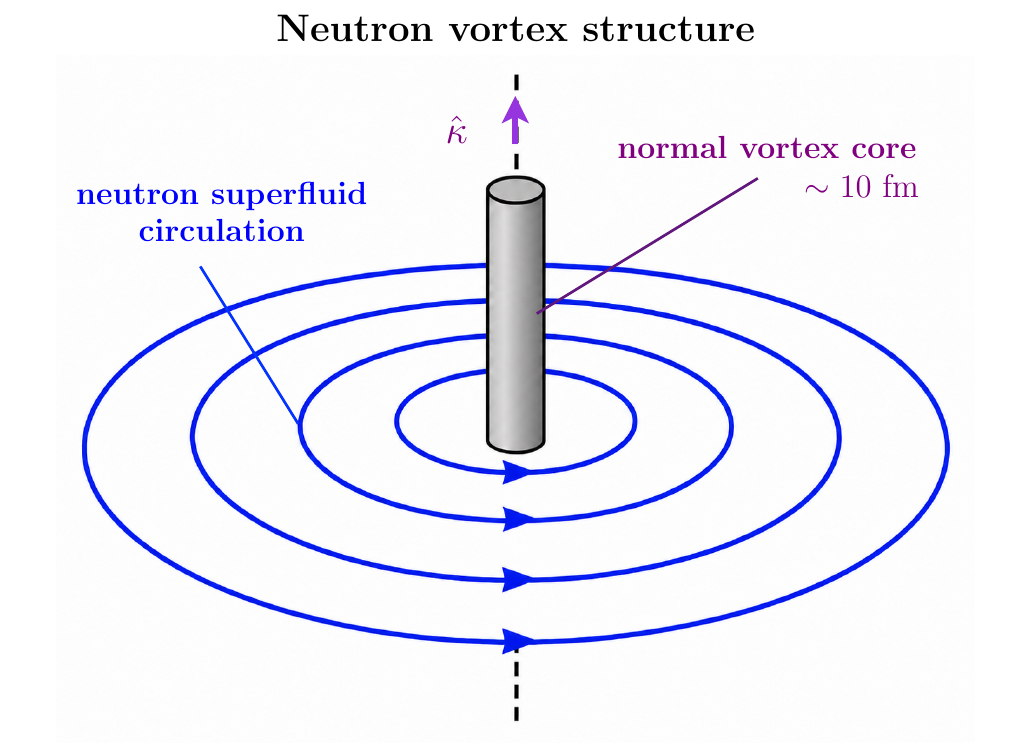}
        \caption{Schematic illustration of a vortex. The shaded
          cylindrical region shows the vortex core, where the order
          parameter is suppressed, while the concentric streamlines
          illustrate the circulating superfluid flow around the vortex
          axis, shown by dashed line.}
    \label{vortex}          
    \end{subfigure}
   \caption{Quantized vortices in a rotating superfluid and the
    associated flow around an individual vortex. Arrows on vortices
    and the container indicate circulation and  rotation, respectively.}
\end{figure*}

If a closed contour encloses $N$ singly quantized vortices, the total
circulation is\footnote{Note that this expression is analogous to Amp\`ere's law for the magnetic field
$\bm{B}$ produced by $N$ parallel wires, each carrying current $I$:
\begin{equation}
\oint d\bm{l}\cdot\bm{B}=\frac{4\pi NI}{c} \qquad\text{(cgs units)}.
\end{equation}.
}
\begin{equation}
  \oint d\bm{l}\cdot\bm{v}_s=N\kappa.
\end{equation}
For a superfluid rotating uniformly at angular velocity
$\Omega=2\pi\nu$, the number of vortices within a circular region of
radius $R$ is (for fiducial values typical of a neutron star)
\begin{equation}
N=\frac{2\Omega}{\kappa}\pi R^2=n_v\pi R^2
\simeq 2.0\times10^{17}
\left(\frac{R}{10\,\mathrm{km}}\right)^2
\left(\frac{\nu}{10\,\mathrm{Hz}}\right),
\end{equation}
where the constant vortex areal density is given by the Feynman
relation
\begin{equation}
n_v=\frac{2\Omega}{\kappa}.
\label{eq:n_quantum}
\end{equation}
Like-signed vortices repel one another and therefore form an
approximately uniform triangular array in equilibrium, as depicted
in Fig.~\ref{bucket}.
The mean areal spacing is
\begin{equation}
 d_v=\left(\frac{2}{\sqrt{3}\,n_v}\right)^{1/2}
\simeq 4\times10^{-3}
\left(\frac{\nu}{10\,\mathrm{Hz}}\right)^{-1/2}\,\mathrm{cm}.
\label{eq:d_v}
\end{equation}

Near the center of a vortex, the velocity grows as $d^{-1}$ until the
fluid becomes normal for $d\simeq\xi$, where $\xi$ is the superfluid
\textit{coherence length}---the characteristic length scale over which
the paired fermions are correlated \citep{Abrikosov:Fundamentals}. In a neutron star,
the coherence length is typically about $10\,\mathrm{fm}=10^{-12}\,
\mathrm{cm}$; see Fig.~\ref{vortex}.
Although the
superfluid velocity increases as $d^{-1}$ close to a vortex, the vorticity averaged over a large number of vortices is
\begin{equation}
\left\langle\nabla\times\bm{v}_s\right\rangle
=n_v\bm{\kappa}=2\bm{\Omega},
\end{equation}
which corresponds to uniform rotation at rotation rate $\bm{\Omega}$.   The contribution to the superfluid angular
momentum of a single vortex at a distance $r_v$ from the rotation axis
is~\citep{Baym1983}:
\begin{equation}
L_v=\frac{N_s \hbar}{2}\left(1-\frac{r_v^2}{R^2}\right), \quad N_s=\frac{\rho_s V}{m_n} ,
\end{equation}
where $N_s$ is the number of neutrons represented by the superfluid
mass density in a uniform cylindrical volume $V$ and radius $R$ with respect to the rotation axis. The superfluid
angular momentum is completely determined by the number and
distribution of vortices; the dynamics of the superfluid, and its
coupling to its normal fluid, are controlled by the forces on individual
vortices.

\subsection{Magnus force and vortex drag}

\begin{figure}
  \centering\includegraphics[width=.5\linewidth] {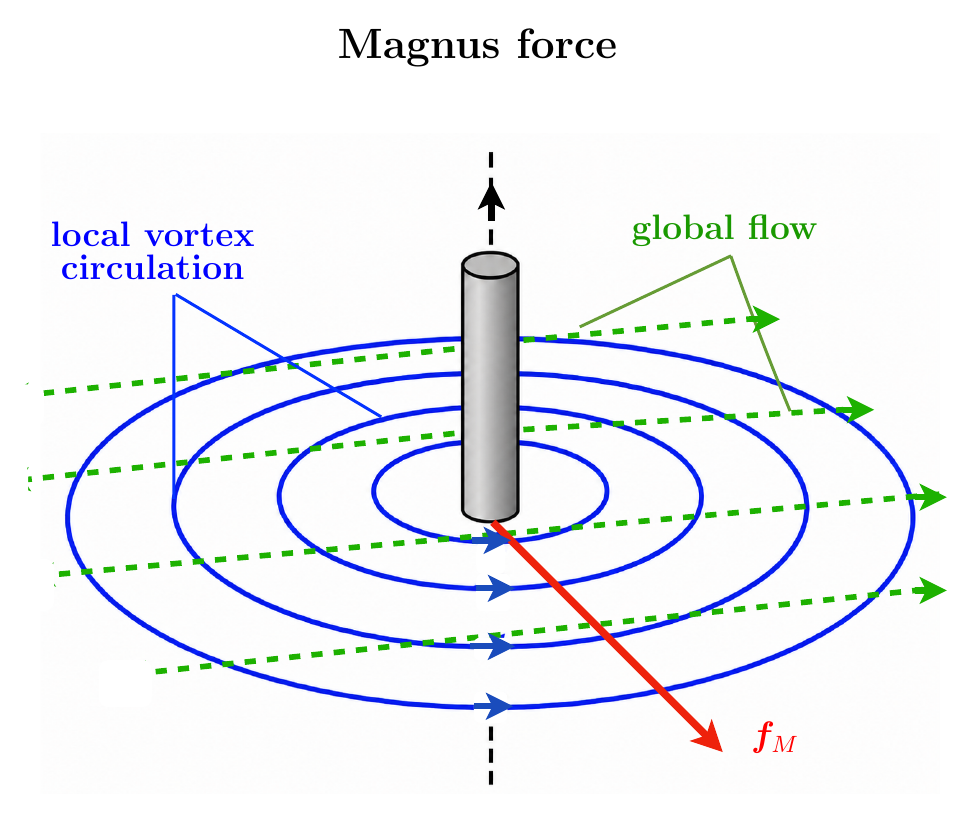}
\caption{
Illustration of the Magnus force on a vortex. The grey cylinder represents a straight vortex line core (circulation $\bm{\kappa}$, oriented along the dashed axis), with its local circulating flow shown in blue. This local circulation is superimposed on a uniform, external global flow $\bm{v}_s$ (green dashed lines). On the side where the two flows add constructively the resulting speed is higher, and on the opposite side, where they add destructively, the speed is lower; by Bernoulli's law this asymmetry in flow speed produces a corresponding asymmetry in pressure, lower on the fast side and higher on the slow side. The resulting net force, the Magnus force $\bm{f}_M$ (red arrow), points from the high- to the low-pressure side, perpendicular to both the vortex line and the global flow, 
giving $\bm{f}_M = -\rho_s\bm{\kappa}\times\bm{v}_s$ in a frame where the vortex is stationary,  and $\bm{f}_M = -\rho_s\bm{\kappa}\times(\bm{v}_s-\bm{v}_v)$ in a frame where the vortex is moving at velocity $\bm{v}_v$.
}
\label{magnus_force}
\end{figure}

The Magnus force is a Bernoulli lift force that arises when a vortex
moves at a velocity different from that of the surrounding superfluid.
The Magnus force per unit length
on a vortex moving with velocity $\bm{v}_v$ is
\be
\bm{f}_{\rm M}=\rho_s\bm{\kappa}\times(\bm{v}_v-\bm{v}_s).
\label{magnus}
\ee
Its physical origin is illustrated in Fig.~\ref{magnus_force}.

If there are no external forces on the vortex, $\bm{f}_{\rm M}=0$ and
$\bm{v}_v=\bm{v}_s$; the vortex moves with the superfluid. In the
rotating vessel of Fig.~\ref{bucket}, the vortices corotate with the
superfluid. If an external force $\bm{f}_{\rm ext}$ acts on a vortex, force
balance requires
\be
\bm{f}_{\rm M}+\bm{f}_{\rm ext}=0.
\label{ftotal}
\ee
The vortex then generally moves at a velocity different from that
of the superfluid. A moving vortex generally  experiences drag. In the neutron star inner crust, a  vortex interacts with nuclei in the lattice and can excite lattice phonons, which in turn couple to the electrons.
If the force exerted by the normal component on the vortex contains both a longitudinal drag term and a transverse (Iordanskii) term, the equation of motion of a straight 
vortex becomes
\be
 \rho_s\bm{\kappa}\times(\bm{v}_v-\bm{v}_s)
 -\eta(\bm{v}_v-\bm{v}_c)
 -\eta'\bm{\kappa}\times(\bm{v}_v-\bm{v}_c)=0,
 \label{dragged_vortex}
\ee
where $\bm{v}_c$ is the velocity of the charged (normal) component, the
nuclear lattice in the neutron-star crust, while $\eta$ and $\eta'$
are the longitudinal and transverse drag coefficients, respectively. The existence and magnitude of the transverse Iordanskii force~\citep{Iordanskii1964} in this
setting remain uncertain, and there is no compelling evidence that $\eta'$ is appreciably different from zero~\citep{Sonin2016,Sergeev2023}.   We therefore neglect this term in the following discussion and set $\eta'=0$.

In the rest frame of the
crust ($\bm{v}_c=0$), let $\bm{v}_s$ be along $\hat{\bm{x}}$ with magnitude 
$v_s\equiv|\bm{v}_s|$,  and let
$\bm{\kappa}$ be along $\hat{\bm{z}}$.  The solution of
Eq.~\eqref{dragged_vortex} is
\be\label{eq:v_v}
\bm{v}_v=v_s\cos\theta(\hat{\bm{x}}\cos\theta-\hat{\bm{y}}\sin\theta),
\qquad \tan\theta\equiv\frac{\eta}{\rho_s\kappa}.
\ee
The vortex moves at drift (dissipation)
angle $\theta$ with respect to $\bm{v}_s$. In the
limit of low drag, $\eta/(\rho_s\kappa)\ll 1$ and
$\theta\simeq\eta/(\rho_s\kappa)$, giving
\be
\bm{v}_v=v_s(\hat{\bm{x}}-\hat{\bm{y}}\theta).
\label{drag-regime}
\ee
The vortex moves nearly with the superfluid, with slow drift along
$-\hat{\bm{y}}$. (The vortex moves with the superfluid for zero drag.) In
the limit of high drag, $\eta/(\rho_s\kappa)\gg 1$ and
$\theta\rightarrow\pi/2$, giving
\be
\bm{v}_v\simeq -\hat{\bm{y}}\left(\frac{\rho_s\kappa}{\eta}\right)v_s. 
\ee
The vortex almost stops, drifting slowly along $-\hat{\bm{y}}$.

\subsection{Pinning force}
\label{subsec:pinning-force}

In terrestrial superfluids such as $^3$He and $^4$He, vortices can pin
to inhomogeneities, typically rough edges in the container. In the
neutron star inner crust, vortices are predicted to pin to the nuclei
that coexist with the superfluid.  The pinning interaction reflects
several competing effects: the change in condensation energy inside a
nucleus, the modification of the vortex-flow kinetic energy,
deformations of the vortex and nucleus as they approach one another, and nuclear
shell effects.  Calculations that compare the energy of a nucleus
inside a vortex to the energy when the nucleus is far away
\citep{EpsteinBaym1988,Donati2003,Donati2006,Avogadro2007b,Avogadro2007a,Avogadro2008}
have been fraught with systematic uncertainties. These calculations
agree that the magnitude of the interaction energy should be several
MeV, occurring over an interaction distance of some $10\,\mathrm{fm}$,
but disagree on the sign of the vortex–nucleus interaction. More
recent calculations using density-functional theory
\citep{Bulgac2013a,Wlazlowski2016} approach the problem dynamically
and find a repulsive interaction under the conditions studied, with a
magnitude of several MeV over a length scale of
$\sim 10\,\mathrm{fm}$. If the interaction is indeed repulsive,
vortices can pin to the interstices of the lattice with a maximum force per
unit length of magnitude $f_p$. (If vortices pin to nuclei, now
disfavored by the most recent calculations, pinning can be much
stronger than in the interstitial case.) In the frame corotating with
the neutron star crust (so $\bm{v}_c=0$ for a pinned vortex),
equilibrium is given by 
\be \rho_s\kappa v_s\leq f_p. 
\ee 
 Dynamical simulations of vortex motion through the lattice~\citep{Link2022}, discussed in more detail below, find that for a repulsive vortex--nucleus interaction,
\begin{equation}
f_p\simeq 2\times10^{-4}
\left(\frac{|E_p|}{4\,\mathrm{MeV}}\right)
\left(\frac{b}{30\,\mathrm{fm}}\right)^{-2}
\mathrm{MeV\,fm^{-2}}
=3\times10^{16}\,\mathrm{dyn~cm^{-1}}.
\label{fp_bcc_repulsive}
\end{equation}
 For an attractive
vortex--nucleus interaction with the same value of $E_p$, the critical pinning force is about an order of magnitude larger than the above estimate. Vortices can pin if $v_s$ is below the critical value \citep{Link2022}~\footnote{The estimates in~\cite{Link2022} assumed a value of $\rho_s$ appropriate for significant 
entrainment of the superfluid by nuclei, as predicted by~\cite{Chamel2012a}.  
Recent calculations that include pairing and the previously omitted geometric/off-diagonal contribution to the superfluid density find a much larger fraction of conduction neutrons than normal-state band theory, with values close to the density of unbound neutrons in the configurations studied~\citep{Almirante2024,Almirante2025,Almirante2026a,Almirante2026b}. We therefore use $\rho_s\simeq10^{14}\,\rm{g\,cm^{-3}}$ as a weak-entrainment benchmark.}.
\be
v_{\rm cr}=\frac{f_p}{\rho_s\kappa}.
\ee

\subsection{Vortex dynamics in the inner crust}

The equation of motion of a single vortex, in a frame corotating with
the crust, for small-amplitude perturbations is
\citep{Link2022}
\begin{equation}
\epsilon_v\partial^2_z\bm{u}(z,t)
+\rho_s\bm{\kappa}\times\left(\partial_t\bm{u}(z,t)-\bm{v}_s\right)
-\nabla_\perp V -\eta\partial_t\bm{u}(z,t)
=0, 
\label{eom}
\end{equation}
where $\bm{u}(z,t)=\hat{\bm{x}}u_x(z,t)+\hat{\bm{y}}u_y(z,t)$ is a
two-dimensional vector giving the shape of the vortex and
$\bm{\kappa}=\kappa\hat{\bm{z}}$.  
 The first term is the force (per unit length) required to bend the vortex, where $\epsilon_v$ is the self-energy per unit length, or tension, of the vortex.
The second term is the Magnus force; $\bm{\kappa}$ is
the circulation vector of magnitude $\kappa=h/2m_n$, and $\bm{v}_s$ is
the external superfluid flow velocity. The third term is the
conservative interaction force (potential $V$) between the vortex and
the lattice; $\nabla_\perp$ is the gradient perpendicular to the
vortex. The fourth term is the drag force between the vortex and the
lattice, assumed here to be linear.

\begin{figure}[t]
\includegraphics[width=\linewidth]{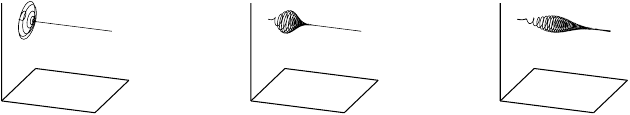} 
\caption{Propagation of a Kelvin wave packet. The wave is helical and
  dispersive.  For short-wavelength modes with $k \simeq 2 \pi b^{-1}$
  the typical group speed in the inner crust of a neutron star is
  $10^{-2}$c.}
\label{wave}
\end{figure}

Setting $V=0$, $\eta=0$, and $\bm{v}_s=0$ to consider free
Kelvin perturbations, and defining the complex transverse displacement
$w(z,t)\equiv u_x(z,t)+iu_y(z,t)$, Eq.~\eqref{eom} becomes
\be
-\frac{\epsilon_v}{\rho_s\kappa}\partial_z^2w(z,t)
=i\partial_tw(z,t),
\label{eom1}
\ee which is mathematically equivalent to the Schr\"odinger
equation. Plane-wave solutions for $w(z,t)$ describe helical waves
with frequencies given by \be \omega_k=\frac{\epsilon_v
  k^2}{\rho_s\kappa}.  \ee These \textit{Kelvin waves} are
dispersive. Their local phase and group velocities are
\be v_{\rm
  ph}=\frac{\omega_k}{k} =\frac{\epsilon_v k}{\rho_s\kappa}, \qquad
v_g=\frac{d\omega_k}{dk}=2v_{\rm
  ph};
\ee
see Fig.~\ref{wave}. 

\begin{figure}[t]
\centering
\includegraphics[width=0.9\linewidth]{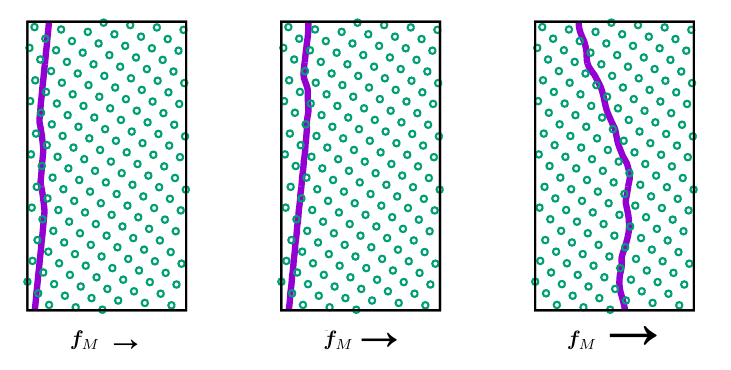}
\caption{Motion of a vortex (depicted as  the thick line) under a uniform
  Magnus force per unit length $\bm{f}_{\rm M}$. The force is gradually increased, as
  indicated by the size of the arrow. One plane of the lattice is shown. For small force (left), the vortex
  assumes a pinned state with kinks, in which it generally follows a lattice
  plane. At intermediate force (center), the kinks move along the
  vortex (and off the numerical grid). At large force (right), the vortex unpins
  completely and moves through the lattice. There is also a component of
  the motion along the superfluid flow (into the page),  not
  shown.}
\label{snapshots}
\end{figure}

The vortex self-energy is $\epsilon_v\sim\rho_s\kappa^2\sim 1\,\mathrm{MeV\,fm^{-1}}$.
For a wavenumber of $k=2\pi b^{-1}$, where $b$ is the lattice spacing
in the inner crust ($\sim 30\,\mathrm{fm}$ in the densest regions), the group
velocity of Kelvin modes is $\sim 10^{-2}$c, with a wave period of
$2\pi/\omega_k\sim 10^{-20}$ s. This wave speed is large because the
vortex is difficult to bend over length scales comparable to the range of the vortex--nucleus potential. In these simulations, the
dynamics of pinning and unpinning of vortices  are determined by the excitation and 
propagation of Kelvin waves. Solving Eq.~\eqref{eom} for a periodic
pinning potential in three dimensions gives the vortex motion shown in
Fig.~\ref{snapshots}.

\begin{figure}[bht]
\centering\includegraphics[width=.8\linewidth]{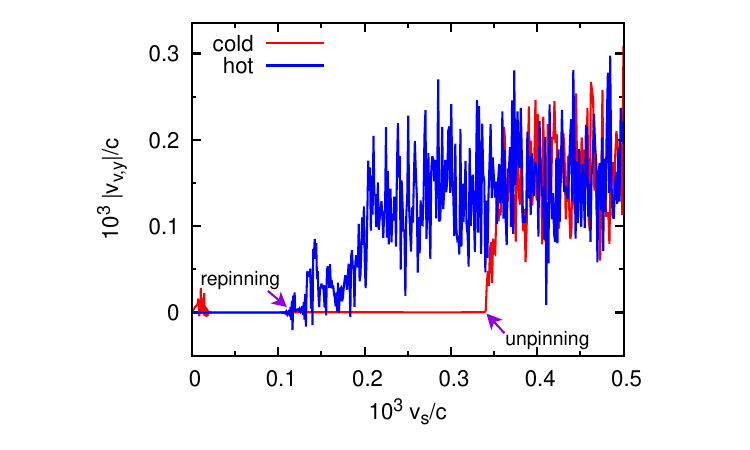}
\caption{Bistability of vortex motion: vortex velocity in the
  direction of an applied Magnus force $\bm{f}_{\rm M}=-\hat{\bm{y}}\,\rho_s\kappa
  v_s$ versus $v_s$. The vortex is pinned at $v_s=0$.
  If $v_s$ is slowly increased, the vortex remains pinned (cold) on the red
  curve. At a critical velocity $v_s=v_{\rm cr}$, the vortex unpins, entering
  the drag regime of Eq.~\eqref{drag-regime}. If $v_s$ 
is now slowly decreased (blue curve), the vortex eventually repins at a value of
$v_s$ that is generally lower than the value at which the vortex
originally unpinned.}
\label{hysteresis}
\end{figure}

Because a vortex is an extended object with internal degrees of
freedom, there are two states of vortex motion for the same applied
force: (i) a pinned, cold state and (ii) a hot, unpinned state, in which
the vortex is heated by the excitation of Kelvin waves as it moves;
see Fig.~\ref{hysteresis}. Transitions between cold and hot states could play a role in neutron-star spin glitches. 

\subsection{Thermal effects}
\label{thermal-effects}

\begin{figure}
\centering
\begin{tabular}{@{}cc@{}}
  \begin{minipage}{0.33\textwidth}
    \includegraphics[width=\textwidth]{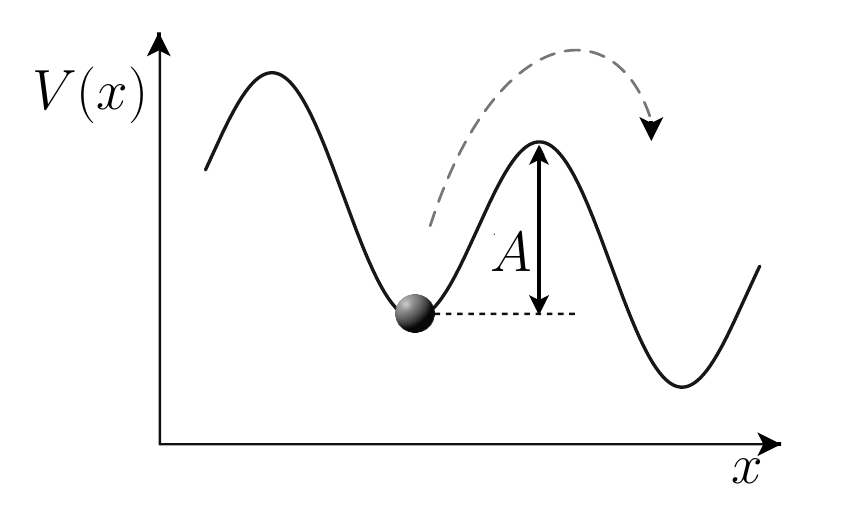} 
  \end{minipage} 
  \begin{minipage}{.65\textwidth}
  \includegraphics[width=\textwidth]{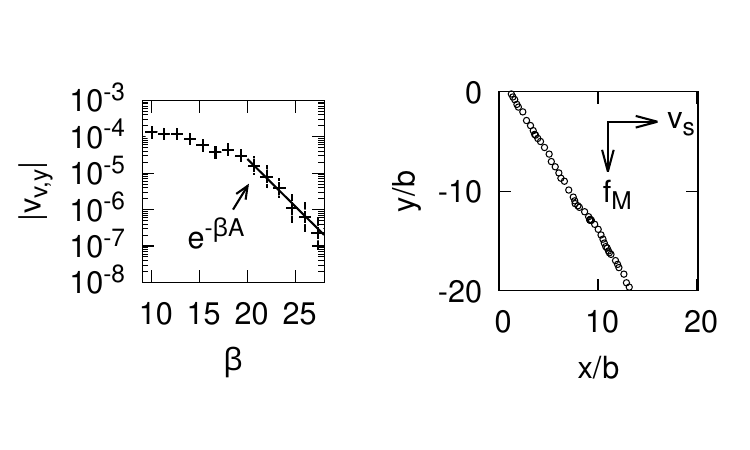}
  \end{minipage}
\end{tabular}
  \caption{\textit{Left}---Schematic of thermal activation of a
  particle in a biased potential. By overcoming an activation barrier
  $A$, the particle moves to the right. The corresponding vortex
  problem is multidimensional. \textit{Middle}---Vortex-creep velocity
  (in units of $c$) in the direction of an applied Magnus force
  $\bm{f}_{\rm M}=-\rho_s\bm{\kappa}\times\bm{v}_s$ versus inverse
  temperature $\beta=1/(k_BT)$ (in $\mathrm{MeV}^{-1}$). At high
  $\beta$ (low $T$), the creep velocity follows an Arrhenius form
  $e^{-\beta A}$. Vertical error bars denote
  statistical uncertainty estimates from an ensemble of simulations. \textit{Right}---Example of vortex motion under thermal activation in the $x-y$ plane. Each circle represents an instantaneous position of the vortex, averaged over its length. 
  The vortex has significant motion along both $\bm{v}_s$ and
   $-\bm{\kappa}\times\bm{v}_s$. }
\label{creep}
  \end{figure}

As a neutron star spins down, the superfluid speed $v_s$ in the frame
of the crust increases, approaching the critical magnitude $v_{\rm cr}$ of
Fig.~\ref{hysteresis}. Before this happens, vortices will be able to
move through thermally activated {\em vortex creep}~\citep{Alpar1984a,Link2014a}; see Fig.~\ref{creep} (left) for a
schematic of this process. Simulations of neutron-star cooling
indicate that the inner-crust temperature is of order
$10\,\mathrm{keV}$ for stars of age $10^3$--$10^4\,\mathrm{yr}$ (see
\cite{YakovlevPethick2004} for a review). Thus, the thermal energy
is about $10^{-2}$ of a typical pinning energy in the inner crust.

  In early work on post-glitch relaxation of the pinned vortex array,
  \cite{Alpar1984a} assumed the creep velocity is
  \be \bm{v}_v=v_0
  \left(\hat{\bm{v}}_s\times\hat{\bm{\kappa}}\right) 
  e^{-\beta  A(v_s/v_{\rm cr})}, \ee
  where $v_0$ is a velocity of order
  $10^7\,\mathrm{cm\,s^{-1}}$, $\beta=1/(k_BT)$ is the inverse
  temperature, and $A(v_s/v_{\rm cr})$ is an activation energy that vanishes
  as $v_s\rightarrow v_{\rm cr}$. This ansatz assumes no motion of pinned
  vortices along $\bm{v}_s$.

The problem of how a pinned vortex moves at finite temperature can be approached in the framework of Langevin dynamics, by adding a stochastic force $\bm{N}(z,t)$ to Eq.~\eqref{eom} \citep{Link2026} with a correlation function,
\be
\left\langle\bm{N}(z,t)\cdot\bm{N}(z^\prime,t^\prime)\right\rangle
\propto\eta k_BT\,\delta(z-z^\prime)\delta(t-t^\prime),
\ee
where $\langle\cdots\rangle$ denotes a thermal average. This force is
uncorrelated in space and time, as in the simpler problem of Brownian
motion. \cite{Link2026} find a vortex creep velocity of the form:\footnote{Recall that the Iordanskii term was ignored in Eq. \eqref{dragged_vortex}. Vortex creep creates an effective Iordanskii term in the vortex equation of motion.}
\be
\bm{v}_v=\cos\theta\left\{
\hat{\bm{\kappa}}\times
\left[\bm{v}_s\times\hat{\bm{\kappa}}\right]\cos\theta
+\left[\bm{v}_s\times\hat{\bm{\kappa}}\right]\sin\theta
\right\}e^S e^{-\beta A(v_s/v_{\rm cr})},
\label{vcreep}
\ee
where $\tan\theta\sim\eta/(\rho_s\kappa)$ and  $e^S$ is a dimensionless ``entropy factor" typical of thermal-activation problems with multiple degrees of freedom, and $A(v_s/v_{\rm cr})$ depends on both the detailed vortex--nucleus interaction and the vortex self-energy per unit length $\epsilon_v$.

In some cases, quantum tunneling could be the dominant mode for vortex motion. Treating a pinned vortex as a system of harmonic oscillators, the temperature $T$ in Eq.~\eqref{vcreep} is replaced by \citep{Link1993},
\be
k_BT\longrightarrow k_BT_{\rm eff}=\frac{\hbar\omega_k}{2}\coth\left(\frac{\hbar\omega_k}{2k_BT}\right),
\ee
where $T_{\rm eff}$ is an effective temperature and $\hbar\omega_k$ is the energy of the dominant Kelvin mode that unpins a pinned vortex segment. 
The effective temperature  $T_{\rm eff}\to T$ for $k_BT\gg\hbar\omega_k/2$, and $T_{\rm eff}\to \hbar\omega_k/2k_B$ for $k_BT\ll\hbar\omega_k/2$. The (crude) estimates of \citet{Link1993} find $\hbar\omega_k\lessapprox 1$ keV, which is significantly less than the typical neutron star temperature 10 keV, and suggests that quantum tunneling is unimportant. A more detailed analysis is needed to obtain the rate of quantum tunneling, which will depend on the detailed vortex–nucleus interaction, which is unknown at present.  

\subsection{The size of spin glitches}

Is the pinning force strong enough to account for large spin glitches?
Suppose the pinned vortices follow the rotation axis of the star of radius
$R$ in the crust of thickness $\Delta R\ll R$.  The angular momentum in
the superfluid available to drive a glitch is the excess above that 
for co-rotation of the superfluid with the crust. The
maximum velocity difference between the pinned superfluid and the crust is
$v_s\simeq v_{\rm cr}=f_p/(\rho_s\kappa)$, and the excess angular momentum in the
inertial frame is 
\begin{equation}
\Delta J_s=\int d^3r\,r\sin\vartheta\,\rho_s v_s
\simeq \pi^2 R^3 \Delta R \,\frac{\bar{f}_p}{\kappa},
\end{equation}
where $\vartheta$ is the polar angle and $\bar{f}_p\equiv\int dr\, rf_p/R$ is the mean pinning force in the crust.\footnote{
The effects of possible nuclear entrainment cancel out of the above estimate,
since $v_s$ scales as $\rho_s^{-1}$.}   At the time of a
glitch, suppose that all of this available angular momentum is given
to the crust plus any other components (\eg, part of the stellar
core) that are tightly coupled together over the timescale of the
glitch. Let the moment of inertia of the tightly coupled components of
the star be $fI$, where $I$ is the total moment of inertia
of the star and $f$ is the fraction of that moment of inertia that is
tightly coupled.  If the core remains coupled during a glitch, then
$f\sim 1$, while if the core is completely decoupled by the glitch
$f\sim 10^{-2}$. (In the latter case, observed glitch recoveries could
represent the response of the core as it recouples.) The glitch magnitude is therefore determined by 
$fI\Delta\Omega_c=\Delta J_s$, giving
\begin{equation}
\frac{\Delta\Omega_c}{\Omega_c}
\simeq 3\times10^{-5} f^{-1}
\left(\frac{R}{10\,\mathrm{km}}\right)^4
\left(\frac{\Omega_c}{10^2\,\mathrm{rad\,s^{-1}}}\right)^{-1}
\left(\frac{I}{10^{45}\,\mathrm{g\,cm^2}}\right)^{-1}
\left(\frac{\Delta R/R}{0.05}\right)
\left(\frac{\bar{f}_p}{10^{16}\,\mathrm{dyn\,cm^{-1}}}\right).
\end{equation}
Pinning is easily strong enough to account for glitches of magnitude
$\Delta\Omega_c/\Omega_c=10^{-6}$ for any $f\le 1$.   The
fraction $f$ can be constrained using both the magnitude and frequency
of glitches in the most prolific glitching pulsars~\citep{Link1999}. A
key question is what mechanism triggers the collective motion of many
vortices, causing these vortices to go from the lower, cold branch of
Fig.~\ref{hysteresis}, to the upper, hot branch.

\subsection{Global dynamics}
\label{global-dynamics}

To connect the vortex microphysics described previously to observed
spin evolution of neutron stars, a global description of vortex motion
is needed. Conservation of vortex lines, expressed in the frame corotating with the crust at angular velocity $\bm{\Omega}$, takes the form \citep{SedrakianWassermanCordes1999,Levin2023}, 
\be\label{vorticity}
\partial_t\bm{\omega}=\curl(\bm{v}_v\times\bm{\omega}),
\ee
where 
\be
\bm{\omega}\equiv 2\bm{\Omega}+\curl{\bm{v}_s},
\ee
is the total vorticity and $\bm{v}_s$ is measured in the frame rotating with the crust. 
The local
vorticity vector $\bm{\omega}$ follows the vortices. If the vortex
array is subject to pinning or drag forces, Eqs.~\eqref{magnus} and \eqref{ftotal} give, 
\be
\rho_s^{-1}\bm{f}=(\bm{v}_v-\bm{v}_s)\times\bm{\omega},
\ee
where $\rho_s^{-1}\bm{f}$ is the force per unit mass acting on the
superfluid. (The velocity difference in the above equation is frame independent).  
In the absence of such a force, the vortices move with the superfluid, $\bm{v}_v=\bm{v}_s$, and the vorticity is advected by the superfluid flow according to Eq.~\eqref{vorticity}. Time independence, $\partial_t \bm{\omega}=0$, follows only for a stationary configuration. If there are pinning or drag forces, the fluid and vortex velocities are different. If the vortex array is along
$\hat{z}$ in cylindrical coordinates
$(\varrho,\phi,z)$ and
uniform, Eq.~\eqref{vorticity} gives, 
\be
\partial_t\omega_z=
-\frac{1}{\varrho} \partial_{\varrho}\left(\varrho \omega_z v_v^{\varrho}\right)
= \frac{1}{\varrho}\partial_\varrho(\varrho\rho_s^{-1}f_\phi),
\ee
so that the vorticity decreases if vortices have a component of their
velocity radially outward, along $\hat{\varrho}$. The vorticity
increases if the vortices are pushed toward the superfluid rotation
axis. The rate at which vortices are forced, and the corresponding change
in $\omega_z$, are determined by the azimuthal force $f_\phi$ density on the
superfluid, implying a torque on the superfluid, and a corresponding opposite
torque on the crust. 

\begin{figure}[t]
\centering
\begin{tabular}{ccc}
\includegraphics[width=.2\linewidth]{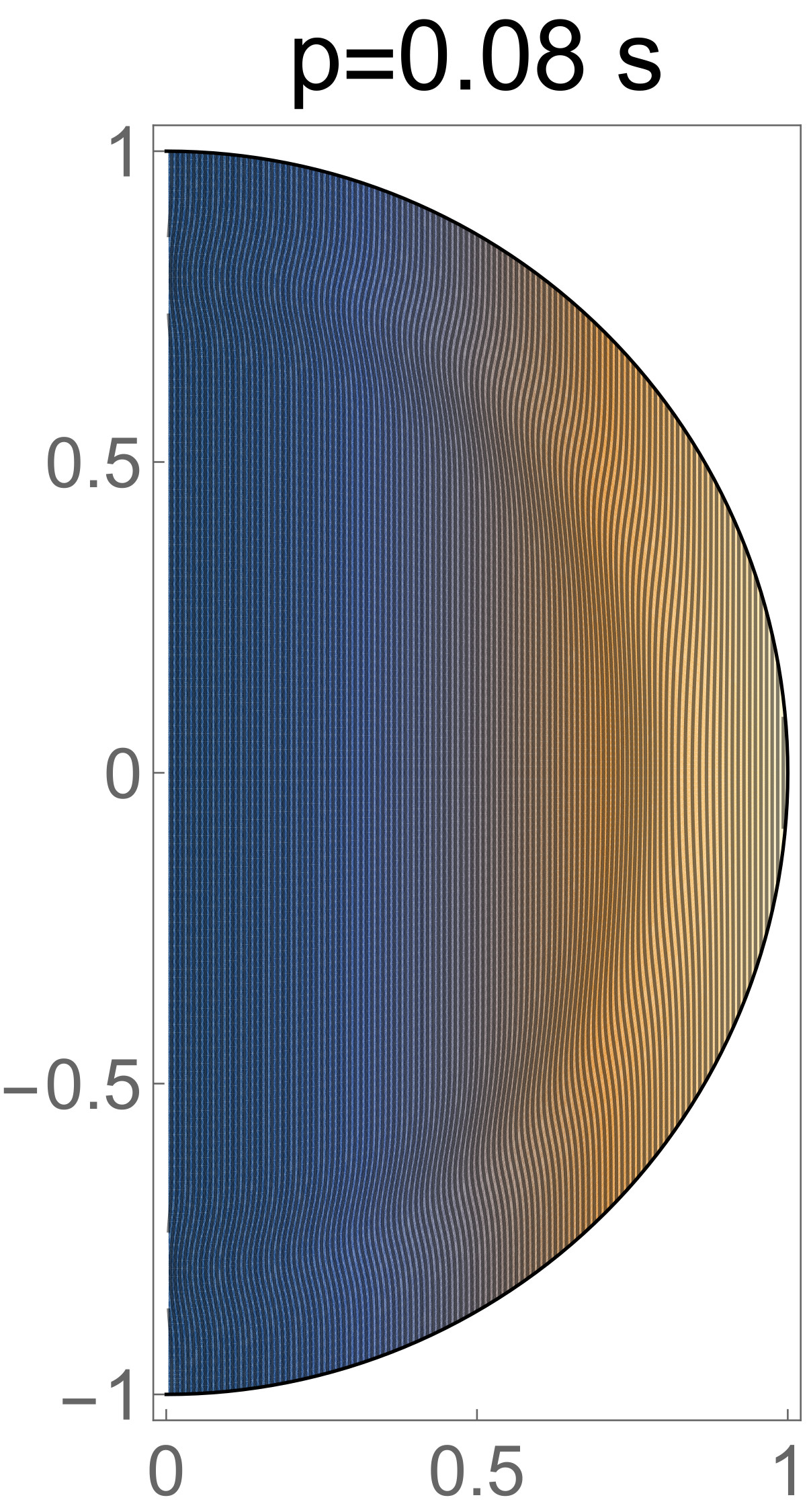} & 
\includegraphics[width=.2\linewidth]{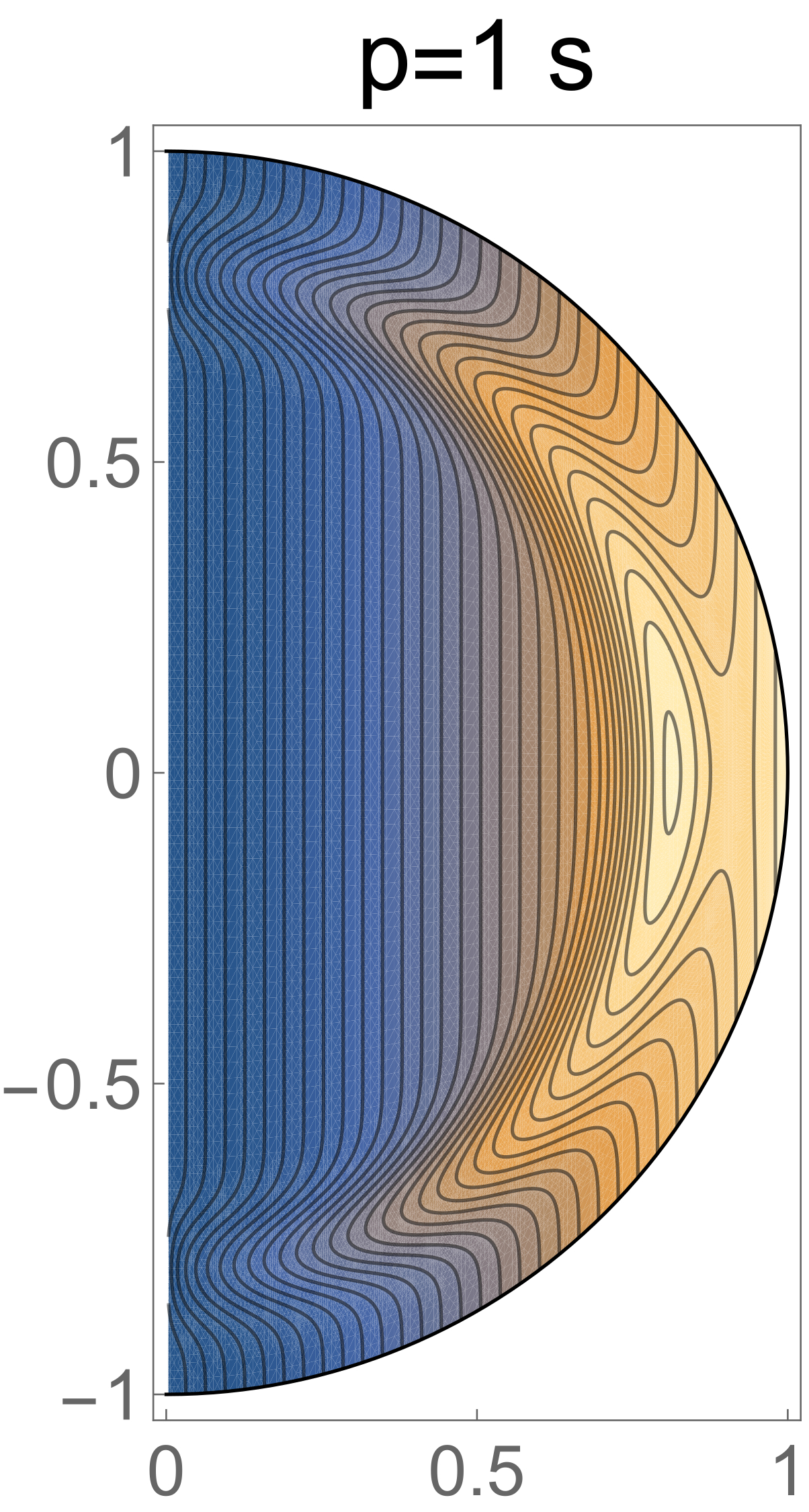} &
\includegraphics[width=.2\linewidth]{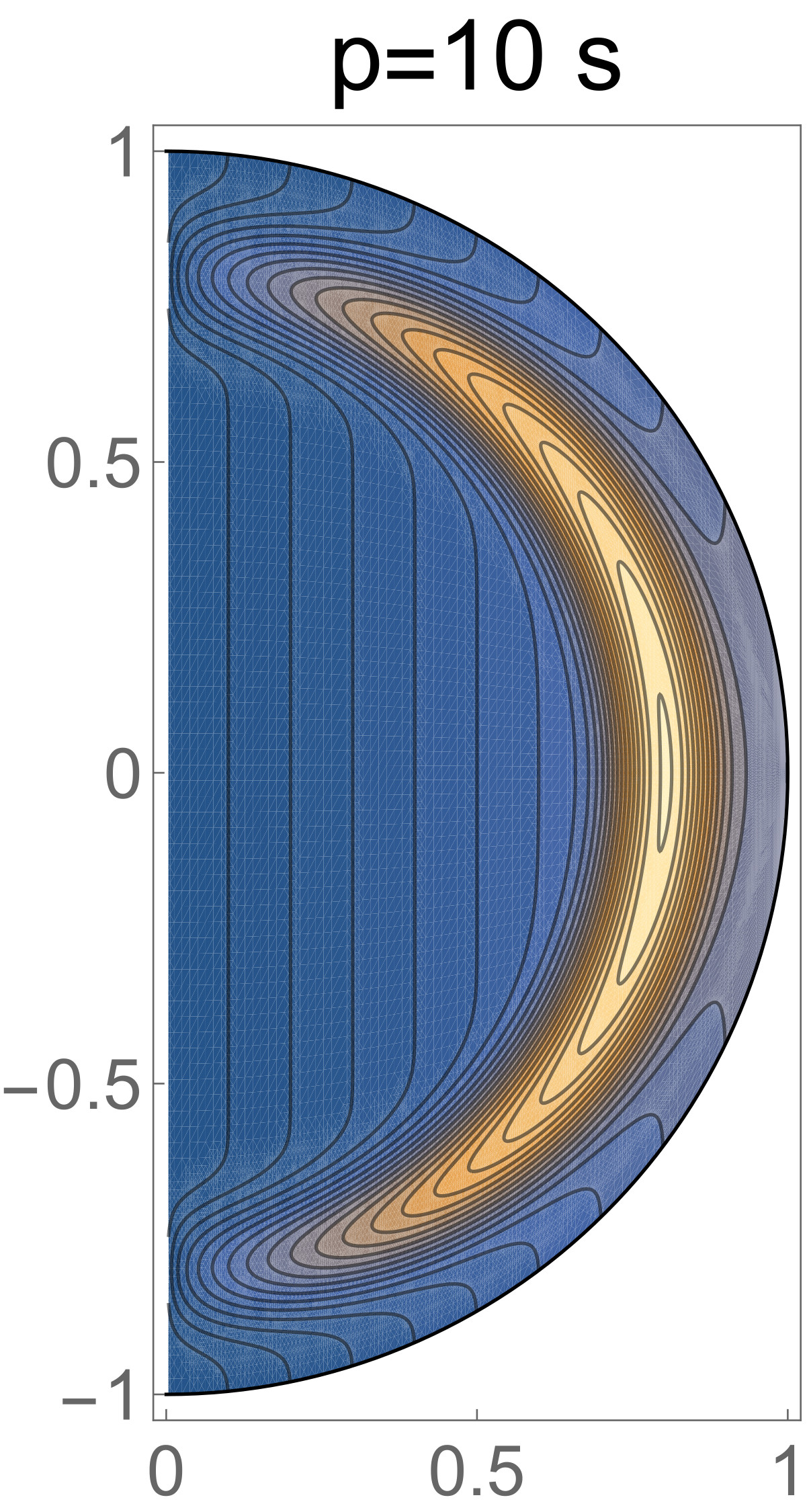} 
\end{tabular}
\caption{A simple model of the evolution of superfluid vorticity in a
  spinning-down neutron star, showing contours of constant circulation
  at different spin periods $P$.  Rotation is about the vertical axis
  with contours following the quantum vortices. The initial spin state
  has a uniform vortex distribution, aligned with the rotation
  axis. The pinning strength is a function only of spherical radius
  $r$, assumed in this example to be distributed as a Gaussian in $r$
  in the outer part of the star. As the star spins down the Magnus
  force increases, significantly deforming the vortex array. Vortices
  bend, reconnect, and are expelled from the star. Eventually, the
  star is left with pinned vortex loops that carry a ``superfluid
  river'' around the equator of the star.  Adapted from
  \citet*{Levin2023}.}
\label{spindown}
\end{figure}

In general, position-dependent forces on the vortex array will cause the
array to bend and twist. 
Suppose that, 
initially, the vortex array is uniform and rectilinear with
$\bm{\omega}=\omega\hat{z}$. The strength of vortex pinning is determined by
the mass density, and so is constant on spherical shells, implying the
vortex velocity will be a function of both $\varrho$ and $z$. As a
simple example, suppose 
$\bm{v}_v=v_v(\varrho,z)\hat{\varrho}$. Eq.~\eqref{vorticity} becomes,
\be
\partial_t\bm{\omega}=\omega\bm{\hat{\varrho}}\,\partial_z v_v(\varrho,z)
-\omega\bm{\hat{z}}\,\frac{1}{\varrho}\partial_\varrho[\varrho v_v(\varrho,z)].
\ee
The initially rectilinear vortex array will deform as the vortices
move, as the vorticity acquires a component along $\hat\varrho$. More generally, there will be twisting of the vortex array
as well, as the vortices move by vortex creep in some regions of the
star, and under drag in other regions. A simple model of what could 
happen is depicted in Fig. \ref{spindown}.

\subsection{Heating by vortex motion}

Vortex creep is intrinsically dissipative. The outward migration of
vortices allows the superfluid to spin down and reduces the
differential rotation stored between the pinned superfluid and the
crust. Though pinning forces could significantly deform the vortex array as discussed in Section \ref{global-dynamics}, we will give simple estimates assuming the vortex array approximates rigid-body rotation, corresponding to a superfluid angular velocity $\bm{\Omega}_s$.  Defining the rotational lag as
$\omega_{\rm lag}\equiv\Omega_s-\Omega_c$ and the outward radial vortex
speed as $v_r$, and writing
$f_{\rm M}\equiv|\bm{f}_{\rm M}|$, the local energy-dissipation rate is
\bea
\dot{q}_{\rm diss}
\sim n_v f_{\rm M}v_r
\simeq n_v\rho_s\kappa\varrho\,\omega_{\rm lag}v_r,
\label{eq:local_creep_dissipation}
\eea
where $\varrho$ is the cylindrical distance from the rotation axis.
Integrating over the pinned region gives the global frictional-heating
rate
\bea
\dot{E}_{\rm diss}
\simeq
\int \omega_{\rm lag}\,|\dot{\Omega}_s|\,dI_s,
\label{eq:creep_heating_global}
\eea
where $dI_s$ is the differential moment of inertia of the superfluid
reservoir. This mechanism may contribute to the thermal evolution of
old neutron stars by converting rotational energy stored in a
persistent superfluid lag into internal heat~\citep{Riper1995,SchaabSedrakian1999,Fujiwara2024}.

The same dissipative dynamics govern the coupling between the
superfluid reservoir and the observed crustal rotation. A minimal
two-component model separates the star into a superfluid component
with moment of inertia $I_s$ and a component consisting of the crust
and charged particles, with moment of inertia $I_c$. Their angular
momenta evolve as
\bea
I_c\dot{\Omega}_c
=
N_{\rm ext}+N_{\rm int},
\qquad
I_s\dot{\Omega}_s
=
-N_{\rm int},
\eea
where $N_{\rm ext}$ is the external electromagnetic braking torque and
$N_{\rm int}$ is the internal torque mediated by vortex motion. In the
creep regime, $N_{\rm int}$ depends on the lag $\omega_{\rm lag}$ through the
creep velocity and therefore inherits the strong temperature and
pinning-energy dependence of the vortex-creep velocity described in Section \ref{thermal-effects}. 

The microphysical picture thus connects three levels of description.
At the smallest scale, nuclear structure and pairing correlations
determine the pinning potential experienced by a vortex. At an
intermediate scale, thermal activation  generates
a slow outward vortex current through the crust. At the macroscopic
scale, this current controls angular-momentum exchange between the superfluid and the crust, post-glitch
relaxation, and frictional heating, providing a direct link between
the microscopic physics of superfluid vortices and observable
neutron-star rotational dynamics~\citep{Alpar1984b,Alpar1984a}.

\subsection*{Open questions and perspectives}
\begin{itemize}
    \item \textit{Pinning strength from first principles.} Microscopic
    calculations of the vortex--nucleus interaction remain sensitive to
    the choice of nuclear energy-density functional and the treatment of
    pairing correlations near the vortex core. Achieving convergence
    between Bogoliubov--de~Gennes, time-dependent density-functional,
    and quantum Monte Carlo approaches is essential for placing the
    maximum angular-momentum reservoir---and hence glitch amplitudes---on a firm quantitative footing.

    \item \textit{Quantum versus thermal creep.} At the low temperatures
    of old neutron stars, quantum tunneling of vortex segments may
    dominate over thermally activated hopping. A systematic treatment of
    the crossover between thermal and quantum creep regimes, and its
    dependence on vortex tension, pinning geometry, and local
    temperature, is still lacking and could substantially affect
    predicted post-glitch relaxation timescales.
    
 \item\textit{Global dynamics.} The combined effects of vortex pinning, vortex drag, and stellar spin-down are likely to significantly deform and twist the vortex array. The configuration of the vortex array is the backdrop for more realistic models of neutron-star spindown, spin glitches, and post-glitch relaxation.

    \item \textit{Creep-induced heating and neutron-star thermal evolution.}
    The frictional heating rate in Eq.~\eqref{eq:creep_heating_global} depends
    on both the lag profile and the superfluid moment of inertia, and
    may be detectable through surface temperature measurements of
    slowly rotating, old pulsars. Comparing predicted heating rates with
    X-ray observations could provide an independent constraint on
    pinning energies and the fractional moment of inertia of the crustal
    superfluid, complementing the information extracted from glitch
    statistics.
\end{itemize}

\section{Flux tubes and other magnetic structures in superconducting neutron-star cores}
\label{sec:flux_tubes}

\subsection{Magnetic flux structures and proton superconductivity}

As the neutron-star core cools below the critical temperature for
proton pairing, the protons become superconducting. It is
important to distinguish the thermodynamic equilibrium magnetic state
from the configuration that is actually produced during the cooling
history of the star.

The mesoscopic physics of proton superconductivity --- on scales
larger than the microscopic interaction range but smaller than the
superfluid region --- is efficiently described by Ginzburg--Landau
(GL) theory and its extension to type-II superconductors by
Abrikosov~\citep{Abrikosov:Fundamentals}. The Ginzburg--Landau parameter is defined as
\begin{equation}
  \label{eq:kappa_GL}
 \kappa_{\rm GL}=\frac{\lambda}{\xi_p},
\end{equation}
where $\lambda$ is the magnetic-field penetration depth and $\xi_p$ is
the proton coherence length. For $\kappa_{\rm GL}>1/\sqrt{2}$ the
proton condensate is a type-II superconductor. For an applied field $H$, the equilibrium phase
diagram consists of a Meissner state for $H<H_{c1}$, an Abrikosov
mixed state containing quantized flux tubes for $H_{c1}<H<H_{c2}$, and
a normal state for $H>H_{c2}$. For $\kappa_{\rm GL}<1/\sqrt{2}$ the
condensate is of type I. In an ideal bulk system the superconducting
and normal phases are separated by the thermodynamic critical field
$H_{\rm cm}$.  In a finite system with conserved magnetic flux,
however, type-I matter can form an intermediate state consisting of
macroscopic normal domains embedded in a superconducting background.

\begin{figure*}[hbt]
    \centering
    \begin{subfigure}[t]{0.45\textwidth}
        \centering
        \includegraphics[height=5.2cm,width=7.cm]{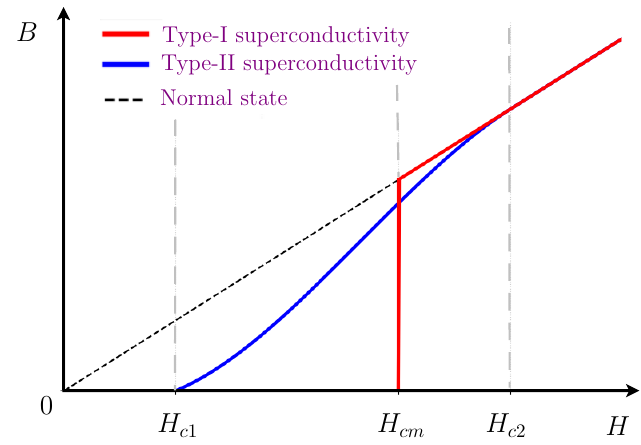}
        \caption{Schematic induction $B$ curves as a function of the
          applied magnetic-field intensity
          $H$ for type-I (red) and type-II (blue) superconductors. For a type-I superconductor,
          the Meissner state minimizes the free energy for $H<$ $H_{c
            m}$, while the system
          becomes normal above the thermodynamic critical field $H_{c
            m}$.
          In a type-II superconductor, the Meissner state is favored
          for $H<H_{c 1}$,
          whereas for $H_{c 1}<H<H_{c 2}$ magnetic flux penetrates in
          the form of quantized
          flux tubes, producing the mixed state; superconductivity is
          destroyed for $H>H_{c 2}$.
          }
        \label{fig:magnetization}
    \end{subfigure}%
    \hfill
    \begin{subfigure}[t]{0.45\textwidth}
        \centering
        \includegraphics[height=5.5cm,width=6.8cm]{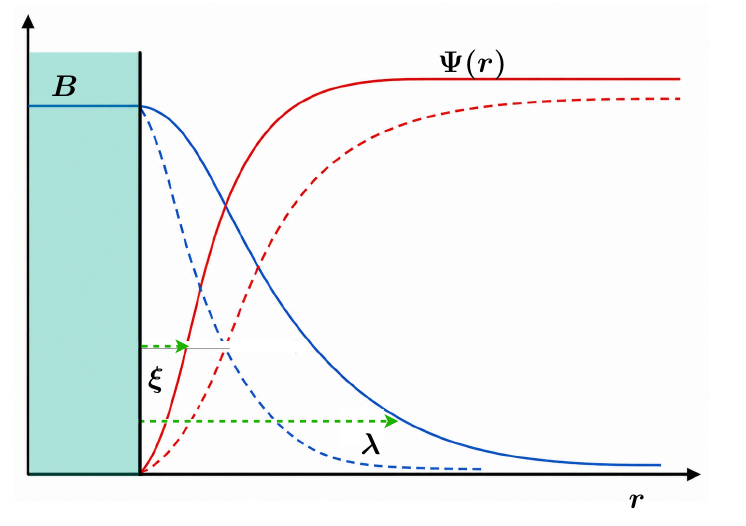}
        \caption{ Schematic profiles of the condensate wave function
          and magnetic field across the interface of a type-II (solid lines) and
          type-I (dashed lines) superconductor (unshaded region) and normal/vacuum
          region (shaded region). In each case, the magnetic
          field penetrates into the superconducting phase over the
          penetration depth $\lambda$, whereas the condensate wave
          function recovers its bulk value over the coherence length
          $\xi$. The relative magnitudes of $\lambda$ and $\xi$
          distinguish the two superconducting types: $\lambda<\xi$ for
          type-I superconductors and $\lambda>\xi$ for type-II
          superconductors.
        }
    \label{fig:GL_profiles}          
    \end{subfigure}
   \caption{Schematic comparison of type-I and type-II superconductors. The left panel 
shows the magnetic induction $B(H)$ across the Meissner, mixed (flux-tube), 
and normal states; the right panel shows the characteristic magnetic-field 
and condensate wave-function $\Psi$ profiles.}
\end{figure*}

The magnetic configuration of a neutron-star core need not coincide
with this equilibrium phase diagram. The superconducting transition
takes place on a cooling timescale that is extremely short compared
with the timescale required to transport magnetic flux over stellar
distances~\citep{Baym1969a}. The electrical conductivity is
very high, and the pre-existing magnetic flux is therefore
approximately conserved during the transition. Global Meissner
expulsion cannot occur instantaneously.

The morphology of the trapped flux depends on the superconducting
type. In a type-II region, the magnetic field is carried by
microscopic quantized flux tubes, each containing one flux quantum
$\Phi_0$~\citep{Baym1969a,SedrakianCluster1995,Jones2006a,Glampedakis2010}.
In a type-I region, isolated Abrikosov tubes are not the stable flux
structures. Conserved flux instead segregates into macroscopic normal
domains, sheets, or tubes in which the local field is of order
$H_{\rm cm}$, separated by superconducting regions.  Complete Meissner
expulsion is recovered only if the magnetic flux has sufficient time
and mobility to leave the superconducting region.

For a type-II proton superconductor, the lower critical field is
approximately~\citep{Abrikosov:Fundamentals}
\bea\label{eq:Hc1}
H_{c1} \simeq \frac{\Phi_0}{4\pi\lambda^2}\ln\kappa_{\rm GL},
\eea
above which flux-tube formation becomes energetically favorable.
The corresponding upper critical field of an
isolated proton superconductor is
\bea\label{eq:Hc2}
H_{c2}=\frac{\Phi_0}{2\pi\xi_p^2},
\eea
at which the Larmor radius (magnetic length) of a charged Cooper pair
becomes comparable to the coherence length and superconductivity is
destroyed. Close to $H_{c2}$ the GL equations can be linearized and
they admit only the normal-state solution once $H \ge H_{c2}$.  In
neutron-star matter these isolated-superconductor estimates are
modified by the coexistence of the proton condensate with the neutron
superfluid, which will be discussed below.

The distinction between type-I and type-II behavior is illustrated
schematically in Figs.~\ref{fig:magnetization} and
\ref{fig:GL_profiles}. The first shows the magnetic induction and the
critical fields separating the Meissner, mixed flux-tube, and normal
states, while the second shows the different spatial scales over which
the condensate and magnetic field recover their bulk values, as
determined by $\xi_p$ and $\lambda$.

In a charged superfluid, the canonical momentum of a Cooper pair
contains both kinetic and electromagnetic contributions.  The
gauge-invariant proton superfluid velocity is
\be
 \bm{v}_p
 =
 \frac{\hbar}{2m_p^*}\nabla\chi_p
 -\frac{e}{m_p^*c}\bm{A},
\ee
where $\chi_p$ is the phase of the proton condensate and $m_p^*$ is the
effective proton mass.  A proton flux tube is a topological defect
around which the condensate phase has a nonzero winding number,
\be
 \oint \bm{dl}\cdot\nabla\chi_p=2\pi n,
 \qquad n\in\mathbb{Z}.
\ee
Because the proton current is screened by the magnetic field,
$\bm{v}_p$ decays exponentially on the scale of the London penetration
depth $\lambda$.  Choosing a contour sufficiently far from the flux-tube
axis therefore gives
\be
 \oint\bm{dl}\cdot\bm{v}_p
 =
 0
 =
 \frac{n\pi\hbar}{m_p^*}
 -\frac{e}{m_p^*c}\Phi,
\ee
where $\Phi=\oint\bm{A}\cdot\bm{dl}$ is the magnetic flux carried by the
flux tube.  It follows that the flux is quantized.  For a singly
quantized flux tube, $n=1$, the flux quantum is
\be
 \Phi_0=\frac{\pi\hbar c}{e}
 \simeq 2.07\times10^{-7}\ {\rm G\,cm^2}.
\ee
The areal density of flux tubes is consequently $n_\Phi=B/\Phi_0$.
For a triangular flux-tube lattice, the mean separation is
\bea
 d_\Phi= \left(\frac{2\Phi_0}{\sqrt{3}B}\right)^{1/2}
 \simeq 5\times10^{-10}\left(\frac{10^{12}\ {\rm G}}{B}\right)^{1/2}
 {\rm cm}.
\eea
This spacing is many orders of magnitude smaller than the intervortex
distance in the rotating neutron superfluid.  The disparity arises
because the number densities of the two types of defects are controlled
by very different macroscopic quantities: the flux-tube density is
fixed by the magnetic induction, $n_\Phi=B/\Phi_0$, whereas the neutron
vortex density is fixed by the rotation rate,
$n_v=2\Omega/\kappa$.  Equivalently, a neutron vortex is electrically
neutral and supports an unscreened velocity field that falls off as
$1/r$, while the current surrounding a proton flux tube is screened
exponentially on the scale $\lambda$.  For representative pulsar values
$B\sim10^{12}\ {\rm G}$ and $\Omega\sim10^2\ {\rm s^{-1}}$, the ratio
$ {n_\Phi}/{n_v}= B\kappa/2\Omega\Phi_0 \sim 10^{13}\text{--}10^{14}
$ shows that approximately this many proton flux tubes occupy the area
associated with a single neutron vortex.

The magnetic field around an isolated flux tube is localized on the
scale $\lambda$. In the London limit, in which spatial variations of the magnitude of the superconducting order parameter are neglected outside the vortex core and only its phase and the electromagnetic field are allowed to vary, the field profile is
\bea
B_\Phi(r)
\simeq
\frac{\Phi_0}{2\pi\lambda^2}
K_0\left(\frac{r}{\lambda}\right),
\eea
where $K_0$ is a modified Bessel function. The proton order parameter
is suppressed within a core of radius $\sim\xi_p$, while the
circulating supercurrent and magnetic field extend to distances of
order $\lambda$. The line energy, or tension, of a flux tube is
\citep{Easson1977}
\bea
\epsilon_\Phi
\simeq
\left(
\frac{\Phi_0}{4\pi\lambda}
\right)^2
\ln\left(\frac{\lambda}{\xi_p}\right),
\eea
which governs the resistance of the tube to bending and its response
to external stresses.

The coexistence of proton flux tubes and neutron vortices couples
magnetic and rotational degrees of freedom. In a two-fluid
neutron--proton system, the mass currents are related to the
superfluid velocities through the entrainment matrix
\citep{Sedrakian1980,Alpar1984b,Chamel2006}
\bea
\begin{pmatrix}
\bm{j}_n \\[4pt]
\bm{j}_p
\end{pmatrix}
=
\begin{pmatrix}
\rho_{nn} & \rho_{np} \\[4pt]
\rho_{np} & \rho_{pp}
\end{pmatrix}
\begin{pmatrix}
\bm{v}_n \\[4pt]
\bm{v}_p
\end{pmatrix},
\eea
where $\bm{j}_n$ and $\bm{j}_p$ are the neutron and proton
mass-current densities, $\bm{v}_n$ and $\bm{v}_p$ are the corresponding
superfluid velocities, $\rho_{nn}$ and $\rho_{pp}$ are the diagonal
entrainment coefficients, and $\rho_{np}$ is the off-diagonal
entrainment coefficient. The diagonal and off-diagonal elements satisfy
the sum rules
\bea
\rho_{nn}+\rho_{np}=\rho_n,
\qquad
\rho_{pp}+\rho_{np}=\rho_p,
\eea
where $\rho_n$ and $\rho_p$ are the total neutron and proton mass
densities. In terms of the nucleon effective masses $m_n^*$ and
$m_p^*$, the matrix elements may be written as
\bea
\rho_{nn}=\rho_n\frac{m}{m_n^*},
\qquad
\rho_{pp}=\rho_p\frac{m}{m_p^*},
\qquad
\rho_{np}
=
\rho_p\left(1-\frac{m}{m_p^*}\right) = \rho_n\left(1-\frac{m}{m_n^*}\right),
\eea
where $m$ is the bare nucleon mass and we do not distinguish between
the bare masses of the neutron and proton.  Galilean invariance
requires the expressions obtained from the neutron and proton sum
rules to give the same off-diagonal coefficient $\rho_{n p}$. The
off-diagonal coefficient $\rho_{np}$ vanishes in the absence of
neutron--proton interactions and quantifies nondissipative
entrainment: the neutron mass current depends on both
$\boldsymbol v_n$ and $\boldsymbol v_p$, and similarly for the proton
mass current. The relations given above apply within the standard
nonrelativistic Landau--Fermi-liquid convention for the effective
masses, with the Galilean-invariance sum rules imposed.

Neutron vortices carry the circulation quantum given by
Eq.~\eqref{eq:circulation}. Although neutrons are electrically neutral,
the entrainment of the proton condensate by the neutron superflow
magnetizes each vortex. Integrating the entrainment-induced proton
current around a closed contour encircling the vortex yields a
non-quantized magnetic flux attached to each neutron vortex,
\bea
 \Phi_n= k_{\rm ent}\Phi_0,
  \qquad
  k_{\rm ent}\equiv\frac{\rho_{np}}{\rho_{pp}}
  =\frac{m_p^*}{m}-1,
\eea
where $k_{\rm ent}$ is the entrainment coefficient. Since $m_p^*<m$ in the dense
outer core, $k_{\rm ent}$ is negative, meaning that the entrainment-induced flux
is anti-parallel to the vortex circulation. The magnitude of $k_{\rm ent}$ is
typically of order $0.2$--$0.4$ in the outer core, so the induced flux
per neutron vortex is a non-negligible fraction of $\Phi_0$.

This magnetization has several important consequences. First, it
provides a microscopic channel through which electrons and protons
scatter off neutron vortices, generating the mutual friction that
couples the neutron superfluid to the charged component in the core
\citep{Alpar1984b}. The magnetic field of a neutron vortex in the
scenario of \cite{Alpar1984b} is screened beyond the penetration depth
$\lambda$ and the average induced field is vanishingly small compared
to the typical fields expected in neutron stars. 

\subsection{Vortex cluster model}

\begin{figure}[t]
\centering\includegraphics[width=.8\linewidth]{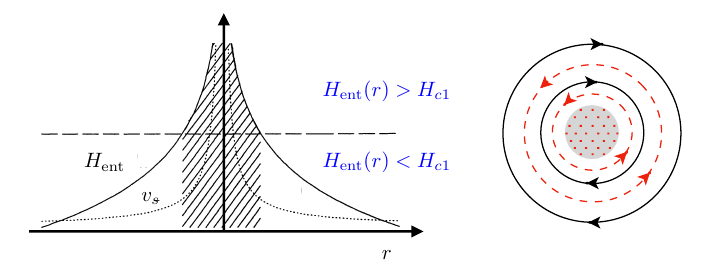}
\caption{Schematic structure of a vortex cluster. {\it Left}---Radial
  profiles of the entrainment-induced magnetic intensity
  $H_{\rm ent}(r)$ and neutron-superfluid velocity are shown together
  with the lower critical field. The hatched region identifies the
  interval in which the magnetic intensity $\vert H_{\rm ent}(r)\vert > H_{c1}$
  and proton flux-tube formation is energetically favorable. {\it
    Right}--- Solid concentric circles show the neutron-superfluid
  circulation around a neutron vortex, while dashed circles indicate
  the entrained proton-superfluid flow generated by the strong
  neutron-proton interaction. For typical proton effective masses, the
  entrainment coefficient is negative, so the neutron and proton
  currents circulate in opposite directions. The shaded region marks
  the domain in which the entrainment-induced magnetic intensity
  exceeds the lower critical field and the formation of proton flux
  tubes is energetically favored. A triangular flux-tube lattice (not to
  scale) is shown schematically within this region. For a Vela-like
  pulsar, the neutron intervortex spacing is of order
  $ d_v \simeq 4\times10^ {-3}$~cm, the cluster radius is typically
  $0.1 d_v $, and the inter-flux-tube spacing is of order
  $10^{-10}$~cm.}
\label{fig:Cluster}
\end{figure}

An alternative description of the magnetic structure associated with
neutron vortices is provided by the vortex-cluster
scenario~\citep{Sedrakian1983}. The essential ingredient is
neutron--proton entrainment. In an entrained two-superfluid mixture,
the electric current carried by the proton condensate is not determined
solely by the proton superfluid velocity. In the frame of the normal
electron component it has the form
\begin{equation}
 {\boldsymbol j}_p
 =
 \frac{e}{m_p}
 \left(
   \rho_{pp}{\boldsymbol v}_p
   +
   \rho_{pn}{\boldsymbol v}_n
 \right),
 \label{eq:proton-current-entrainment}
\end{equation}
where $\rho_{pp}$ is the diagonal proton superfluid-density
coefficient and $\rho_{pn}$ is the off-diagonal entrainment
coefficient. The second term in Eq.~\eqref{eq:proton-current-entrainment}
is present even when the proton condensate has no independent phase
circulation. A quantized neutron vortex therefore acts as a source of
electric current through the circulating neutron velocity \eqref{eq:v_rn}.
This is the microscopic origin of the magnetic structure in the
vortex-cluster model.

It is useful to distinguish the magnetic-field intensity
${\boldsymbol H}$ generated by this source current from the magnetic
induction ${\boldsymbol B}$ that remains after the proton
superconductor has responded. Amp\`ere's equation for the
entrainment-current contribution reads
\begin{equation}
 \boldsymbol{\nabla}\times{\boldsymbol H}_{\rm ent}
 =
 \frac{4\pi}{c}{\boldsymbol j}_{\rm ent}
 =
 \frac{4\pi e}{m_pc}\rho_{pn}{\boldsymbol v}_n .
 \label{eq:curl-H-entrainment}
\end{equation}
For a straight neutron vortex, cylindrical symmetry implies that
${\boldsymbol H}_{\rm ent}=H_{\rm ent}(r)\hat{\boldsymbol\kappa}_n$.
Substitution of Eq.~\eqref{eq:v_rn}  into
Eq.~\eqref{eq:curl-H-entrainment}, followed by radial integration,
gives
\begin{equation}
 {\boldsymbol H}_{\rm ent}(r)
 =
 \hat{\boldsymbol\kappa}_n
 \frac{\Phi_n}{2\pi\lambda^2}
 \ln\left(\frac{d_v}{r}\right),
 \label{eq:H-vortex-cluster}
\end{equation}
where $d_v$ is the neutron intervortex distance defined in
Eq.~\eqref{eq:d_v}. The boundary condition $H_{\rm ent}(d_v)\simeq0$
expresses the fact that the field associated with an individual vortex
is screened on scales comparable to its vortex
cell. Equation~\eqref{eq:H-vortex-cluster} is valid outside the
microscopic vortex core; its logarithmic increase at small $r$ must be
cut off at a distance of order the relevant coherence length -- $\max \left(\xi_p, \xi_n\right)$. For the
sign of $\rho_{pn}$ usually found in microscopic calculations, $k_{\rm ent}<0$,
the entrainment-induced field is directed opposite to the
neutron-vortex circulation.

The important point is that the source
${\boldsymbol H}_{\rm ent}(r)$ is not an optional magnetic field added
to the system. Once the neutron superfluid rotates by forming
quantized vortices and the entrainment coefficient is nonzero, the
circulating neutron flow necessarily produces the current appearing
in Eq.~\eqref{eq:curl-H-entrainment}. Rotation therefore generates a
magnetic-field intensity around every neutron vortex even in the
absence of a pre-existing fossil field. The proton condensate must
respond to this source. At sufficiently large radii, where
\begin{equation}
 |H_{\rm ent}(r)|<H_{c1},
\end{equation}
the response is a Meissner countercurrent, which screens the
entrainment current without introducing proton phase singularities.
Closer to the neutron-vortex axis, however,
$|H_{\rm ent}(r)|$ increases logarithmically. When
\begin{equation}
 |H_{\rm ent}(r)|\geq H_{c1},
 \label{eq:cluster-criterion}
\end{equation}
complete Meissner screening is no longer the state of lowest Gibbs
free energy. In a type-II proton superconductor, the energetically
favored response is then the appearance of quantized proton flux
tubes.

This statement is stronger than a phenomenological assumption that
proton flux tubes happen to become pinned to neutron vortices. Within
the equilibrium vortex-cluster model, the flux tubes occur because
the vortex-free proton state ceases to minimize the Gibbs free energy
in the region where Eq.~\eqref{eq:cluster-criterion} is satisfied.
The radius $r_{\rm cl}$ of this region is determined by
\begin{equation}
 |H_{\rm ent}(r_{\rm cl})|=H_{c1}.
 \label{eq:cluster-radius-definition}
\end{equation}
Using the expression \eqref{eq:Hc1} for $H_{c1}$ one obtains
\begin{equation}
 r_{\rm cl}
 =
 d_v
 \exp\left[
  -\frac{2\pi\lambda^2 H_{c1}}{|\Phi_n|}
 \right]
 =
 d_v
 \left(\frac{\xi_p}{\lambda}\right)^{
  1/(2|k_{\rm ent}|)} .
 \label{eq:cluster-radius}
\end{equation}
Although this radius is small compared with $d_v$, it can reach a macroscopic fraction of the neutron intervortex distance,
\begin{equation}
 r_{\rm cl}\sim10^{-2}\text{--}10^{-1}d_v,
 \qquad
 \lambda\ll r_{\rm cl}\ll d_v .
\end{equation}
Thus, although the cluster occupies only a small part of the neutron
vortex cell, its transverse dimension is many orders of magnitude
larger than the microscopic penetration depth. The resulting
configuration, illustrated in Fig.~\ref{fig:Cluster}, consists of a
neutron vortex surrounded by a dense, approximately triangular
lattice of proton flux tubes. The proton flux tubes are coaxial with
the neutron vortex, with their orientation selected by the sign of
the entrainment-induced field.

The number of flux tubes associated with one neutron vortex can be
estimated from
\begin{equation}
 N_{\Phi}^{\rm cl}
 \simeq
 \frac{1}{\Phi_0}
 \int_{r<r_{\rm cl}} B(r)\,dS
 \sim
 \frac{\pi r_{\rm cl}^2\langle B_{\rm cl}\rangle}{\Phi_0}
 \label{eq:number-cluster-fluxtubes}
\end{equation}
and can reach $ N ^{\rm cl}_{\Phi}\sim10^{12}\text{--}10^{13}.$

The local induction inside the cluster can be of order
$10^{14}\,{\rm G}$, whereas averaging over the much larger neutron
vortex cell gives a rotation-induced macroscopic field of order
$10^{11}$--$10^{12}\,{\rm G}$~\citep{SedrakianCluster1995}. The latter
can be comparable to the fossil field of an ordinary pulsar, although
this comparison is obviously model- and object-dependent. In a
realistic star the total magnetic structure must include both the
fossil flux and the rotation-induced contribution. Existing fossil
flux tubes may be redistributed or concentrated around neutron
vortices rather than being newly nucleated from a completely flux-free
state.

The word ``formation'' should therefore be understood primarily in
the thermodynamic sense. The vortex-cluster calculation establishes
that a flux-tube state has lower Gibbs free energy than a locally
vortex-free proton condensate wherever
$|H_{\rm ent}|>H_{c1}$. It does not by itself determine the time
required to reach this state or the nucleation barrier that must be
overcome. In a newly formed neutron star, clusters may be established
during the transition to proton superconductivity. In an older star,
their realization may instead involve the motion, capture, and
rearrangement of flux tubes already associated with the fossil
magnetic field.

The vortex-cluster picture relies on several assumptions. First, the
proton condensate must be a type-II superconductor, so that the
response above $H_{c1}$ is an Abrikosov flux-tube lattice. If the
proton condensate is of type I, the same entrainment source remains
present, but the preferred response is expected to be a normal-proton
magnetic domain coaxial with the neutron vortex rather than a bundle
of individual flux tubes~\citep{Sedrakian1997}. Second, the
entrainment must be sufficiently strong that the maximum field near
the neutron-vortex core exceeds $H_{c1}$. Introducing a cutoff
$r\sim\xi_p$ in Eq.~\eqref{eq:H-vortex-cluster}, this condition may
be written approximately as
\begin{equation}
 2|k_{\rm ent}|
 \ln\left(\frac{d_v}{\xi_p}\right)
 >
 \ln\left(\frac{\lambda}{\xi_p}\right).
 \label{eq:cluster-existence-condition}
\end{equation}
Thus, nonzero entrainment guarantees a rotation-induced magnetic
source, but an arbitrarily weak entrainment coefficient does not
necessarily guarantee a finite flux-tube cluster.

Third, the derivation assumes a locally straight neutron vortex,
approximately cylindrical symmetry, and well-separated neutron
vortex cells. It also employs the London approximation outside the
proton and neutron coherence-length scales. Fourth, it assumes that
the system can reach or remain sufficiently close to thermodynamic
equilibrium. Strong nucleation barriers, pinning, or an inherited
magnetic topology may delay or prevent complete relaxation to the
minimum-energy cluster configuration. Finally, magnetic flux tubes
cannot terminate freely inside the superconducting core. Their
large-scale continuation, closure, or connection to the background
fossil field must be consistent with the global magnetic topology of
the star.

Subject to these assumptions, the cluster is not an additional
phenomenological structure imposed on a neutron vortex. It is the
equilibrium response of the proton condensate to the magnetic source
that inevitably accompanies neutron circulation in an entrained
neutron--proton superfluid. The neutron vortex and its surrounding
bundle of proton flux tubes then form a strongly magnetized composite
object. Relativistic electrons scatter from the collective magnetic
field of the bundle rather than from the much weaker field of an
isolated entrainment-magnetized neutron vortex. This can substantially
reduce the electron relaxation time, enhance the effective drag, and
increase the mutual-friction coupling between the neutron superfluid
and the charged component of the star
\citep{SedrakianCluster1995}.

\subsection{Corrections to the critical fields and the
type-I/type-II boundary}
\label{sec:critical-fields-coupled}

The standard one-component GL criterion,
$\kappa_{\rm GL}=1/\sqrt{2}$, provides the natural starting point for
classifying proton superconductivity as type I or type II.  In
neutron-star matter, however, the proton condensate coexists with a
neutron superfluid and may couple to it through density--density and
derivative, or current--current, interactions.  These couplings can
modify the surface energy between normal and superconducting regions
and thereby shift the boundary between type-I and type-II behavior.

The magnitude of this effect remains model dependent.  In the limit
of approximate isospin symmetry, a sufficiently strong attractive
coupling between neutron and proton condensates may favor type-I
superconductivity and an intermediate-state domain structure
\citep{Buckley2004}.  Neutron-star matter is, however, strongly
isospin asymmetric and contains only a small proton fraction.  In
microscopic pairing models the neutron--proton condensate coupling
then vanishes at mean-field level and remains relatively small beyond
mean field, so that the usual one-component classification is not
generically invalidated \citep{Alford2005}.

More general two-component GL theories nevertheless show that density
and gradient couplings can shift the transition region and replace the
sharp one-component boundary by a broader range of parameters with
nonstandard flux-tube behavior \citep{AlfordGood2008,Haber2017}.  In
particular, the transitions into and out of the flux-tube phase need
not occur continuously at the conventional fields $H_{c1}$ and
$H_{c2}$.  They may instead become first-order transitions at shifted
fields, conventionally denoted by
\be
 H'_{c1}<H_{c1},
 \qquad
 H'_{c2}>H_{c2}.
\ee
Clustered or spatially inhomogeneous flux-tube states may therefore
occur close to the nominal type-I/type-II boundary.

Entrainment also modifies the upper critical field through the
coupling of the proton condensate to the neutron superfluid.  Within
GL theory, this correction can change $H_{c2}$ by several tens of
percent \citep{SinhaSedrakian2015}.  Because the proton coherence
length is controlled by the density-dependent proton pairing gap,
$H_{c2}(\rho)$ is generally largest near the crust--core interface
and decreases toward the stellar center.  Strong internal magnetic
fields may consequently destroy proton superconductivity first in
selected density layers, producing a partially superconducting core.
The resulting spatial distribution of type-I, type-II, and normal
regions in magnetars is discussed below in Sec.~\ref{sec:superconductivity_magnetars}.

\subsection{Vortex--flux-tube interactions and magnetic evolution}

Proton flux tubes and neutron vortices interact through several
mechanisms. Long-range interactions arise from the magnetic and
hydrodynamic fields of the entrained proton currents; shorter-range
effects occur when the two defects cross and their cores overlap.
Depending on their relative orientation, flux tubes may pin to neutron
vortices, resist vortex motion, or be cut through by vortices
\citep{Muslimov1985,Sauls1989,Srinivasan1990}. The response
  of flux tubes under the force exerted by vortices is determined by
  the drag force of the former \citep{Gusakov2019}.  The geometry is
not universal: poloidal, toroidal, and mixed magnetic fields lead to
complicated relative orientations of vortex and flux-tube arrays
\citep{Lander2013,Graber2015}. Stable neutron-star magnetic equilibria
are expected to contain both poloidal and toroidal components in a twisted-torus configuration~\citep{Braithwaite2009}. Such
mixed-field geometries can produce a complicated spatial topology of
type-I, type-II, and normal regions in neutron-star cores
\citep{Das2025,Das2026}.

The global evolution of flux tubes remains an open problem. Their
motion is driven by buoyancy, tension, drag, electron scattering,
vortex interactions, and stresses from the evolving magnetic field. If
flux tubes are tightly locked to the neutron vortex array, secular
spin-down transports magnetic flux through the core; if vortices
instead cut through flux tubes, the magnetic and rotational evolution
may partly decouple. Distinguishing these regimes is essential for
realistic models of field evolution, precession, glitch generation, and
recovery, and long-term thermal evolution of neutron stars.

\subsection{Superconductivity in magnetar-strength fields}
\label{sec:superconductivity_magnetars}

\begin{figure}[t]
\centering\includegraphics[width=.9\linewidth]{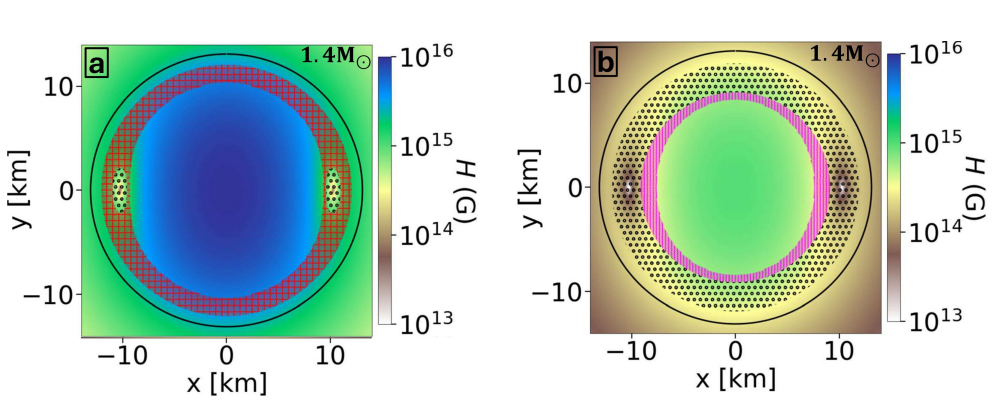}
\caption{Color maps of the poloidal magnetic-field strength $H$ in the
  $x$--$y$ plane for a neutron star with mass $M=1.4\,M_{\odot}$ and
  temperature $T=10^{8}\,\mathrm{K}$ constructed using the DDME2
  equation of state, see \cite{Das2026}. The magnetic axis is aligned
  with the $y$-axis. Panels (a) and (b) correspond to maximum surface
  fields $B_{S,\max}=10^{16}\,\mathrm{G}$ and $10^{15}\,\mathrm{G}$,
  respectively. Red crosses indicate type-II superconducting regions
  with $\kappa_{\rm GL}>1/\sqrt{2}$ and $H_{c1}<H<H_{c2}$, where
  magnetic flux is carried by an Abrikosov flux-tube lattice. Black
  dots mark regions satisfying $\kappa_{\rm GL}>1/\sqrt{2}$ and
  $H<H_{c1}$, corresponding to the equilibrium Meissner state of a
  type-II superconductor. Magenta vertical hatching denotes type-I regions with $\kappa_{\rm GL}<1/\sqrt{2}$ and $H<H_{cm}$, where the equilibrium state is Meissner-like or may contain an intermediate layered-domain structure when magnetic flux is conserved.  Unhatched central-core regions are nonsuperconducting, with $H>H_{c2}$ in the type-II regime or $H>H_{cm}$ in the type-I regime.  Because magnetic-flux expulsion from a neutron-star core is expected to be slow, metastable flux tubes may persist in regions with $H<H_{c1}$, while normal magnetic domains may survive in type-I regions with $H<H_{cm}$.  Adapted from~\cite{Das2026}.} 
\label{fig:B_field}
\end{figure}

In ordinary pulsars, the internal magnetic field is usually assumed to
lie below the upper critical field $H_{c2}$ of the proton
superconductor, so that the type-II flux-tube picture described above
can apply throughout much of the outer core. Magnetars, however, have
surface dipole fields of order $10^{15}$~G, and their internal fields
may be substantially larger. This raises the question of whether proton
superconductivity survives at all under magnetar conditions.

Within GL theory, the upper critical field $H_{c2}$ at which type-II
superconductivity is destroyed is given by
Eq.~\eqref{eq:Hc2}. Microscopically, this equation depends on
$\xi_p$, which in turn depends most sensitively on the  proton
pairing gap. Inclusion of the correction arising from entrainment of
the proton condensate by the neutron superfluid  shows that $H_{c2}$
can be modified by up to several tens of percent
\citep{SinhaSedrakian2015}. The resulting $H_{c2}$ is maximal near the
crust--core boundary and decreases toward the stellar center,
reflecting the density dependence of the proton gap.

As a consequence, magnetars with internal fields in the range
$10^{15}\lesssim B\lesssim 5\times 10^{16}$~G may be only partially
superconducting: the field can quench superconductivity in the inner
core while leaving the outer core  -- the portion of the core near the crust–core interface --
superconducting. Details require specific pairing gaps, magnetic-field configurations, an equation of state, and a temperature distribution. For example, a two-dimensional general-relativistic
analysis of superconductivity in magnetars with toroidal and
twisted-torus magnetic fields~\citep{Das2025,Das2026} revealed a
complicated distribution of type-I and type-II superconducting
regions, see Fig.~\ref{fig:B_field}. For $B\gtrsim10^{16}$~G, 
the proton condensate may be entirely destroyed throughout the core, implying important modifications to the cooling of magnetars.  Neutrino cooling of magnetars (as well as lower field neutron stars) proceeds through several
weak-interaction channels.  The direct Urca process,
$n\to p+e+\bar\nu_e$, together with the inverse reaction
$p+e\to n+\nu_e$, is highly efficient but is allowed only when the
particle Fermi momenta satisfy momentum conservation.  Below this
threshold, neutrino emission can proceed through the slower modified Urca process, $N+n\to N+p+e+\bar\nu_e$, where the additional nucleon $N=n,p$ acts as a spectator and supplies the required momentum. A magnetic field can broaden the momentum-conservation condition and thereby permit or enhance direct Urca emission near or below its zero-field threshold~\citep{YakovlevPethick2004}.

Superfluidity and superconductivity suppress the Urca rates because
the participating quasiparticles must overcome the corresponding
pairing gaps.  At the same time, the breaking and subsequent
recombination of Cooper pairs provides an additional neutrino-emission
channel, schematically
$[nn]\to[nn]+\nu+\bar\nu$ (and analogously for proton pairs), known as
Cooper-pair breaking and formation.  Destruction of proton
superconductivity therefore removes the gap-induced suppression of
reactions involving proton quasiparticles, while the proton
pair-breaking and formation channel itself vanishes
~\citep{SinhaSedrakian2015}.

The spatial distribution of superconducting regions is further
complicated by the geometry of the internal magnetic field. Early
studies used one-dimensional models with prescribed, spherically
symmetric field profiles, which already revealed a layered structure:
type-II behavior in the outer core, a possible transition to type-I
behavior at higher density, and non-superconducting matter in the inner
core for sufficiently strong fields. A first two-dimensional,
general-relativistic analysis using axially symmetric magnetar models
with dominant toroidal fields, solved via the Einstein--Maxwell
equations with the \textsc{XNS} code, revealed qualitatively new
features \citep{Das2025}. The outer cores of
low- to intermediate-mass magnetars sustain superconductivity over
larger volumes than their higher-mass counterparts, and the
distribution of type-I and type-II regions is  complex. Most
strikingly, these two-dimensional models contain non-superconducting
regions with toroidal topology --- doughnut-shaped voids in the
superconducting domain --- a feature absent from one-dimensional
studies and directly linked to the toroidal field geometry.

A more comprehensive general-relativistic study, incorporating both
toroidal and poloidal field geometries and microscopically derived
proton pairing gaps from realistic equations of state, extended this
picture further \citep{Das2026}; see Fig.~\ref{fig:B_field}. The
superconducting topology depends sensitively on the interplay between
field geometry, stellar mass, and the density dependence of the
pairing gap. Poloidal field configurations produce superconducting
shells with connectivity different from the toroidal case, and mixed
twisted-torus geometries yield still more complex domain
structures. An important finding is that superconducting regions in
millisecond pulsars hosting strong internal toroidal fields can
enhance the amplitude of continuous gravitational-wave emission,
providing a potential observational window into the internal
superconducting state. Overall, these results demonstrate that
the magnetic phase diagram of a neutron-star core --- the spatial map
of type-I, type-II, and non-superconducting domains --- is a genuinely
multidimensional, field-geometry-dependent object whose determination
requires self-consistent relativistic modeling.

\subsection{Interface physics}
\label{subsec:Interface_physics}

Two interfaces in neutron-star interiors are especially relevant for
the coupling between microscopic pairing structure and macroscopic
rotational, magnetic, and thermal evolution.

The first is the crust--core boundary~\citep{Pethick1995}. A flux tube, or a neutron vortex
dressed by a flux-tube cluster, approaching this interface experiences
a surface barrier produced by the interaction between its magnetic
field and the Meissner currents induced by the crustal magnetic field.
This barrier impedes the outward motion of vortices during secular
spin-down and allows angular momentum to accumulate in a superfluid
layer near the boundary. Once the barrier is overcome, rapid vortex
transport transfers angular momentum to the normal component and can
trigger a glitch-like event \citep{Sedrakian1999}. This mechanism
is closely related to the broader problem of vortex--flux-tube
interactions in superconducting neutron-star cores and to the magnetic
coupling between the core and the crust
\citep{Alpar1984b,Muslimov1985,Sauls1989,Srinivasan1990}.

The second interface separates the low-density $^1S_0$ neutron
condensate from the high-density $^3P_2$--$^3F_2$ condensate. It
is expected to be located at subnuclear density  $n\sim 0.5\,n_0$,
where the singlet gap closes and the triplet gap opens. (In
some models an unpaired strip  between the singlet and triplet domains
may exist, so this assumption is model dependent). If this
transition is sufficiently sharp, the interface acts as a Josephson
junction between two distinct neutron superfluid phases
\citep{Sedrakian2025}. However, this is not guaranteed: in some models
the singlet gap closes before the triplet gap becomes appreciable,
leaving a crossover or an extended region in which the pairing is weak. A phase difference $\Delta\chi_n=\chi_{2,n}-\chi_{1,n}$ between the two neutron condensates drives a stationary Josephson current,
\bea
j_n = j_{c,n}
\sin(\Delta\chi_n),
\label{eq:neutron_josephson}
\eea
where $j_{c,n}$ is the current amplitude.
Because the neutron and proton condensates are coupled by entrainment,
this neutral supercurrent also induces a charged proton current through
the interface. As the star spins down and vortices migrate outward,
their passage through the interface generates a time-dependent phase
difference. If proton flux tubes are dragged along with the neutron
vortices, the resulting oscillating charged current radiates energy
from the interface. The associated mean radiated power is estimated as
\bea
\langle P_{\star,\Phi}\rangle
\sim
10^{28}
\left(
\frac{\tau}{10^5\,{\rm yr}}
\right)^{-2}
\left(
\frac{R}{10^6\,{\rm cm}}
\right)^5
\left(
\frac{B}{10^{13}\,{\rm G}}
\right)^{3/2}
{\rm erg\,s^{-1}},
\eea
where $\tau$ is the characteristic spin-down time. This power can
exceed standard Ohmic dissipation in the crust and may contribute to
late-time heating during the photon-cooling epoch
\citep{Sedrakian2025}. The Josephson mechanism therefore provides a
link between microscopic pairing structure, vortex motion, magnetic
flux transport, and the long-term thermal evolution of neutron stars.

\subsection{Type-I superconductivity}
\label{subsec:typeI_superconductivity}

Although proton superconductivity in neutron-star cores is often
assumed to be of type II, type-I may exist  in regions where
$\kappa_{\rm GL}\leq 1/\sqrt{2}$; see
Eq.~\eqref{eq:kappa_GL}. Such a regime can occur locally in BCS-based
models of dense matter, especially at higher densities where the
proton coherence length increases and the penetration depth changes
with the proton fraction and effective mass. In addition, couplings
between neutron and proton Cooper-pair condensates can shift the
effective type-I/type-II boundary. In contrast to a type-II
superconductor, magnetic flux does not penetrate a type-I phase in the
form of quantized Abrikosov flux tubes. Instead, if complete Meissner
expulsion is impossible on macroscopic scales, the system enters an
intermediate state consisting of alternating superconducting and
normal domains. The detailed geometry of this state depends on
magnetic-field strength, surface energy, boundary conditions, flux
conservation, and the nucleation history of the superconducting phase
\citep{Sedrakian1997,Buckley2004,Alford2005,Jones2006b}.

The thermodynamic critical field $H_{\rm cm}$ separating the
superconducting and normal phases is related to the condensation
energy density by
\bea
\frac{H_{\rm cm}^2}{8\pi}
\simeq
\frac{1}{2}N_p(p_{F_p})\Delta^2,
\eea
where $N_p(p_{F_p}) = m_p^*p_{F_p}/2\pi^2\hbar^3$ is the proton
quasiparticle density of states at the
Fermi surface per unit volume, per unit energy, and for one spin
projection, $p_{F_p}$ is the proton Fermi momentum,  and $\Delta$ is the proton pairing gap. For typical
microscopic proton gaps this gives characteristic fields of order
$H_{\rm cm}\sim 10^{14}$--$10^{15}$~G, with substantial density and model dependence.

A simple, Landau-type scaling estimate of the transverse size of the
intermediate-state domains is obtained by balancing magnetic and
surface-energy contributions,
\be
d \sim \sqrt{L \delta},\qquad \delta \equiv \frac{8 \pi \sigma}{H_{c m}^2}
\ee
where $\delta$ is the domain-wall parameter and $\sigma$ is the normal--superconducting surface energy per unit area.  Assuming  that $\delta$ is of
order the coherence length, $\delta \sim \xi_p$, 
\be d \sim \sqrt{L\xi_p},
\ee
where $L$ is the macroscopic length scale of the superconducting
region and $\xi_p$ is the proton coherence length. Numerically,
\bea
d \simeq 7.1\times 10^{-4}
\left(\frac{L}{5\times10^{5}\ {\rm cm}}\right)^{1/2}
\left(\frac{\xi_p}{10\ {\rm fm}}\right)^{1/2}
{\rm cm}.
\eea
For the typical range $\xi_p\simeq 10$--$100$~fm and
$L\simeq 5\times10^{5}$~cm, this gives
$d \simeq 7\times10^{-4}\text{--}2.2\times10^{-3}\ {\rm cm},$
up to geometry-dependent numerical factors of order unity.
Flux conservation fixes the relative
widths of superconducting and normal layers 
\begin{equation}
  f_N=\frac{B}{H_{cm}},
  \qquad
  \frac{d_N}{d_N+d_S}=\frac{B}{H_{cm}},
  \qquad
  \frac{d_S}{d_N}=\frac{H_{cm}}{B}-1,
\end{equation}
where $f_N$ is the normal-domain volume fraction and $B$ is the
average magnetic induction. A more precise relation requires a
specified domain geometry and an explicit energy minimization.  For
$B\sim10^{12}\,\mathrm G$ and $H_{cm}\sim10^{14}\,\mathrm G$, the
superconducting layers are approximately two orders of magnitude wider
than the normal domains in this simple laminar geometry.

The dynamical coupling of a type-I superconducting core to neutron
vortices differs qualitatively from the type-II flux-tube picture. One
possible configuration is a coaxial normal proton domain attached to
each neutron vortex by entrainment-induced magnetic currents. Motion of
this combined vortex--normal-domain structure generates transverse
electric fields in the normal region and leads to Ohmic dissipation.
An alternative configuration contains larger normal domains that
accommodate several neutron vortices; in this case, dissipation arises
from scattering of normal protons off vortex-core quasiparticles. Both
limiting models lead to mutual-friction coefficients that can be much
smaller than those expected when neutron vortices are strongly pinned
to a dense array of type-II flux tubes. Consequently, type-I
superconductivity has been discussed as a possible way to reconcile
core proton superconductivity with long-period Eulerian precession of
isolated pulsars~\citep{Link2003,Sedrakian2005,Jones2006b}.

The microscopic origin of type-I behavior remains
unsettled. \citet{Buckley2004} argued that a strong attractive
coupling between neutron and proton condensates could favor type-I
behavior in neutron-star matter. \citet{Alford2005}
reexamined this mechanism and emphasized that neutron-star matter is
strongly isospin asymmetric; in their pairing model the
neutron--proton condensate coupling vanishes at mean-field level and
remains small beyond mean field, so that the standard type-I/type-II
criterion is not generically overturned. Later work showed that
additional density and gradient couplings can nevertheless shift the
transition region between type-I and type-II behavior
\citep{AlfordGood2008,Haber2017}. Thus, type-I superconductivity
should not be viewed as a universal property of neutron-star cores,
but as a possible local phase whose occurrence depends sensitively on
the density dependence of the pairing gap, entrainment, effective
masses, and the multi-component structure of dense matter.

\subsection*{Open questions and perspectives}

\begin{itemize}
    \item \textit{Location and nature of the type-I/type-II transition.}
    Microscopic calculations of $\kappa_{\rm GL}$ in dense matter
    depend sensitively on the nucleon--nucleon interaction model, the
    treatment of medium polarization, and the entrainment coefficient.
    The density at which the proton superconductor crosses from
    type-II to type-I behavior --- and whether such a transition
    occurs at all --- remains uncertain. Convergence between
    Ginzburg--Landau approaches, microscopic many-body calculations,
    and constraints from neutron-star observations are needed to
    resolve this question.

    \item \textit{Flux-tube dynamics and magnetic-field evolution.}
    Whether proton flux tubes are dragged outward by neutron vortices
    during spin-down, or whether vortices cut freely through the
    flux-tube array, determines whether magnetic and rotational
    evolution are coupled or decoupled on secular timescales. A
    self-consistent treatment of this problem requires quantitative
    knowledge of the vortex--flux-tube interaction energy, the
    pinning and cutting rates, and the back-reaction on both the
    superfluid and the magnetic-field topology.

    \item \textit{Superconducting topology in magnetars.}
    Two-dimensional general-relativistic models have revealed
    non-trivial superconducting domain structures, including toroidal
    non-superconducting voids, that depend on field geometry and
    stellar mass. Extending these models to include realistic
    temperature profiles, time-dependent field evolution, and
    feedback from the rotational dynamics remains to be done, and
    could significantly affect predictions for magnetar cooling,
    X-ray luminosity, and gravitational-wave emission.

  \item \textit{Josephson effect and internal heating.}  The recently
    proposed Josephson mechanism at the $^1S_0$--$^3P_2$ interface
    relies on the assumption that the transition between pairing
    phases is sufficiently sharp to support coherent pair tunneling,
    and that proton flux tubes co-move with neutron vortices on
    secular timescales. Neither assumption has been rigorously
    justified from first principles.  Establishing the conditions
    under which the Josephson effect operates, and quantifying its
    contribution to late-time neutron-star heating relative to other
    dissipation channels, are important open problems.
\end{itemize}

\section{Precession, Tkachenko modes, and long-term rotational variability}
\label{sec:tkachenko_precession}

\subsection{Free precession in a superfluid neutron star}

A rigid body's rotation vector, in the state of minimum energy for fixed angular momentum,  coincides with its angular momentum vector. If excited out of this state, the body with rotational angular frequency $\Omega$ will wobble (or precess) at an angular frequency
\be
\Omega_{\rm pr}=\epsilon\Omega\simeq\frac{\Delta I}{I_1}\Omega,
\ee
where $\Delta I$ is the difference in moment of inertia between the major and minor principal axes of inertia and $I_1$ is the moment of inertia of the principal axis.  

\begin{figure}[t]
\centering\includegraphics[width=.7\linewidth]{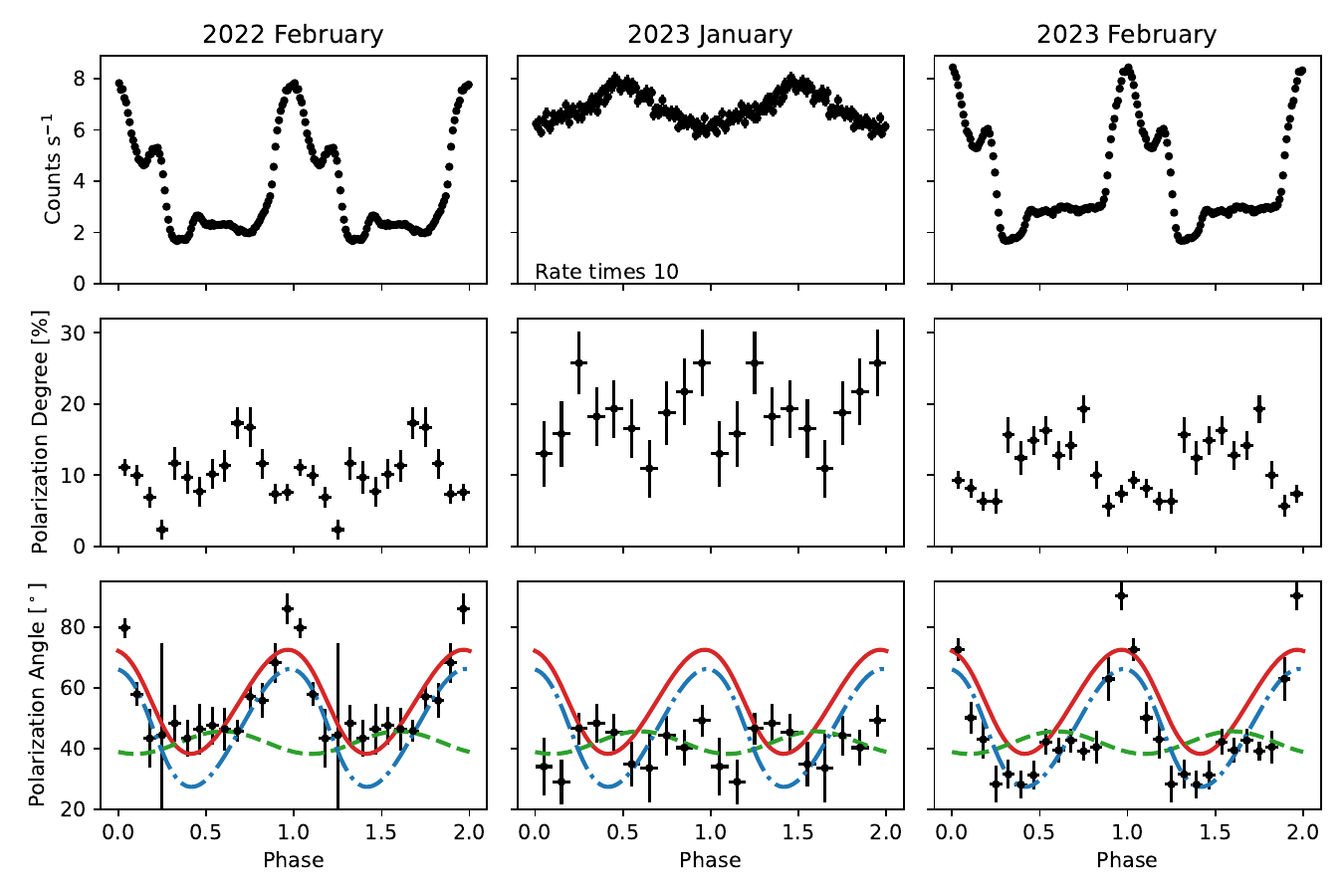}
\caption{Imaging X-ray Polarimetry Explorer (IXPE) observations of Hercules X-1 as a function of spin phase, showing count rate, polarization degree, and polarization angle (from~\citet{Heyl2024}). The strong periodicity in these quantities has been interpreted as evidence for precession of the neutron star. }
\label{fig:Herx1}
\end{figure}

Evidence for slow wobble (precession) of the rigid NS crust has been
seen in Her X-1 \citep{Heyl2024}; see Fig. \ref{fig:Herx1}, and in
the magnetars SGR 1806-20 \citep{Makishima2024}, 4U
0142+61~\citep{Makishima2014} and XTE J1810-197
\citep{Desvignes2024}. Within the precession interpretation the
observations imply $\epsilon\simeq 7\times 10^{-7}$ (Her X-1), $\epsilon\sim 4.5\times 10^{-4}$ (SGR 1806-20),
$\epsilon\sim 2\times 10^{-4}$ (4U 0142+61), and $\epsilon\sim 
10^{-7}-10^{-6}$ (XTE J1810-197). The stellar deformation could be
sustained by crust rigidity \citep{Horowitz2009} or magnetic stresses
\citep{Wasserman2022}.

The quantum fluids in the neutron-star interior fundamentally alter the precession dynamics relative to those of a rigid body.
 In a minimal description, one separates the star
into a crust-plus-charged component with moment of inertia $I_c$ and a
neutron superfluid with moment of inertia $I_s$. If vortices are mobile
and the mutual-friction coupling between the two components is weak,
the star admits a slow precession mode analogous to the classical
rigid-body mode. Its frequency remains controlled primarily by the
stellar deformation. However, dissipation associated with mutual
friction damps this motion, and the slow precession can survive over
many cycles only if the coupling between the normal and superfluid
components is sufficiently weak~\citep{Shaham1977,SedrakianWassermanCordes1999,Jones2001}.
To linear order in $\epsilon$, the asymptotic
behavior of this mode in the weak  ($\beta,\beta'\to 0$)
and strong ($\beta\to 0,\,\beta'\to 1$) couplings is given 
by~\citep{SedrakianWassermanCordes1999}
\bea
\Omega_{{\rm pr},1}^{\rm weak} \simeq
\epsilon\Omega\left[1+(I_s/I_c)\left(\beta^{\prime}+i
    \beta\right)\right],\quad \Omega_{{\rm pr},1}^{\rm strong} \simeq
\epsilon \Omega ~(I_c/I_s)\left[\left(\beta^{\prime}-1\right)+i \beta\right].
\eea
A second, intrinsically superfluid, precession mode appears because
the system possesses two rotational degrees of freedom. In the limit
where a substantial fraction of the superfluid vorticity is pinned to
the crust, the superfluid angular momentum is approximately fixed in
the body frame. The precession frequency  to leading
order in $\epsilon$ and assuming $I_s/I_c\gg\epsilon$ is given by 
\bea
\Omega_{{\rm pr},2}^{\rm weak} \simeq\Omega\left\{-1+(I/I_c)\left(\beta^{\prime}+i \beta\right)\right\},\quad
\Omega_{{\rm pr},2}^{\rm strong} \simeq\Omega\left\{
  (I_s/I_c)+\epsilon+(I/I_c)\left[\left(\beta^{\prime}-1\right)+i
    \beta\right]\right\}
\eea
with $I = I_c+I_s$.
Strongly pinned superfluidity, corresponding to the limiting strong
coupling case $\beta = 0$ and $\beta' =1$ would lead to a much faster
precession than the year-scale periods inferred in candidate
precessing pulsars, as in this case
\bea
\Omega_{{\rm pr},2}^{\rm strong} 
\simeq \left(\epsilon+\frac{I_s}{I_c}\right) \Omega \simeq 
\frac{I_s}{I_c} \Omega .
\eea
This observation, originally emphasized by \citet{Shaham1977}, poses a
well-known tension between long-period precession and the presence of
strongly pinned vortices in the star.
\begin{figure}[t]
\centering\includegraphics[width=.6\linewidth,height = 7.3cm]{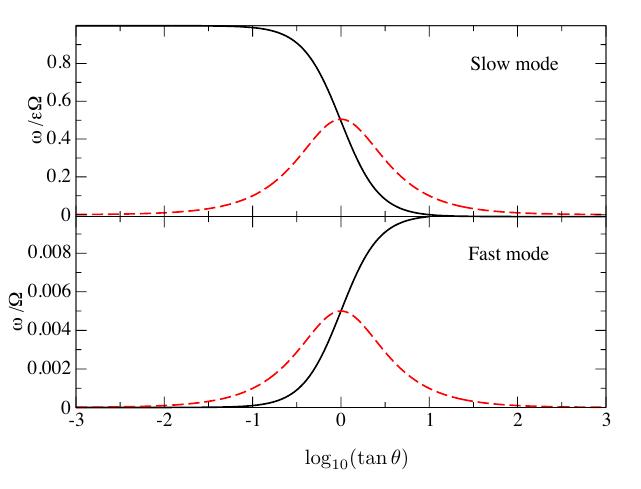}
\caption{ Eigenfrequencies of the precessional modes (solid curves) and their damping rates (dashed curves) for a compact star containing a superfluid component~\citep{SedrakianWassermanCordes1999}, plotted as a function of the dissipation angle, represented as
  $\log_{10}(\tan\theta)$. The solutions shown neglect the transverse (Iordanskii) friction force, i.e., the term $\propto \eta'$ in Eq.~\eqref{dragged_vortex}. The slow mode, shown in the upper panel,
  is normalized to the classical precession frequency
  $\epsilon\Omega$, where $\epsilon$ is the stellar ellipticity. The
  fast mode, shown in the lower panel, has a frequency that is a
  fraction of the stellar rotation frequency $\Omega$.  Adapted from
  \cite{Sedrakian2016}.  }
\label{fig:precession_modes}
\end{figure}

This tension can be reduced if vortex pinning is imperfect, if only a
small fraction of the superfluid is pinned, or if the relevant region
of the star is weakly coupled to the crust. Detailed two-component
models including imperfect pinning and dissipation show that
long-period precession is not excluded, but is possible only within a
restricted range of mutual-friction and pinning parameters
\citep{SedrakianWassermanCordes1999,Wasserman2003,Akgun2006}.
Figure~\ref{fig:precession_modes} illustrates the precessional
eigenmodes and their damping of a neutron star with a superfluid
component, characterized by dissipation angle $\theta$, see
Eq.~\eqref{eq:v_v}.  As seen in the upper panel, in the weak-coupling
limit between the superfluid and the charged component of the star,
the slow branch approaches the classical free-precession mode. As the
coupling increases, its oscillation frequency is suppressed and its
damping becomes comparable to or larger than the real frequency; the
absolute damping rate is maximal near $\tan\theta\simeq 1$, where drag
and Magnus forces are comparable. The slow branch therefore ceases to
represent long-lived free precession in the strong-coupling regime.
The lower panel shows the fast precession mode. The damping vanishes
asymptotically in both the weak- and strong-coupling limits, while its
characteristic frequency scale is set by $\Omega$, rather than by
$\epsilon\Omega$.  In the asymptotic strong-coupling limit, this
branch approaches an undamped fast-precession mode.  The observational
interpretation is nevertheless non-unique: some timing and
pulse-profile modulations attributed to free precession may instead
arise from magnetospheric state switching or related variations in the
external torque~\citep{Jones2012,Ashton2017,Jones2017}. It remains an open question whether or not underdamped, long-period precession can occur
when pinned vortices move through the thermal activation process
described in Section~\ref{thermal-effects}.

A qualitatively distinct situation arises in magnetars, where the
strong internal field can quench proton superconductivity in part of
the core. In non-superconducting regions, the $P$-wave neutron
superfluid couples to the stellar plasma through scattering of protons
off quasiparticles confined in neutron vortex cores by the strong
nuclear force. The resulting crust--core coupling timescales span from
a few seconds in the deep core to several minutes near the crust--core
interface. Two important consequences follow: first, oscillation
models that assume a completely decoupled core superfluid are
difficult to reconcile with this rapid coupling; second, magnetar
precession is strongly damped and, if observed, would likely require
sustained or recurrent excitation within this
model~\citep{Sedrakian2016}. Of course, these conclusions are model
dependent given the uncertainties in the magnitude and density
dependence of the gap, as well as in other model inputs.

\subsection{Vortex lattice elasticity and Tkachenko modes}

A uniformly rotating neutron superfluid contains an array of singly
quantized vortices with areal density
\bea
n_v = \frac{2\Omega}{\kappa},
\qquad
\kappa=\frac{\pi\hbar}{m_n},
\eea
where $\kappa$ is the circulation quantum for a neutron Cooper pair (see Section~\ref{sec:vortex_creep}). Each vortex resists bending on account of its self-energy (tension), of order $\rho_s\kappa^2$. Over scales much larger than the intervortex spacing,  the vortex lattice resists shear motion and supports collective oscillations known as Tkachenko modes \citep{Tkachenko1966, Baym1983,Haskell2011}.  The
corresponding shear modulus of the vortex array is of order the vortex self-energy per unit length times the vortex areal density \citep{Baym1983,Haskell2011}:
\begin{equation}
 \mu_v
 =
 \frac{\rho_s\kappa^2 n_v}{16\pi}
 =
 \frac{\rho_s\kappa\Omega}{8\pi}.
 \label{eq:vortex_shear_modulus}
\end{equation}
The Tkachenko-wave speed is defined by
\begin{equation}
 c_T^2\equiv\frac{\mu_v}{\rho_s}
 =\frac{\kappa\Omega}{8\pi}.
 \label{eq:Tkachenko_velocity}
\end{equation}
Tkachenko modes, which are transverse, shear oscillations of the vortex lattice, are characteristic collective excitations of a rotating superfluid. These are analogous to lattice phonons in a solid but occur in the vortex array rather than in the nuclear crystal~\citep{Tkachenko1966,Ruderman1970,Sonin2016}.

In the simplest incompressible-fluid limit, Tkachenko modes have an
approximately linear dispersion relation,
\bea
\omega_T
\simeq
c_T k ,
\label{eq:Tkachenko_linear}
\eea
where $k$ is the wave number. However, neutron-star matter is
compressible, and vortex-lattice oscillations couple to ordinary sound
waves, substantially modifying the long-wavelength behavior. A
commonly used interpolation formula is
\bea
\omega_T^2
=
\frac{c_T^2 c_s^2 k^4}
     {4\Omega^2+c_s^2 k^2},
\label{eq:Tkachenko_dispersion}
\eea
where $c_s$ is the sound speed. For $c_s k\gg 2\Omega$ one recovers
the approximately linear Tkachenko spectrum of
Eq.~(\ref{eq:Tkachenko_linear}). In the opposite long-wavelength
regime, $c_s k\ll 2\Omega$, the mode becomes much softer,
\bea
\omega_T
\simeq
\frac{c_T c_s}{2\Omega}\,k^2 .
\label{eq:Tkachenko_soft}
\eea
The soft $k^2$ scaling is particularly important for neutron stars,
because global-scale modes with $k\sim R^{-1}$ can then have periods
of hundreds of days.

The hydrodynamics of neutron stars requires a two-fluid treatment in
which the neutron superfluid coexists with a charged conglomerate of
protons and electrons. The vortex-lattice displacement couples to the
relative motion of these fluids, to compressibility, and to chemical
coupling between composition and pressure perturbations. A further
ingredient is mutual friction, which dissipates relative motion through
the interaction of vortices with the ambient normal component.

\cite{Noronha2008} studied long-wavelength Tkachenko waves in a
two-component setting including mutual friction and shear viscosity.
They found that modes propagating perpendicular to the spin axis are
weakly damped when the coupling between the superfluid and normal
components is small. In the strong-coupling regime, the oscillation
frequencies are reduced, but the modes may still remain weakly damped
for small and moderate values of the shear viscosity.

\cite{Haskell2011} extended the analysis by including compressibility
and chemical coupling in a two-fluid neutron--proton model. These
effects can alter the mode spectrum appreciably.  For rapidly rotating
pulsars with spin frequencies above roughly $100$~Hz, parts of
parameter space no longer support low-frequency Tkachenko modes;
instead, the relevant excitations shift toward modified sound-wave
branches at much higher frequencies.

The damping and survivability of Tkachenko modes therefore probe the
same microphysics that enters glitch recovery and rotational coupling:
mutual friction, shear viscosity, entrainment, and the effective
coupling of the neutron superfluid to the charged component. The
existence of very long-lived Tkachenko oscillations would favor
regions of the star in which dissipation is sufficiently weak and the
vortex lattice remains coherent on global scales.

\subsection{Tkachenko modes as a source of long-term timing variability}

Tkachenko oscillations induce periodic distortions of the vortex
lattice, which in turn modulate the local superfluid velocity and the
spatial density of vortex lines. Since vortex density is directly
related to the coarse-grained rotation of the superfluid, these modes
can generate small oscillatory perturbations of the rotation rate.
Through coupling to the crust and charged component, they may manifest
as quasiperiodic variations in the observable spin frequency and
spin-down rate.

A useful order-of-magnitude estimate follows from
Eq.~\eqref{eq:Tkachenko_linear} in the incompressible limit:
\begin{equation}
c_T \simeq 7.1 \times 10^{-2}\left(\frac{\nu}{10 \mathrm{~Hz}}\right)^{1 / 2} \mathrm{~cm} \mathrm{~s}^{-1} ,
\end{equation}
where $\nu$ is the spin frequency.
For the simple global incompressible estimate $P_T \sim R / c_T$,
\begin{equation}
P_T \simeq 164\left(\frac{R}{10 \mathrm{~km}}\right)\left(\frac{\nu}{10 \mathrm{~Hz}}\right)^{-1 / 2} \text { days. }
\end{equation}
Thus, for canonical neutron-star parameters Tkachenko modes have
periods of the order of months to years. Including compressibility
replaces this estimate by the softer mode 
Eq.~(\ref{eq:Tkachenko_soft}), which can yield even longer
periods, provided the condition $c_sk\ll 2\Omega$ is fulfilled.
These periods naturally overlap with the $100$--$1000$~day
modulations observed in pulsar timing data
\citep{Ruderman1970,Noronha2008,Haskell2011}.

This makes Tkachenko modes an attractive alternative, or complement,
to free precession in explaining long-term rotational variability. The
two mechanisms are physically distinct. Precession is a global wobble
of the stellar rotation axis relative to the body frame and may affect
pulse profile geometry directly. Tkachenko modes are internal
oscillations of the superfluid vortex array and primarily modulate the
exchange of angular momentum between the superfluid and the crust. In
practice their observable signatures may overlap, because both can
produce quasiperiodic variations in pulse arrival times and spin-down
rates. 

Tkachenko modes in a neutron star can exist only if there are large regions in the star in which vortices are unpinned and free to move. The force density in a Tkachenko mode of amplitude $u$ and wavenumber $k$ is $f_T\sim\mu_v k^2\, u=(\mu_v/R)(kR)^2(u/R)$. If vortex pinning occurs with a force per unit length $f_p$, the pinning force density is $n_vf_p$. For $f_p=10^{16}$ dyn cm$^{-1}$ (see Section \ref{subsec:pinning-force}), the ratio is $f_T/(n_vf_p)\sim 10^{-14}(kR)^2(u/R)$; the rigidity force of the vortex lattice is completely swamped by the pinning force.

\subsection{Oscillation modes of superfluid neutron stars}

Superfluidity modifies the entire oscillation mode spectrum of neutron
stars, not only the low-frequency Tkachenko modes discussed above.
The theoretical study of these effects has a rich history, spanning
from early analytic work using the tensor virial method to
state-of-the-art general-relativistic two-fluid calculations.

An analytically tractable entry point to the problem is provided by
the tensor virial method~\citep{Chandrasekhar1969} applied to self-gravitating superfluid
ellipsoids~\citep{Sedrakian2000}.   Earlier work by \cite{Lindblom1994} showed that
superfluid hydrodynamics introduces additional oscillation modes, with
analytical solutions for simplified uniform models revealing modes
with no ordinary-fluid counterpart. For more realistic stellar models,
however, their numerical computations found the lowest-frequency modes
nearly indistinguishable from those of an ordinary fluid, blurring the
separation between ordinary and genuinely superfluid modes.
 This distinction becomes transparent
within the tensor-virial treatment of  oscillation mode spectrum of 
Maclaurin (axisymmetrical), Jacobi (triaxial), and Roche (tidally
affected) ellipsoids, generalized to a
two-fluid system consisting of a neutron superfluid and a normal
fluid, including the new effects of mutual gravitational
attraction and mutual friction between the components~\citep{Sedrakian2000}. 
The tensor virial perturbation equations separate exactly into 
center-of-mass/co-moving and relative/counter-moving sectors 
(for an inviscid normal component),
making the physical nature of the two families explicit. The
oscillation modes of superfluid  spheroids fall naturally
into two generic classes: co-moving modes, in which the two fluids
oscillate in phase, and relative modes, in which they oscillate
counter to one another. In the idealized case of an inviscid normal
component these two sectors decouple  completely. The co-moving modes are
then identical to those of a single-fluid star and are undamped,
whereas mutual friction acts only on the modes associated with relative motion of the two components. In this idealized model, the latter do
not emit gravitational waves because their motion produces no net
mass-current perturbation. Normal-fluid viscosity mixes the co-moving
and relative sectors and modifies the damping and secular stability of
the modes. Although this clean separation relies on the simplifying
assumptions of the ellipsoidal model and is not expected to remain
exact in a realistic stratified and compressible star, the
tensor-virial treatment provides a particularly transparent physical
interpretation of the two-fluid mode spectrum and an important
analytic benchmark for more realistic numerical calculations.

The general-relativistic formalism for superfluid neutron-star
oscillations was developed by \cite{Comer1999}, who
derived the equations governing the linear perturbations of a
two-fluid star in full general relativity and computed quasi-normal
modes numerically for simplified polytropic models.
When the two fluid components have
different adiabatic indices, each frequency of the single-fluid
spectrum splits into two, confirming the existence of a distinct
superfluid mode branch in the relativistic setting. Their analysis
also demonstrated that $w$-modes  --- modes that arise from coupling of fluid motions to the spacetime metric --- are primarily spacetime
oscillations, largely unaffected by the superfluid dynamics.
This framework was subsequently extended by Andersson, Comer, and
Langlois to include an outer envelope of ordinary fluid matter,
junction conditions at the core--envelope interface, and a
systematic study of how entrainment modifies the quasi-normal
mode spectrum, revealing a series of avoided crossings between
ordinary and superfluid branches as the entrainment parameter is
varied \citep{Andersson2002}.

The central structural result established by these studies is that a
two-fluid superfluid star supports two families of fluid pulsation
modes for each multipole, already apparent in the ellipsoidal,
incompressible fluid approximation. The first family --- the ordinary
or co-moving modes --- has the two fluids moving approximately
together, and their frequencies are close to those of a single-fluid
star. The second family --- the superfluid or counter-moving modes ---
has the fluids oscillating out of phase.  In the absence of dissipative effects, such as shear viscosity of the normal fluid, these sectors largely decouple.
 In general, however, the counter-moving branch retains
restoring forces associated with the equation of state, chemical
coupling, and composition gradients. Entrainment shifts its
frequencies and can produce avoided crossings with ordinary modes. An
observationally identified superfluid-mode frequency has the potential to constrain
 the degree of dissipation and entrainment, but the
inference will inherit the uncertainties associated with the equation
of state, stellar composition and pairing microphysics.

\cite{Rau2018} carried out a comprehensive study of compressional
($p$- and $g$-) modes in cold two-superfluid neutron stars, using a
two-fluid formalism that accounts for leptonic buoyancy due to the
presence of muons in the core, with an analogous treatment for the
superfluid inner crust. Their calculation,
performed in the Cowling approximation with full general-relativistic
background, showed that $g$-modes driven by leptonic composition
gradients survive in the superfluid star, with frequencies that
depend sensitively on the stellar mass, the nuclear compressibility,
and the strength of neutron--proton entrainment. For the $p$-mode
spectrum, the two fluids behave as if uncoupled except in the regime
of large entrainment. A particularly significant finding is the
existence of nearly resonant $p$--$g$ mode pairs, which could drive
nonlinear $p$--$g$ instabilities even at zero temperature. Such
instabilities, if present, could excite $g$-modes to large amplitudes
during binary inspiral, producing observable tidal phase shifts in
gravitational-wave signals from neutron-star mergers.

Inertial modes, restored by the Coriolis force, are also split into
ordinary and superfluid branches in a rotating two-fluid star.  The
ordinary $r$-modes are of particular interest because they are driven
unstable by gravitational radiation reaction via the
Chandrasekhar--Friedman--Schutz (CFS) mechanism
\citep{Chandrasekhar1970,Friedman1978}.  Historically, the
gravitational-radiation-driven secular instability of rotating stars
was first identified by Chandrasekhar for Maclaurin spheroids
\citep{Chandrasekhar1970} and later formulated in general terms by
\cite{Friedman1978} using canonical energy and angular momentum.
The same CFS mechanism was subsequently shown by \cite{Andersson1998}
to make $r$-modes generically unstable in rotating relativistic stars.
\cite{Lindblom2000} showed that for $r$-modes undergoing CFS 
instability mutual friction provides an
additional damping channel for these modes: for a typical range of
core superfluid parameters, the characteristic mutual-friction damping
timescale is of order $10^4$~s, far too long to suppress the CFS instability. However, within a small fraction of the allowed parameter
space, mutual friction damping times are short enough ($\lesssim 5$~s) to stabilize the $r$-modes completely. 

At finite temperatures below but near the superfluid transition,
additional temperature-dependent superfluid modes
appear. \cite{Kantor2017}  showed that when muons are
present in the core, an infinite sequence of superfluid $r$-modes
emerges whose frequencies vary with temperature; avoided crossings
between normal and superfluid branches at specific resonance
temperatures lead to strongly enhanced dissipation, substantially
suppressing the $r$-mode instability near those temperatures. This
resonance damping mechanism has been proposed as a possible resolution
of the paradox that rapidly rotating neutron stars in low-mass X-ray
binaries are observed to spin at rates well below the theoretical CFS
instability limit.  The more complete picture of inertial modes in a
rotating superfluid star --- including entrainment, finite-temperature
effects, and the full spectrum beyond $r$-modes --- was worked out by
\cite{Dommes2018}, who derived dispersion relations in the
short-wavelength limit and provided an approximate analytic treatment
of the superfluid $r$-mode.

\subsection{Torsional crust modes and superfluidity}

Torsional oscillations of the neutron-star crust constitute a further
class of modes sensitive to the microphysics of dense matter and, in
particular, to superfluidity. The original motivation came from the
suggestion that magnetar giant flares can excite global seismic
oscillations of the solid crust, with low-order toroidal modes having
frequencies in the range relevant for the observed quasi-periodic
oscillations (QPOs) in soft gamma repeaters~\citep{Duncan1998}.
Subsequent calculations of toroidal shear waves in realistic crust
models showed that the observed low-frequency QPOs can be broadly
consistent with crustal torsional modes, while higher radial overtones
probe the crust thickness, compactness, and magnetic-field corrections
more sensitively~\citep{Piro2005,Samuelsson2007,Watts2007}. The
identification of magnetar QPOs with crustal shear modes has also been
used to constrain the nuclear symmetry energy, the crust thickness, and
even to distinguish ordinary neutron-star crusts from more exotic
compact-star models~\citep{Watts2007,Steiner2009,Sotani2024}. 
If the magnetic field of the stellar interior is highly tangled, the crust becomes dynamically unimportant, and the oscillation frequencies are determined by the effective shear modulus of the tangled field, rather than that of the crust~\citep{Link2016}.

A particularly important superfluid effect is entrainment in the inner
crust. In this region, neutron-rich nuclei form an elastic lattice
immersed in a sea of unbound neutrons. Although these neutrons are
expected to be paired and superfluid, they need not all move
independently of the lattice. In the band-theory description of
dripped neutrons, Bragg scattering from the periodic nuclear potential
leads to nondissipative entrainment, so that the density of conduction
neutrons can be substantially smaller than the density of unbound
neutrons~\citep{Chamel2012a,Chamel2017}. This would increase the
effective inertia associated with the motion of the elastic component
and can shift the frequencies of crustal torsional modes. Relativistic
calculations of axial modes including an elastic crust and an
interpenetrating superfluid component indeed show that the
torsional-mode frequencies depend on the entrainment parameter and on
how much of the inner-crust neutron fluid participates in the motion
of the lattice~\citep{Samuelsson2009,Passamonti2011}.

It now appears that the entrainment effect is not nearly as large as that calculated by \citet{Chamel2012a} and \citet{Chamel2017}.
 Calculations based on superfluid hydrodynamics found much
weaker entrainment than implied by normal-state band theory, because the
neutron superfluid can flow through the nuclear
clusters~\citep{Martin2016}.  \cite{Watanabe2017a} argued that pairing
suppresses band-structure effects and gives new life to crustal glitch
models. More recent Hartree-Fock-Bogoliubov and
linear-response calculations by \cite{Almirante2024,Almirante2025} further
challenge the strong-entrainment picture: for slab and rod phases, and
subsequently for the crystalline phase, they found that the superfluid
density can remain close to the density of unbound neutrons when
pairing and the geometric/off-diagonal contribution to the superfluid
density are included~\citep{Almirante2026a,Almirante2026b}. In this
interpretation, normal-state band theory overestimates the entrainment of
dripped neutrons, and the effective superfluid reservoir of the inner
crust is much larger than in the original estimates by \cite{Chamel2012a,Chamel2017}. Thus, in
crustal asteroseismology the superfluid gap, the effective neutron mass,
and the entrainment coefficient should be regarded as correlated
microscopic inputs rather than as independently fixed quantities. 

The crustal interpretation is further affected by uncertainties in the
deep inner crust. Nuclear pasta phases near the crust–core transition can
soften the elastic response and change the frequencies of shear modes,
especially if the pasta region has a reduced or vanishing shear
modulus~\citep{Gearheart2011,Sotani2011,Tews2017}. These effects are
conceptually distinct from superfluid entrainment, but they are entangled
observationally because both modify the effective shear speed and inertia
of the oscillating crust. Consequently, any attempt to infer neutron-star
parameters from torsional frequencies must control not only the stellar
compactness and crust thickness, but also the symmetry-energy
dependence of the crust composition, the possible extent of pasta
phases, and the superfluid entrainment of dripped neutrons.

In magnetars, however, purely crustal torsional modes are not isolated
normal modes of the star. The magnetic field couples the elastic crust
to the fluid core, where Alfv\'en waves can propagate along
magnetic-field lines. Levin showed that a mechanical crustal mode can
lose energy rapidly by launching Alfv\'en waves into the core and that
the core may support an Alfv\'en continuum at the relevant
frequencies~\citep{Levin2006,Levin2007}. This led to the view that the
observed QPOs should be interpreted as global magneto-elastic
oscillations rather than as purely crust-confined shear modes. Toy
models and relativistic magneto-elastic simulations demonstrated the
importance of continuum edges, turning points, gap modes, resonant
absorption, and phase mixing in shaping the
spectrum~\citep{Glampedakis2006,Sotani2008,Gabler2011,Gabler2012,vanHoven2010,vanHoven2012}. 
In such models, modes whose frequencies lie inside the Alfv\'en continuum
are strongly damped, whereas longer-lived QPOs can be associated with
continuum edges, turning points, or discrete gap modes.

Superfluidity and superconductivity in the core modify this
magneto-elastic problem further. If the core neutrons are superfluid,
only the charged component is directly tied to the magnetic field, while
the neutron superfluid couples to it through entrainment and mutual
friction. This changes the effective Alfv\'en speed and can shift the QPO
spectrum. Calculations of superfluid magnetars have shown that partial
decoupling of the neutron component can leave observable imprints on the
magneto-elastic frequencies and can help produce coherent oscillations
with appreciable surface amplitudes~\citep{Gabler2013,Gabler2016,
Passamonti2016}. The role of proton superconductivity, vortex-flux-tube
interactions, mutual friction, and the density dependence of pairing gaps
remains less settled. These effects determine how efficiently crustal
motion, neutron superfluid motion, and core Alfv\'en dynamics communicate
with each other, and hence whether superfluidity shifts modes into or out
of the Alfv\'en continuum or changes their damping times.

A qualitatively different response to a giant flare is provided by the
global fluid $f$-mode.  Unlike the predominantly axial torsional and
magneto-elastic oscillations discussed above, the lowest-order
quadrupolar $f$-mode is a polar oscillation of the entire star, with a
characteristic frequency of order a few kHz and a strong coupling to
gravitational radiation.  
\cite{Levin2011} compared the excitation of $f$-modes and
torsional modes by the magnetic reconfiguration associated with a
magnetar giant flare.  They found that torsional modes can be excited
efficiently, whereas only a small fraction of the released magnetic
energy is transferred directly to the low-order $f$-modes.  The
suppression is particularly strong for an internal magnetic
rearrangement, whose characteristic Alfv\'en timescale is much longer
than the $f$-mode period, so that the fluid responds approximately
adiabatically.  A rapid reconfiguration of the external magnetosphere
can excite the $f$-mode more effectively, but the resulting
gravitational-wave signal was still found to be weak for canonical
magnetar parameters.  Thus giant flares provide a common excitation
mechanism for low-frequency torsional/magneto-elastic oscillations and
high-frequency global $f$-modes, but the efficiency with which the
available magnetic energy is transferred to the two sectors is very
different.

\subsection*{Open questions and perspectives}

\begin{itemize}

   \item \textit{Precession versus Tkachenko modes as the cause of long-period variations in spin rate.}
    Observed long-period periodicities, together with
    correlated pulse-profile changes in some cases, are consistent with
    both precession and Tkachenko-mode models. Distinguishing between
    them requires quantitative predictions for the pulse-profile
    modulation patterns specific to each mechanism. Developing
    self-consistent models that account for both mechanisms
    simultaneously, and testing them against high-cadence timing data,
    remains an open problem.

    \item \textit{Excitation and saturation of Tkachenko modes.}
    The mechanism whereby Tkachenko modes might be excited in neutron stars has not been studied in detail. 
    Glitches, crustal cracking, or vortex
    avalanches may inject energy into the vortex lattice on short
    timescales, but the subsequent nonlinear evolution, mode
    competition, and saturation amplitude are poorly understood.
    A quantitative theory of Tkachenko mode excitation would allow
    predictions for the modulation amplitude and coherence time of
    long-term timing variations. Tkachenko oscillations are quenched in regions of the star with vortex pinning.

    \item \textit{Superfluid $r$-mode instability and gravitational waves.}
    Whether the CFS instability of $r$-modes is suppressed in
    accreting neutron stars remains unresolved. The answer depends on
    the mutual-friction coefficient, the temperature profile, the
    presence of resonance-damping by superfluid modes at specific
    temperatures, and the equation of state. A firm theoretical
    prediction for the $r$-mode instability window in superfluid stars,
    confronted with the observed spin distribution of neutron stars in
    low-mass X-ray binaries, could provide strong constraints on pairing
    gaps and mutual-friction parameters.

  \item \textit{Magnetar precession and rapid crust–core coupling.}
    The observed long-period variability in magnetars is in tension
    with rapid crust--core coupling in magnetars with
    $B\gtrsim 10^{15}$~G, driven by proton scattering off neutron
    vortex quasiparticles in non-superconducting
    regions~\citep{Sedrakian2016}.
    If this mechanism efficiently 
    damps free precession, any observed long-period precession would require continuous or recurrent driving, for example by ongoing seismic or magnetic activity.
     Identifying such a source, and predicting the expected
    periodicity and amplitude, is an important open problem connecting
    magnetar activity, internal field structure, and rotational
    dynamics.
  \end{itemize}

\section{Quantum vorticity in quark matter}
\label{sec:quark_vorticity}

\begin{figure}[t]
  \centering
  \includegraphics[width=.47\linewidth]{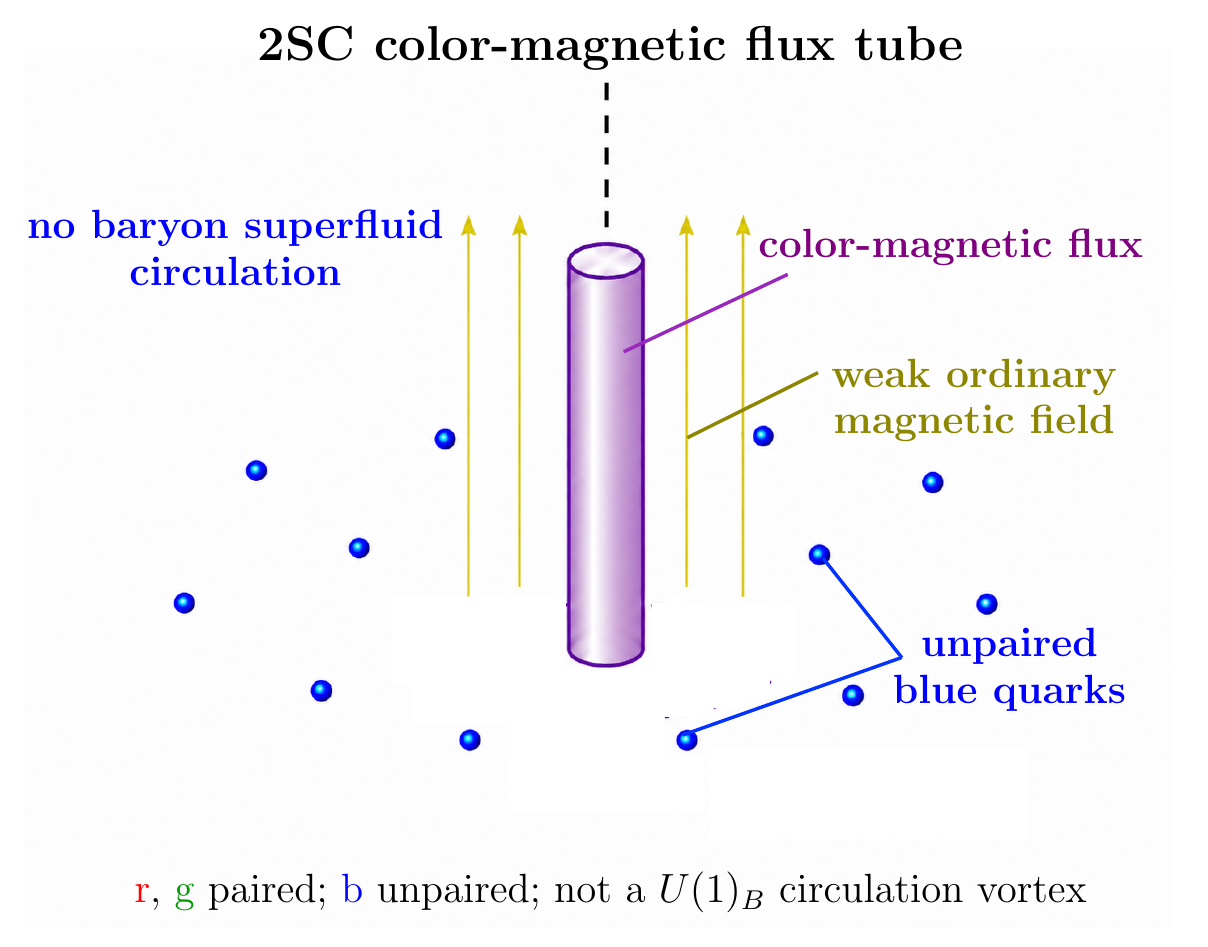}
  \includegraphics[width=.49\linewidth]{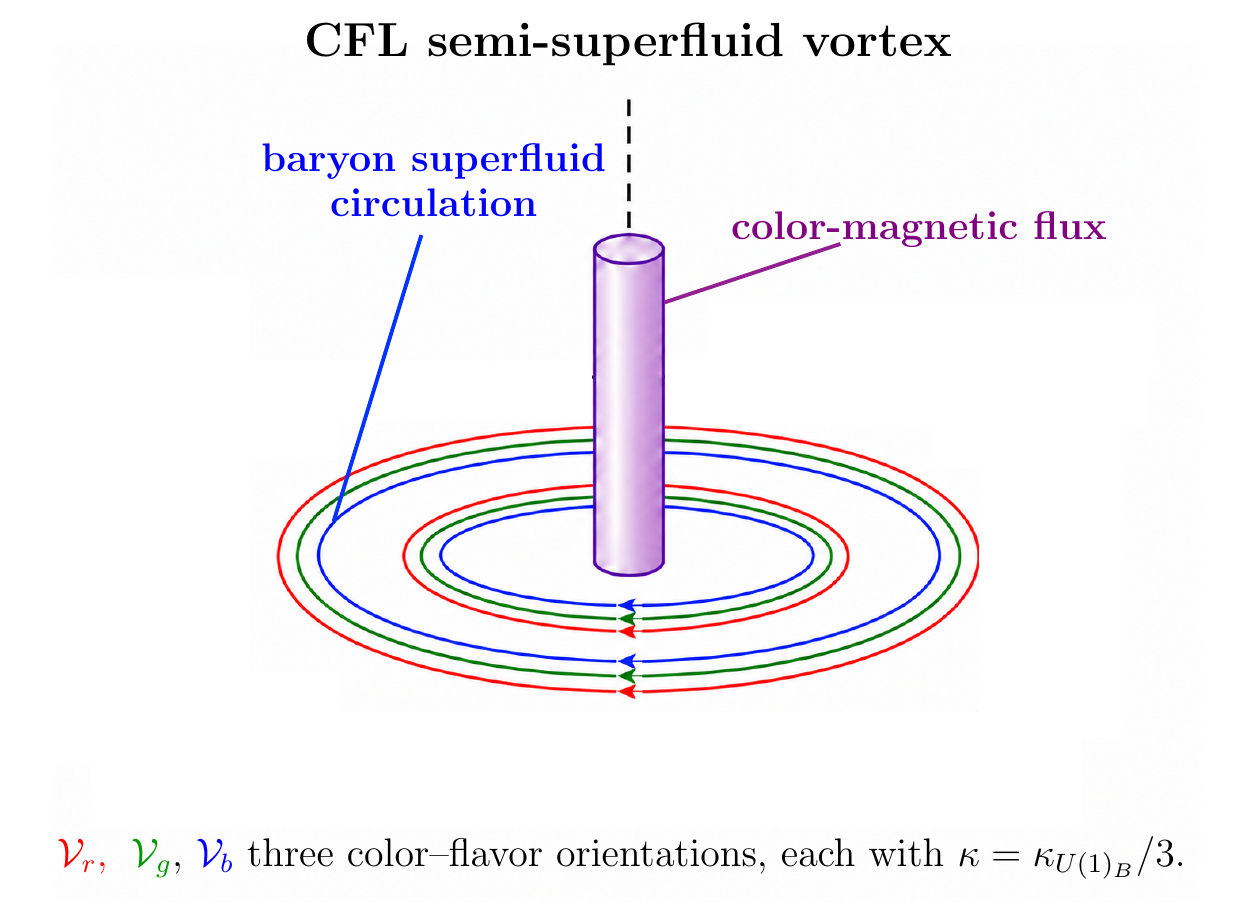}
  \caption{Schematic comparison of line defects in the 2SC and CFL phases. {\it Left:} A 2SC color-magnetic flux tube. Conventional 2SC matter is not a baryon superfluid, so the defect carries no  quantized baryon circulation; it contains predominantly color-magnetic flux with a smaller ordinary magnetic component. The flux tube is not topologically protected, i.e.,  
  it can in principle be continuously unwound through a rearrangement of the condensate and gauge fields. Once formed, however, an energy barrier may make it long-lived in a metastable state.  {\it Right:} A minimal CFL
    semi-superfluid vortex. CFL is a baryon superfluid, and its
    elementary vortices carry both color-magnetic flux and one third
    of the circulation of a conventional $U(1)_B$ vortex. The labels
    $\mathcal V_r$, $\mathcal V_g$, and $\mathcal V_b$ denote three
    possible color--flavor orientations of the elementary vortex. The CFL vortex
     is topologically protected: the phase winds
nontrivially around the vortex, and this winding cannot be removed by any smooth local deformation of the condensate.     
     }
\label{fig:topological_defects_in_QCD}
\end{figure}

If deconfined quark matter occurs in the inner cores of mature neutron
stars, it is expected to be sufficiently cold that attractive
quark--quark interactions favor paired, color-superconducting
states~\citep{Alford2008}.  The possible line defects then depend
strongly on the pairing pattern.
Figure~\ref{fig:topological_defects_in_QCD} summarizes the two best
studied possibilities, the two-flavor color-superconducting (2SC)
phase and the color-flavor-locked (CFL) phase.  The most important
distinction for neutron-star rotation is simple: conventional 2SC
matter is not a baryon superfluid, whereas CFL matter is.
Consequently, the line defect shown in the 2SC panel is primarily a
magnetic flux tube, while the CFL defect is a genuine rotational
vortex carrying quantized superfluid circulation.

Color superconductivity differs from an ordinary metallic
superconductor because quarks carry both electric charge and color
charge.  The paired medium therefore responds to both electromagnetic
and gluonic fields (the latter mediating the strong force).  In a color superconductor the ordinary photon
mixes with a gluon: one linear combination remains long ranged, while
the orthogonal combination is screened by the condensate
\citep{Alford2000}.  We refer to localized flux of this screened mixed
field as \emph{color-magnetic flux}.  Because the screened field also
contains a small electromagnetic component, a color-magnetic flux tube
can carry a weaker ordinary magnetic field, as indicated in the left
panel of Fig.~\ref{fig:topological_defects_in_QCD}.  The long-ranged
combination is usually called the rotated electromagnetic field.
Thus, ordinary stellar magnetic fields need not be expelled from quark
matter in the same way as from a conventional electromagnetic
superconductor.

For a genuine superfluid component --- here the CFL
phase --- the circulation is quantized,
\begin{equation}
  \oint_C \bm p\cdot d\bm\ell=2\pi n\hbar,
  \qquad n\in\mathbb Z,
\end{equation}
where $\bm p$ is the canonical momentum appropriate to the phase.  In
a rotating star the macroscopic angular velocity is then reproduced,
on average, by an array of quantum vortices, as discussed for baryonic matter
in Section~\ref{sec:vortex_creep}.  With this qualitative picture in
mind, we discuss the 2SC and CFL phases separately below and return
afterward to the symmetry-breaking patterns that explain their
different defect structures.

\subsection{2SC phase}
\label{subsec:2SC_vorticity}

In the 2SC phase, up and down quarks of two colors, conventionally
called red and green, form Cooper pairs, while the blue quarks remain
unpaired in the simplest version of the phase. Electrons  maintain electric neutrality. The resulting condensate is a
color superconductor, but conventional 2SC matter is \emph{not} a
baryon superfluid. It therefore does not need an array of quantized
circulation vortices in order to rotate.

It can, however, support magnetic flux structures. Because
electromagnetic and color fields mix in the superconducting state,
these defects carry predominantly color-magnetic flux together with a
smaller ordinary magnetic component, as shown in the left panel of
Fig.~\ref{fig:topological_defects_in_QCD}. More formally, one linear
combination of the photon and a gluon remains unscreened
(rotated electromagnetic field), whereas the
orthogonal combination acquires a Meissner mass \citep{Alford2000}.
Color-magnetic flux tubes associated with this massive combination
have been studied in Ginzburg--Landau descriptions of 2SC matter
\citep{Bailin1984,Sedrakian2001,Alford2010}.

Whether such flux tubes are present depends on the magnetic state of
the 2SC medium. If the massive mixed sector is type II, the
corresponding lower critical field can be very large, of order
$10^{17}$~G in representative estimates \citep{Alford2010}. This does
not necessarily exclude flux tubes from neutron-star cores. During the
transition into the 2SC phase, magnetic flux may become trapped between
growing superconducting domains. Subsequent compression of the residual
normal regions can locally raise the field above the threshold for
flux-tube formation, leading to Abrikosov-like color-magnetic tubes.
Their abundance, stability, and spatial organization are not yet
determined from first principles, but they provide a natural mechanism
by which a 2SC core can retain magnetic substructure.

These flux tubes may be dynamically important even though they are not
rotational vortices. The 2SC phase contains gapless or weakly gapped
fermionic excitations, including electrons and unpaired blue quarks.
When such particles encircle a color-magnetic tube, they acquire
Aharonov--Bohm phases. Aharonov--Bohm scattering of these charged excitations from the flux tubes produces a drag force
on moving flux tubes and modifies transport in the quark core
\citep{Alford2010}. Their mobility, coupling to normal excitations, and
collective drift under magnetic stresses may therefore influence
magnetic-field evolution over stellar timescales. At a qualitative
level, this is analogous to the frictional coupling of electrons to
proton flux tubes in a hadronic superconducting core, although the
underlying fields are different.

\subsection{CFL phase}
\label{subsec:CFL_vorticity}

The CFL phase is qualitatively different. At sufficiently high density,
up, down, and strange quarks of all three colors participate in
pairing. In the idealized limit all quark quasiparticles are gapped and
the phase is electrically neutral without a large electron population
\citep{Alford2008}. Most importantly here, the CFL condensate is a
baryon superfluid. A rotating CFL core must therefore contain quantized
vortices \citep{Eto2014}.

The elementary CFL vortex is, however, more complicated than a neutron
vortex. Its baryon-phase winding is fractional, with the remaining
winding compensated by a color gauge transformation.
Consequently, the vortex carries both
baryon-superfluid circulation and localized color-magnetic flux, as
illustrated in the right panel of
Fig.~\ref{fig:topological_defects_in_QCD}. Such defects are commonly
called \emph{non-Abelian semi-superfluid vortices}
\citep{Balachandran2006,Eto2014}. The term ``non-Abelian'' refers to
their internal color--flavor orientation; for the present discussion
the essential point is the coexistence of quantized circulation and
color-magnetic flux. Because the circulation is tied to a nontrivial
phase winding, the CFL vortex is topologically protected, like the superfluid vortices described in Sec.~\ref{superfluid-rotation}.

An ordinary $U(1)_B$ vortex, i.e., one obtained by winding only the
overall baryon-number phase of the CFL condensate by $2\pi$, with no
accompanying color-gauge winding, carries one full circulation
quantum. It is, however, energetically disfavored relative to three
elementary CFL vortices: non-Abelian (``semi-superfluid'') vortices
that each carry only a fraction of the circulation,
\begin{equation}
 \kappa_{\rm CFL}=\frac{1}{3}\kappa_{U(1)_B},
 \label{eq:fractional_CFL_circulation}
\end{equation}
together with a compensating color-magnetic flux. Because vortex
tension scales as the square of the winding number, three such
$1/3$-quantized vortices cost only a third of the energy of a single
$n=1$ ordinary vortex; a singly quantized Abelian vortex can therefore
lower its energy by splitting into this triplet, whose circulations
add up to the original value \citep{Nakano2008,Alford2016}. The
fundamental rotational defect in CFL matter is thus neither purely
hydrodynamic nor purely magnetic, but a composite object.

A rotating CFL core is expected to form a lattice of these
semi-superfluid vortices \citep{Balachandran2006,Sedrakian2008}. Their
long-range interaction is repulsive, as in ordinary superfluids, so an
approximately regular array is favored in equilibrium
\citep{Nakano2008}. Their color-magnetic flux is localized near the
vortex cores and should not be confused with the macroscopic magnetic
field associated with the unscreened rotated electromagnetic field.
CFL matter may therefore allow the long-ranged rotated field to
penetrate the bulk while still containing localized color-magnetic
structures tied to its rotational vortices.

A central issue for hybrid stars is how hadronic vortices connect to
vortices in a CFL core. If the hadronic outer core contains neutron
superfluid vortices and proton magnetic flux tubes, then a
hadron--quark interface must rearrange both circulation and magnetic
flux. Early analyses proposed ``colorful boojums,'' junction regions
at which several hadronic defects merge and convert into color-carrying
CFL vortices \citep{Cipriani2012}. Later work found that, under
appropriate assumptions, the circulation of a single hadronic vortex
can match that of a single non-Abelian CFL vortex, allowing direct
vortex continuity without a boojum
\citep{Chatterjee2019,Alford2019}. The precise interface structure
therefore depends on the realization of quark--hadron continuity,
color-flux neutrality, and any intermediate phases between nuclear and
CFL matter.

These interface questions are not only topological. Vortex
connectivity determines how angular momentum is transmitted between a
hadronic shell and a quark core. If CFL vortices are pinned, strongly
impeded at the interface, or forced to reorganize into junction
structures, they may affect long-term spin evolution, post-glitch
relaxation, and the coupling of the inner core to the rest of the star
\citep{Cipriani2012,Chatterjee2019,Alford2019}. At present these
implications remain qualitative, but they identify vortex dynamics in
CFL matter as a potentially important ingredient in the rotational
phenomenology of hybrid stars.

The physical distinction between the two panels of
Fig.~\ref{fig:topological_defects_in_QCD} can be stated more formally
through the corresponding symmetry-breaking patterns. Neglecting
electromagnetism and small quark masses, the 2SC phase has schematically
\be
SU(3)_C\times SU(2)_L\times SU(2)_R\times U(1)_B
\longrightarrow
SU(2)_C\times SU(2)_L\times SU(2)_R\times U(1)_{\widetilde B}.
\ee
Here $SU(3)_C$ is the gauge symmetry of QCD that rotates the three
quark colors into one another; $SU(2)_L\times SU(2)_R$ is the chiral
flavor symmetry of the two light quarks, acting independently on
left- and right-handed $u$ and $d$ fields (exact only in the
massless-quark limit); and $U(1)_B$ is the global symmetry whose
conserved charge is baryon number. The 2SC condensate pairs quarks of
only two of the three colors, leaving a residual $SU(2)_C$ that
freely rotates the two paired colors into each other while treating
the third as distinct. It also breaks both $U(1)_B$ and the color
generator $T_8=\lambda_8/2$ individually, with $\lambda_8 =
\mathrm{diag}(1,1,-2)/\sqrt3$ the diagonal Gell-Mann matrix
distinguishing the third color from the other two, but leaves
invariant one particular combination of the two,
$\widetilde B \equiv B - \tfrac{2}{\sqrt3}\,T_8$; this surviving
symmetry is the $U(1)_{\widetilde B}$ appearing above. A baryon-like
$U(1)$ symmetry therefore remains unbroken, which is the formal reason
why conventional 2SC matter is not a baryon superfluid.

For three approximately massless flavors, the CFL pattern is
\be
SU(3)_C\times SU(3)_L\times SU(3)_R\times U(1)_B
\longrightarrow
SU(3)_{C+L+R}\times Z_2 .
\ee
Here baryon number is broken, and CFL matter is consequently a
superfluid.

In the ideal CFL limit, the elementary CFL vortex can occur in different internal color--flavor
orientations, conventionally denoted by $\mathcal V_r$, $\mathcal V_g$,
and $\mathcal V_b$. These symbols label three equivalent orientations
of the same type of minimal vortex, rather than three different
circulation quanta. Each has
\be
 \kappa_{\mathcal V_r}
 =\kappa_{\mathcal V_g}
 =\kappa_{\mathcal V_b}
 =\frac{1}{3}\kappa_{U(1)_B},
\ee
while the associated color-magnetic flux has a different internal
orientation \citep{Balachandran2006,Nakano2008,Eto2014}. Thus the three
labels in Fig.~\ref{fig:topological_defects_in_QCD} should be read as
alternative realizations of one elementary CFL vortex. A full Abelian
$U(1)_B$ vortex may split into three such vortices, whose fractional
circulations add to one circulation quantum.

\subsection{Alternatives}
\label{subsec:quark_vorticity_alternatives}

The idealized 2SC and CFL phases do not exhaust the possible pairing
patterns of dense quark matter at neutron-star densities. Both
phases are subject to physical stresses that disfavor simple,
homogeneous pairing: Fermi-momentum mismatches among the $u$, $d$,
and $s$ quarks, induced by the strange-quark mass, together with the
additional requirements of $\beta$ equilibrium and charge neutrality.

More exotic two-flavor phases can alter the simple 2SC picture. If
additional condensates form on top of the conventional 2SC state and
break baryon-number symmetry, rotational vortices may appear. Studies
of two-flavor quark--hadron continuity have discussed superfluid
two-flavor phases supporting non-Abelian Alice strings or related
vortex configurations~\citep{Fujimoto2021}. Such states interpolate
between conventional 2SC matter and genuinely superfluid quark phases,
but their realization in neutron-star matter remains model dependent.

Another important possibility is \emph{crystalline color
superconductivity}, in which Cooper pairs carry nonzero momentum and
the gap parameter varies periodically in space~\citep{Anglani2014}.
Such phases are simultaneously superfluid and rigid: they break
baryon-number symmetry, allowing rotational vortices, while also
breaking translational symmetry and generating a crystalline
condensate structure.

This coexistence of superfluidity and rigidity has direct implications
for quantum vorticity. Rotational vortices embedded in a crystalline
color superconductor may be pinned by the spatial modulation of the
pairing gap, especially along lines or surfaces where the condensate is
already suppressed. Estimates of the shear modulus indicate that some
crystalline phases could be much more rigid than the conventional
neutron-star crust, and rough estimates suggest that vortex pinning can
be substantial~\citep{Mannarelli2007,Anglani2014}. These observations
motivated the proposal that a crystalline quark-matter shell could
participate in glitch-like phenomena by storing and suddenly releasing
angular momentum. The detailed structure of vortices in realistic
crystalline phases is far from being fully understood, however, and a
quantitative glitch theory based on quark-matter pinning has not yet
reached the level of maturity of crustal-vortex models.

A second important class of alternatives arises from meson condensation
inside CFL matter. In particular, a CFL+$K^0$ phase may be favored when
stress from the strange-quark mass is relieved by kaon condensation. The
additional breaking of a global symmetry permits vortices associated
with the kaonic condensate. Such vortices can carry electric charge and
may become superconducting along their cores, allowing current-carrying
loops or ``vortons'' in certain circumstances~\citep{Kaplan2002}.
Although their astrophysical abundance is uncertain, they provide an
example of how the topological defect content of quark matter can become
considerably richer once secondary condensates are included.

Spin-one color-superconducting phases offer a further possibility,
especially when conventional cross-flavor pairing is disfavored by
large Fermi-surface mismatches. In the color-spin-locked state,
same-flavor quarks pair in a spin-one channel, and model studies
suggest that all quarks may acquire gaps, albeit much smaller than in
the dominant spin-zero phases~\citep{Alford2003,Alford2008}. The
vortex content and magnetic response of these phases depend
sensitively on the specific symmetry-breaking pattern. Their possible
impact on compact-star physics can be significant, particularly because
some spin-one phases may alter magnetic screening~\citep{Schmitt2003} and suppress
otherwise rapid quark direct-Urca cooling~\citep{Schmitt2006}. Their rotational defect
structure is less established than in CFL matter, but they remain
plausible candidates for nonstandard vorticity in quark cores of
hybrid stars.

To summarize, if quark matter cores exist in compact stars, there may
be a multitude of quantum line defects with quite different structure
and physical properties. This is in contrast to the nucleonic core
where the physics of quantum vorticity is better
established. Thus, in ideal 2SC matter the primary line defects are
color-magnetic flux tubes rather than rotational
vortices~\citep{Alford2010}; in CFL matter the fundamental rotational
defects are non-Abelian semi-superfluid vortices carrying both
circulation and color flux~\citep{Eto2014}; and in less symmetric
phases, such as crystalline or meson-condensed color superconductors,
additional families of vortices and new pinning mechanisms may
emerge~\citep{Mannarelli2007,Kaplan2002,Anglani2014}. The resulting
defect networks can influence angular-momentum transport,
magnetic-field and thermal evolution, dissipative coupling, and
potentially the timing phenomenology of hybrid stars.

\subsection*{Open questions and perspectives}

\begin{itemize}

\item \textit{The ground state of stressed two-flavor quark matter.}
At neutron-star densities, the strange-quark mass, $\beta$ equilibrium,
and electric and color neutrality produce mismatches between the quark
Fermi surfaces. It remains unclear whether the resulting state is
conventional 2SC, a gapless or crystalline phase, or a phase with
secondary pairing of the nominally unpaired quarks. Since ideal 2SC
does not break baryon-number symmetry, an important question is whether
the actual stressed phase realized in stars is a superfluid and, if so,
what rotational vortices it supports.

\item \textit{Microscopic dynamics of quark-matter defects.}
The tensions, core structures, interaction energies, and transport
coefficients of 2SC color-magnetic flux tubes and CFL semi-superfluid
vortices are still poorly constrained. Analogues of the problems
encountered for nucleonic vortices---pinning, drag, creep, cutting,
reconnection, and excitation of internal vortex modes---should be
derived microscopically for quark matter and incorporated into a
coarse-grained hydrodynamic description.

\item \textit{Vortex continuity at the hadron--quark interface.}  It
  is not known whether hadronic neutron vortices connect individually
  to quark vortices, or whether several defects must merge and
  reorganize through boojum-like junctions. A consistent treatment
  must conserve circulation and satisfy the gauge-flux matching and
  color-neutrality conditions appropriate to the phases on the two
  sides.

\item \textit{Macroscopic and observable consequences.}
A global theory is needed to determine whether quark-matter defects can store  angular momentum, participate in glitches, support Tkachenko-like collective modes, or influence magnetic-field evolution
and long-term rotational coupling. Connecting microscopic defect
dynamics to cooling, timing irregularities, and post-glitch relaxation
would provide possible observational tests of color-superconducting
matter in neutron-star cores.

\end{itemize}

\section{Conclusions and outlook}
\label{sec:conclusions}

Pairing in neutron-star matter gives rise to a hierarchy of quantum
phenomena whose consequences extend from microscopic energy gaps to
stellar-scale rotational dynamics. The basic pairing pattern is broadly
established: $^1S_0$ neutron superfluidity is expected in the inner crust
and low-density outer core, proton $^1S_0$ superconductivity in part of
the core, and neutron triplet pairing at higher density. Nevertheless,
the quantitative values of the relevant gaps remain uncertain because of
medium polarization, self-energy effects, and the incomplete knowledge of
the interaction in high-density partial waves. These uncertainties feed
directly into neutron-star cooling, transport, and dynamical coupling.

The rotational response of these paired phases is governed by topological
defects. Neutron vortices provide the superfluid with angular momentum,
while their pinning, unpinning, and thermally-activated creep enable
dissipative exchange with the crust and charged component. Proton
superconductivity adds magnetic flux tubes in type-II regions, or
alternating normal and superconducting domains in a type-I phase.
Interactions among vortices, flux tubes, and phase boundaries are central
to models of mutual friction, glitch dynamics, magnetic-field evolution,
and the possible persistence of long-period precession.

Several newer developments broaden this picture. The possibility of a
Josephson effect at an $S$-wave--$P$-wave neutron-superfluid interface
connects pairing-phase structure with charged currents, radiation, and
late-time heating. If vortex pinning is absent in large regions 
of a neutron star,  Tkachenko modes of the vortex lattice provide an
alternative or complementary route to long-term timing variability,
alongside free precession, but their excitation, damping, and nonlinear
survival in realistic neutron stars remain open questions.

If deconfined quark matter is present, its defect content depends
strongly on the pairing pattern. Conventional 2SC matter supports
color-magnetic flux structures but is not, by itself, a baryon
superfluid, whereas CFL matter breaks baryon number symmetry and
rotates through non-Abelian, semi-superfluid vortices carrying both
circulation and color flux. How these defects connect to hadronic
neutron vortices and proton magnetic structures at a phase boundary
remains unresolved and may affect the rotational coupling of
hybrid-star cores.

Future progress requires further integration of microscopic many-body
theory, mesoscale vortex and flux-tube dynamics, and global multifluid
stellar modeling. Improved calculations of pairing gaps, entrainment,
pinning energies, and dissipative coefficients are needed, as are
simulations that connect these quantities to glitches, precession,
thermal evolution, and magneto-rotational coupling. Neutron stars thus
remain exceptional laboratories in which nuclear pairing phenomena,
normally studied at femtometer scales, reveal themselves through
observable variations in spin and emission over astrophysical times.

\section*{Acknowledgements}

This article is, in part, based upon work from COST Action SCALES,
CA24139, supported by COST (European Cooperation in Science and
Technology).  A.~S.  has been supported by the Polish NCN Grant
No. 2023/51/B/ST9/02798 and, in part, by the collaborative research
Grant No. 24RL-1C010 provided by the HESC of the Republic of Armenia.
B.~L. is supported by United States NSF Grant No. 2607233.

\bibliographystyle{Harvard}
\input{article_v5.bbl}

\end{document}